\documentclass[prb,twocolumn]{revtex4-1}

\usepackage{esint,bm}

\usepackage{hyperref}

\begin{document}



\title {Orbital Magnetization in Solids: Boundary contributions as a non-Hermitian effect}



\author{K. Kyriakou}
\author{K. Moulopoulos}


\affiliation{University of Cyprus, Department of Physics, 1678, Nicosia, Cyprus}


\date{\today}

\begin{abstract}
The theory of orbital magnetization is reconsidered by defining additional quantities that incorporate a non-Hermitian effect due to anomalous operators that break the domain of definition of the Hermitian Hamiltonian. As a result, boundary contributions to the observable are rigorously and analytically taken into account. In this framework, we extend the standard velocity operator definition in order to incorporate an anomaly of the position operator that is inherent in band theory, which results in an explicit boundary velocity contribution. Using the extended velocity, we define the electrons' intrinsic orbital circulation and we argue that this is the main quantity that captures the orbital magnetization phenomenon. As evidence of this assertion, we demonstrate the explicit relation between the $n \text{th}$ band electrons' collective intrinsic circulation and the approximated, evaluated with respect to Wannier states, local and itinerant circulation contributions that are frequently used in the modern theory of orbital magnetization. A quantum mechanical formalism for the orbital magnetization of extended and periodic topological solids (insulators or metals) is re-developed without any Wannier-localization approximation or heuristic extension [Caresoli, Thonhauser, Vanderbilt and Resta, Phys. Rev. B $\mathbf{74}$, 024408 (2006)]. It is rigorously shown that, as a result of the non-Hermitian effect, an emerging covariant derivative enters the one-band (adiabatically deformed) approximation $\mathbf{k}$-space expression for the orbital magnetization. In the corresponding many-band (unrestricted) $\mathbf{k}$-space formula, the non-Hermitian effect contributes an additional boundary quantity which is expected to give locally (in momentum space) giant contributions whenever band crossings occur along with Hall voltage due to imbalance of electron accumulation at the opposite boundaries of the material.
\end{abstract}


\maketitle


\section{Introduction} \label{s1}
Boundary effects are ubiquitous in condensed matter systems. However, how these effects influence bulk quantities such as the bulk orbital magnetization \thinspace \textbf{M} \thinspace seems to be still unclear \cite{{b1},{b2},{b3}}. Circular dichroism measurements have confirmed the existence of surface states with non-trivial orbital moment textures in \textbf{k}-space \cite{{b4},{b5}} due to Orbital Rashba Effect \cite{b6}, while gigantic orbital magnetization values are predicted to occur in the vicinity of band crossings at the surfaces of sp alloys \cite{b7}. A simple and direct method to link boundary properties with bulk quantities, if found, would conceptually give a direct realization of a bulk-boundary correspondence in a general sense. Hints of such a link have appeared but they have not yet been combined in a single theoretical framework for condensed matter systems. In the chemists' community the link between boundary effects and  \textquotedblleft{bulk}\textquotedblright \thinspace quantities seems to have been studied in detail and is formalized as surface integrals (fluxes) of certain generalized currents in the so-called atomic theorems \cite{{b8},{b9},{b10}} that determine atom properties viewed as parts (fragments) of a molecule; for example, the atomic dielectric polarization \cite{b11} and atomic magnetic susceptibility \cite{b12} have been determined within that method. In the mathematical physics community the connection between boundary effects and bulk quantities can be attributed to anomalous operators that break the domain of definition of the Hamiltonian operator, thereby leaving residues either in the Ehrenfest theorem \cite{{b13},{b14},{b15}} or in the Hellmann-Feynman \cite{b16} \thinspace theorem; these can be converted into space coordinate surface integrals (for 3D systems) over the system's boundaries. In this paper we rigorously take into account these boundary residues as non-Hermitian effects in order to model the boundary contributions to the orbital magnetization of non-interacting electrons. 

In general, anomalous operators act on states that belong within a given Hilbert space, where the Hamiltonian is assumed Hermitian and the system is closed, and they produce states that are outside this given Hilbert space; this leads to emergent non-Hermiticity in the Hamiltonian which is precisely the above mentioned boundary residue.

One of the most common examples of such an anomalous operator (that leaves a boundary residue in the Ehrenfest theorem) is the position operator \thinspace $\mathbf{r}$ \thinspace whenever periodic boundary conditions at the ends of the system are adopted for the wavefunctions. In Solid State Physics one usually bypasses this kind of anomaly as in \thinspace Ref.\onlinecite{b17} \thinspace by redefining a proper (periodic) operator for the electrons' position that does not leave any boundary residue and by working with its expectation value. In this work we deal with this problem in a direct way, that is we maintain the standard electrons' expectation value $\left\langle\mathbf{r}\right\rangle$ as defined within the Schr\"{o}dinger picture (despite the fact that the electrons' position expectation value $\left\langle\mathbf{r}\right\rangle$ becomes undefined within the Bloch representation in the thermodynamic limit, its displacement $\Delta\!\left\langle\mathbf{r}\right\rangle$ after a finite time interval is always a well-defined quantity as shown in Appendix \ref{a1}) and simply extend the standard velocity operator by adding to it an extra operator term that takes into account the non-Hermitian effect of the Hamiltonian operator. The expectation value of the added operator term is determined entirely from the boundaries of the system and it rigorously gives a boundary velocity contribution for the electron (although formalized in a bulk framework). 

Therefore, having in mind the evolution of the quantum state under consideration as well as the position operator expectation value within the Schr\"{o}dinger picture, we are led to define the velocity operator in an extended form as, ${\mathbf{v}_{ext}=\mathbf{v}+\mathbf{v}_{b}}$ \thinspace where \thinspace $\mathbf{v}$ \ is the standard velocity operator as given in the literature (which can be viewed as a bulk property) and \thinspace $\mathbf{v}_{b}$ \thinspace is the added boundary operator term that takes into account the non-Hermitian effect. In this fashion, the extended velocity operator expectation value $\left\langle\mathbf{v}_{ext}\right\rangle$ is always equal to the rate of change of the electrons' position expectation value \ ${\displaystyle{\left\langle {\mathbf{v}_{ext}}\right\rangle = \frac{d}{dt}\!\left\langle {\mathbf{r}}\right\rangle}}$ \ irrespectively of the system's size or the boundary conditions to be imposed on the wavefunction; it should be noted that the latter equality is not guaranteed if the boundary velocity operator is not taken into account, and this has been the source of paradoxes\cite{b15}.

The above boundary velocity \thinspace ${ \mathbf{v}_b }$ \thinspace expectation value, can be used as a probe with respect to transport properties that are carried by the system's boundaries. However, although the boundary velocity expectation value \thinspace ${ \left\langle \mathbf{v}_b \right\rangle }$ \thinspace  is well defined and not zero within Bloch representation in the thermodynamic limit, the expectation value of certain observable operators involved
 in orbital magnetization calculations in the literature,
can be undefined, e.g. the position operator expectation value $\left\langle\mathbf{r}\right\rangle$ and the circulation operator expectation value $\left\langle\mathbf{r} \times \mathbf{v} \right\rangle$. Such subtle behaviors, as well as relevant consequences with respect to the modern theory of orbital magnetization, are presented in Sec.\ref{s2c} and summarized in Table \ref{T1}.

Orbital magnetization is the quantity to be crucially affected by the above non-Hermitian effect and it is this observable that is the focus of our treatment. Before we start, let us note that, although in conventional materials the orbital magnetization is only of the order of a few per cent of the total magnetization, in materials with topologically nontrivial band structures the electrons' collective orbital magnetization can be larger than spin magnetization which has been confirmed in experiments \cite{{b19},{b20},{b21}}, owing to large orbital magnetization contribution arising from the effective reciprocal space monopoles near the band crossings.

Nowadays, the so-called modern theory of orbital magnetization \thinspace $\mathbf{M}$ \thinspace has been argued to have come to a mature stage \cite{b18}. Three main methods for deriving the bulk orbital magnetization formula in the context of modern theory are currently widespread: a quantum mechanical method with direct calculation of circulating currents for trivial band insulators in the presence of boundaries \cite{{b22},{b23}}, a semiclassicall wave packet approximation method \cite{{b24},{b25},{b26}} and one that takes the derivative of free energy with respect to magnetic fields under periodic boundary conditions \cite{{b27},{b28}}. In the first of the above methods two incompatible features had to be overcome in order for the magnetization to be a genuine bulk property, namely adoption of periodic boundary conditions (PBCs) and usage of the circulation operator \thinspace $\mathbf{r}\times\mathbf{v}$ \thinspace in the Bloch representation. This was done with the aid of the Wannier representation which can be rigorously employed in normal insulators with zero Chern number. 

Furthermore, it has been argued that bulk behavior of observables in crystalline materials is ensured when computing within PBCs. In spite of this belief, and contrary to what has been stated in the literature \cite{{b29},{b30}}, the system by construction has a \textquotedblleft{terminated}\textquotedblright  \thinspace boundary surface (assuming a 3D material), the one on which PBCs are imposed; boundary contributions due to non-Hermitian effects are therefore generally not ruled out, especially whenever observables incorporate anomalous operators, such as the position operator that enters the expressions for the electron's magnetic and electric dipolar moment.

In the spirit of re-examining the orbital magnetization formula within a quantum mechanical theoretical framework that takes into account boundary effects and at the same time relaxes the Wannier-localization approximation, we were motivated to define a circulation operator that contains the extended velocity operator in the form \thinspace $\displaystyle{\frac{1}{2}(\mathbf{r}\times\mathbf{v}_{ext}-\mathbf{v}_{ext}\times\mathbf{r}})$, in order to analytically determine the orbital magnetization of a system of effectively non-interacting electrons (i.e. in a density functional theory framework). Although this circulation operator takes into account boundary contributions as a consequence of the extended velocity operator  \thinspace $\mathbf{v}_{ext}$, its expectation value is still problematic in the Bloch representation within PBCs and it becomes undefined for extended systems in the thermodynamic limit (see Appendix \ref{a1}). 

In spite of the undefined expectation value of the latter circulation operator in periodic and extended systems, and to our surprise, we found out that it can always be decomposed into two distinct parts, namely, an intrinsic one that has a definite value and an extrinsic one that carries the undefined value. The intrinsic one has an intensive and bulk behavior that properly counts the local and circulating probability micro-currents embodied in the (generally) extended wavefunction's structure with boundary contributions being explicitly taken into account due to  the non-Hermitian effect. 

Specifically, the expectation value of the intrinsic orbital circulation is found to have the following properties: (i) it does not depend on the system's size and has a finite value within PBCs in the Bloch representation in the thermodynamic limit, (ii) it carries information about the electrons' orbital circulating probability micro-currents which are encoded as structured wavefunction in real space (for free electrons and plane waves it becomes zero), (iii) its value does not depend on the position origin (as long as the shift of the origin can be attributed to a unitary transformation of the wavefunction) and (iv) it takes into account boundary contributions as a consequence of the non-Hemitian effect.

Although we do not use any Wannier states in this work, we nevertheless demonstrate how an explicit relation between the electrons' $n \text{th}$ band collective intrinsic circulation (evaluated with respect to Bloch eigenstates) and a starting point formula of the modern theory of orbital magnetization (namely, the electrons' collective circulation evaluated with respect to Wannier states) can be established. This is accomplished by using the standard velocity, the newly defined boundary velocity and the intrinsic circulation and by assuming that each Bloch eigenstate satisfies the periodic gauge. In this respect, we expand each Bloch eigenstate into the basis of localized bulk Wannier states and localized surface orbitals, and as a result the $n \text{th}$ band electrons' collective intrinsic circulation (initially evaluated with respect to Bloch eigenstates) acquires two distinct contributions which are the same as the ones given in \thinspace Ref.\onlinecite{b22}, that is, the collective local circulation contribution (LC) plus the collective itinerant circulation contribution (IC), the latter being, in our formulation, explicitly attributed to the new boundary velocity.
It is important to re-emphasize that, using the relation between the boundary and the standard velocity, the IC can be recast in a form that can be evaluated as a bulk property.  

In this framework we propose that the intrinsic circulation is the proper quantity that encodes the electrons' intrinsic orbital behavior in periodic (or moderately disordered) and extended systems, without the need of any approximation, and as such it must be employed in a rigorous quantum mechanical theoretical framework for calculating the orbital magnetization.

In the fashion described above, we exploit the intrinsic orbital circulation in order to model the orbital magnetization of non-interacting electrons and as such we use it to derive two quantum mechanical formulas, one as an $\mathbf{r}$-space and another one as a \textquotedblleft{reciprocal}\textquotedblright \enspace $\mathbf{k}$-space formula, both being relaxed from any Wannier-localization approximation. 

The $\mathbf{r}$-space formula is derived for an extended system within PBCs over the terminated boundaries, giving therefore the bulk orbital magnetization.   
  
In the derivation of the $\mathbf{k}$-space expression we relax the PBCs constraint, and as a consequence, certain interesting features emerge. 
Namely, a covariant derivative appears in the one-band (adiabatically deformed) approximation formula for the orbital magnetization as an emerging operator, and survives due to the non-Hermitian effect that is attributed to the anomalous momentum gradient operator \thinspace $ {\partial}_\mathbf{k}$ \thinspace that enters the static (off-diagonal) Hellmann-Feynman theorem that we derive in  Appendix \ref{a3}. 
In the many-band (unrestricted) formula the non-Hermitian effect contributes an additional boundary quantity which explicitly depends on the off-diagonal matrix elements of the boundary velocity operator \thinspace $\mathbf{v}_{b}$ \thinspace as well as on a new boundary momentum gradient operator \thinspace $\mathbf{k}_{b}$ \thinspace (defined in \thinspace Eq.~(\ref{apc13})).
The latter additional boundary quantity, is expected to give locally (in momentum space) giant orbital magnetization contributions (due to its structure) whenever band crossings occur along with Hall voltage as a consequence of boundary conditions that may generally break the standard Born-von K\'{a}rm\'{a}n periodicity.

The theoretical method that we propose can be employed either for calculating the built-in orbital magnetization of solids in the absence of external fields \cite{{b22},{b23}} or for calculating the induced orbital magnetization as a response to external fields, e.g. to an electric field \cite{b31}. In this work we determine the built-in magnetization in solids when time reversal symmetry is assumed to be broken, either from a staggered magnetic field that averages to zero over the unit cell, or through spin-orbit coupling to a background of ordered local moments.  

We have organized the paper as follows. In Sec.\thinspace \ref{s2} we define the electrons' boundary velocity and then the extended velocity operators, as well as the electrons' intrinsic and extrinsic orbital circulations with the aid of the extended velocity operator. In Sec.\thinspace \ref{s3} using the electrons' intrinsic circulation we derive two quantum expressions for the bulk orbital magnetization of non-interacting electrons, one as an $\mathbf{r}$-space and the other as a \textquotedblleft{reciprocal}\textquotedblright \enspace $\mathbf{k}$-space formula. We summarize and conclude in Sec.\thinspace \ref{s4}. Some details concerning analytical manipulations and derivations are given in three Appendices. 

\section{Definitions} \label{s2}

\subsection{Extended velocity operator}
By taking into account the evolution of the state under consideration, and by demanding that the velocity operator expectation value must always be equal with the rate of change of the electrons' expectation value $\displaystyle\frac{d}{dt}\!\left\langle {\mathbf{r}}\right\rangle$, it is necessary to define the velocity operator in an extended theoretical framework as,
\begin{equation} \label{e1}
\mathbf{v}_{ext}=\mathbf{v}+\mathbf{v}_{b}
\end{equation}
where, 
\begin{equation} \label{e2}
\mathbf{v}=\frac{i}{\hbar}\left[ H(\mathbf{r},t),\mathbf{r} \right]
\end{equation}
is the standard velocity operator and 
\begin{equation} \label{e3}
\mathbf{v}_{b}=\frac{i}{\hbar}\!\left( {H(\mathbf{r},t)}^+ -H(\mathbf{r},t) \right)\!\mathbf{r}
\end{equation}
is the boundary velocity operator.

The introduction of this new operator $\mathbf{v}_b $ is rather naturally motivated by \thinspace Ref.\onlinecite{b13}\textendash\onlinecite{b14} \thinspace and its expectation value is not zero only whenever the position operator becomes anomalous due to the non-Hermitian effect, in which case there are paradoxes first noted in \thinspace Ref.\onlinecite{b15}.  

For closed systems \thinspace ${\left\langle\Psi(t)\vert\Psi(t)\right\rangle=1}$, the Hamiltonian is Hermitian \thinspace ${{H(\mathbf{r},t)}^+ = H(\mathbf{r},t)}$ \thinspace with respect to the states that belong within the domain of its definition and these states form the given Hilbert space. The non-Hermitian effect emerges whenever the state \thinspace $\mathbf{r}\Psi(\mathbf{r},t)$ \thinspace does not belong within the given Hilbert space, that is \thinspace
${ 
{H(\mathbf{r},t)}^+ 
(\mathbf{r}\Psi(\mathbf{r},t) )
\neq  
{H(\mathbf{r},t)} \,
(\mathbf{r}\Psi(\mathbf{r},t)) 
}$, \smallskip which is a characteristic property of all wavefunctions $\Psi(\mathbf{r},t)$ that satisfy PBCs over the system boundaries.

Although the expectation value of the boundary velocity operator \thinspace Eq.~(\ref{e3}) \thinspace given by 
\begin{equation} \label{e3b} 
\left\langle \mathbf{v}_{b} \right\rangle = 
\frac{i}{\hbar} 
( 
\left\langle 
H(\mathbf{r},t) \Psi(t) \vert\, \mathbf{r} \Psi(t) 
\right\rangle 
-
\left\langle
\Psi(t) \vert \, 
H(\mathbf{r},t) \mathbf{r} \Psi(t) 
\right\rangle 
)  
\end{equation} 
is by definition a bulk quantity, due to space-volume integration (assuming a 3D system) in position representation, it can always and equivalently be evaluated as a boundary quantity due to the structure (and symmetry) of the integrands that allows an integration by parts.

In this respect, by working in position representation, for real scalar and vector potentials and after a straightforward integration by parts, the expectation value of \thinspace Eq.~(\ref{e3}) \thinspace is given in the form 
\begin{equation} \label{e4}
\left\langle\mathbf{v}_{b}\right\rangle=
-\oiint_S \mathbf{r} \, (\, \mathbf{J}_{pr}(\mathbf{r},t)\! \cdot \!d\mathbf{S} \, )
+
\frac{i\hbar}{2m}
\oiint_S \left|\Psi(\mathbf{r},t)\right|^2\!d\mathbf{S}
\end{equation}
with $S$ being the terminated boundary surface of the system  where the boundary conditions are imposed, and \thinspace ${  \mathbf{J}_{pr}(\mathbf{r},t)\!=\!\text{Real}[\Psi(\mathbf{r},t)^{\displaystyle *}\mathbf{v}\,\Psi(\mathbf{r},t)] }$  \thinspace is the standard local probability current density (for a spinless electron). 

The general form of \thinspace Eq.~(\ref{e4}) \thinspace can be further reduced for periodic systems. Specifically, by assuming a Bloch wavefunction \thinspace ${ \displaystyle  \Psi_{\mathbf{k}}(\mathbf{r}, t) = \frac{1}{\sqrt{N}} \, e^{ \displaystyle i \mathbf{k.r}} u_{\mathbf{k}}(\mathbf{r}, t) }$ \thinspace 
and cell normalization convention 
\thinspace
${\displaystyle
\left\langle \Psi_{\mathbf{k}}(t)\vert \Psi_{\mathbf{k}}(t)\right\rangle
= 
\left\langle u_{\mathbf{k}}(t)\vert u_{\mathbf{k}}(t)\right\rangle_{cell}
=1}$  
\thinspace 
(where $N$ is the total number of the unit cells enclosed within the volume $V$ of the system), then, 
\thinspace Eq.~(\ref{e3b}) \thinspace truncates into the form 
\begin{equation} \label{e3b1}  
\left\langle \mathbf{v}_{b} \right\rangle =
\frac{i}{\hbar}
\left( 
\left\langle 
H_k \, u_{\mathbf{k}}(t) \vert \, \mathbf{r}  
\, u_{\mathbf{k}}(t) 
\right\rangle _{cell} 
-
\left\langle
u_{\mathbf{k}}(t) \vert \, 
H_k \, \mathbf{r}  \, u_{\mathbf{k}}(t) 
\right\rangle _{cell} 
\right)
\end{equation} 
where \thinspace
${H_{k} }$
\thinspace
is given by
\thinspace ${ H_{k} = e^{ \displaystyle -i \mathbf{k.r}}H(\mathbf{r}, t)e^{ \displaystyle i \mathbf{k.r}}  }$.
In deriving \thinspace Eq.~(\ref{e3b1}) \thinspace
we have used the fact 
that \thinspace ${ u_{\mathbf{k}}(\mathbf{r}, t)  }$ \thinspace fallls wiithin the domain of definition of the Hamiltonian  ${H_k}$, that is  \thinspace
${\left\langle 
H_{k} \,  u_{\mathbf{k}}(t) \vert \, u_{\mathbf{k}}(t) \right\rangle 
-
\left\langle u_{\mathbf{k}}(t) \vert \, 
H_{k} \,  u_{\mathbf{k}}(t) \right\rangle = 0 }$.
By then exploiting the symmetry of the integrands and performing integration by parts, \thinspace Eq.~(\ref{e3b1}) \thinspace takes the simplified form
\begin{equation}  \label{e3b2}
\left\langle \mathbf{v}_{b} \right\rangle =
-
\oiint_{cell} \!\!\! \mathbf{r} \, (\, \mathbf{J}_{pr}(\mathbf{r}, t,  \mathbf{k})\! \cdot \!d\mathbf{S} \, ) 
\end{equation}
that is valid for periodic systems.

The first term of \thinspace Eq.~(\ref{e4}) \thinspace can be seen as a position-weighted probability flux through the boundaries of the system, while the second and purely imaginary part, cancels a possible imaginary remnant part of the standard velocity operator expectation which is given by
\begin{equation} \label{e5}
\left\langle\mathbf{v}\right\rangle=\iiint_V \mathbf{J}_{pr}(\mathbf{r},t)dV
-\frac{i\hbar}{2m} \oiint_S \left|\Psi(\mathbf{r},t)\right|^2\!d\mathbf{S}.
\end{equation} 
By adding \thinspace Eq.~(\ref{e4}) \thinspace and \thinspace Eq.~(\ref{e5}), that is \thinspace
${  
\left\langle\mathbf{v}_{ext}\right\rangle =
\left\langle\mathbf{v}\right\rangle
+
\left\langle\mathbf{v}_{b}\right\rangle 
}$
we see that ${ \left\langle\mathbf{v}_{ext}\right\rangle }$
is always a real quantity as expected (see discussion below).

The boundary velocity operator can also be useful in the study of materials with strong spin-orbit coupling interaction if a modification of its expectation value form is made, that is, by taking into account the 
spin dependence of the standard velocity operator (as an outcome of the non-relativistic limit of the Dirac equation)
 \thinspace ${ \displaystyle \mathbf{v}=\frac{1}{m}\mathbf{\Pi} + \frac{\hbar}{4m^2c^2} \bm{\sigma} \times \nabla{V}(\mathbf{r}) }$ \thinspace that enters the local probability current density \thinspace
${ \displaystyle \mathbf{J}_{pr}(\mathbf{r},t)\!=\! \text{Real}[\Psi(\mathbf{r},t)^{\dagger} \mathbf{v}\,\Psi(\mathbf{r},t)] }$ \thinspace which now must be evaluated with respect to spinors. 

With the aid of \thinspace Eq.~(\ref{e1}) \textendash \thinspace (\ref{e3}), the extended velocity operator can be recast in the form
\begin{equation}\label{e6}
\mathbf{v}_{ext}=\frac{i}{\hbar}({H(\mathbf{r},t)}^+\mathbf{r} -\mathbf{r}H(\mathbf{r},t)),
\end{equation}
and the equality ${\displaystyle{\left\langle {\mathbf{v}_{ext}}\right\rangle = \frac{d}{dt}\!\left\langle {\mathbf{r}}\right\rangle}}$ \thinspace holds irrespectively of the position operator behavior (hence irrespective of the boundary conditions). By the definition as given in \thinspace Eq.~(\ref{e6}) and by working in the position representation \thinspace $\mathbf{r}^+\!=\mathbf{r}$, we can easily deduce that the extended velocity operator is always a Hermitian operator \thinspace $\mathbf{v}_{ext}^+\!=\mathbf{v}_{ext}$ \thinspace and its expectation value is always real, in agreement with a summation of \thinspace Eq.~(\ref{e4}) \thinspace and  (\ref{e5}) \thinspace without the need of any specific boundary conditions to be imposed, which is also valid even for open systems where the Hamiltonian is not a Hermitian operator. 

A simple and intuitive criterion to demonstrate the necessity of introducing the extended velocity operator is as follows: Consider a stationary and extended plane wave state of a free electron of mass $m$ with well defined momentum $\hbar \mathbf{k}$ in a finite volume $V.$ The system is assumed to be closed, that is the electrons' wavefunction is normalized to unity at every instant $t$ within the volume $V$, ${\left\langle\Psi(t)\vert\Psi(t)\right\rangle=1}$. In this fashion, the electrons' displacement $\Delta{\left\langle\mathbf{r}\right\rangle}$ must always be smaller than (or equal to) the systems' size. Using the standard velocity definition \thinspace  ${\displaystyle\mathbf{v}=\frac{i}{\hbar}\left[ H(\mathbf{r},t),\mathbf{r} \right]}$ \thinspace the elctrons' displacement acquires the value \thinspace ${\displaystyle \Delta{\left\langle\mathbf{r}\right\rangle}=\frac{\hbar \mathbf{k}}{m}t}$ \thinspace which will eventually lead the electron out of the system. This paradox is bypassed within the extended velocity operator definition, as it turns out that the boundary velocity contributes an equal magnitude and opposite sign than the bulk electrons' velocity $ \left\langle \mathbf{v}\right\rangle $  resulting in zero displacement ${\displaystyle \Delta{\left\langle\mathbf{r}\right\rangle}=0}$ at every instant $t$ for the assumed stationary state. In fact, the extended velocity operator guarantees that every stationary state (irrespectively of the static potentials) will always produce zero displacement for the electron, that is \thinspace ${\displaystyle \frac{d}{dt}\!\left\langle {\mathbf{r}}\right\rangle=\left\langle\mathbf{v}_{ext}\right\rangle=\left\langle\mathbf{v}\right\rangle+\left\langle\mathbf{v}_{b}\right\rangle=0}$, as expected from the trivial fact that the position operator expectation value is a static quantity with respect to any stationary state. 

In this fashion, we can develop a simple and direct method to link boundary effects with bulk properties as a form of a bulk-boundary correspondence in a general sense for every stationary state, namely ${\left\langle\mathbf{v}\right\rangle_n=-\left\langle\mathbf{v}_{b}\right\rangle_n}$ where $n$ indexes the Hamiltonian eigenstate; this is an example, therefore, of a bulk formulation that properly takes into account boundary currents that are rigorously related to the bulk band structure.

There are two important features of the extended velocity operator \thinspace $\mathbf{v}_{ext}$ \thinspace that can be deduced from its off-diagonal matrix elements with respect to the (generally time-dependent) Hamiltonian instantaneous eigenstates $\left|n(t)\right\rangle$. These are derived by direct application of Eq.~(\ref{e6}) and Eq.~(\ref{e1}) and are given by
\begin{eqnarray} \nonumber \label{e7}
\left\langle m(t)\vert\mathbf{v}\vert n(t)\right\rangle + && \, \left\langle m(t)\vert\mathbf{v}_b\vert n(t)\right\rangle= \\*[2pt] && 
\displaystyle \frac{i}{\hbar}(E_m(t)-E_n(t))\left\langle m(t)\vert\mathbf{r}\vert n(t)\right\rangle 
\end{eqnarray}
where, the off-diagonal matrix elements of the boundary velocity operator  are explicitly calculated (after a straightforward integration by parts) as
\begin{eqnarray} \label{e7b} \nonumber
\left\langle m(t)\vert\mathbf{v}_b\vert n(t) \right\rangle &=&  
-\frac{1}{2}\oiint_S 
\mathbf{r}
\left( 
(\mathbf{v} \, \psi_{m})^{ \dagger} \psi_{n}
 + \psi_{m}^{ \dagger} \, \mathbf{v} \, \psi_{n} 
\, \right) \! \cdot \! d\mathbf{S} 
\nonumber  \\*
&& + \frac{i\hbar}{2m} \oiint_S 
   \psi_{m}^{ \dagger} \,\psi_{n} \, d\mathbf{S},  
\end{eqnarray}
${ \psi_{n} = \psi_{n} (\mathbf{r},t) = \left\langle \mathbf{r} \vert n(t)\right\rangle }$ are the Hamiltonian's instantaneous  eigenfunctions and ${\mathbf{v}}$ is the velocity operator given by \thinspace Eq.~(\ref{e2}). Eq.~(\ref{e7b}) can be viewed 
as the off-diagonal counterpart of  Eq.~(\ref{e4}). 

The two important features then follow. 
First, the off-diagonal position matrix elements in Eq.~(\ref{e7}) will explicitly be involved in the many-band (unrestricted) formula of the orbital magnetization that we will derive in this article; therefore, boundary contributions due to the off-diagonal boundary velocity matrix elements will explicitly be taken into account.
Second, the off-diagonal position matrix elements in Eq.~(\ref{e7}) are proportional to the electrons' transition dipole moment, therefore the emission and absorption of photons can be rigorously related with boundary properties owing to the off-diagonal boundary velocity matrix elements.  

Generalizing the results of this subsection we point out that, whenever one defines an operator in an extended way \thinspace 
${ \bm{\mathcal{O}}_{ext} }$ \thinspace so that its expectation value \thinspace
${ \left\langle  \bm{\mathcal{O}}_{ext} \right\rangle }$ 
\thinspace is equal with the rate of change of the expectation value of a given Hermitian operator \thinspace $\mathbf{G}$, that is \thinspace ${ \displaystyle 
\left\langle  \bm{\mathcal{O}}_{ext} \right\rangle =
\frac{d}{dt} \left\langle \mathbf{G} \right\rangle }$, the definition of \thinspace ${ \bm{\mathcal{O}}_{ext} }$ \thinspace can be consistently given by the Ehrenfest theorem, as long as a corresponding boundary operator \thinspace $\bm{\mathcal{O}}_{b} $ \thinspace is taken into account. The expectation value of the boundary operator \thinspace
$ \left\langle \bm{\mathcal{O}}_{b} \right\rangle $ \thinspace is extremely sensitive to the boundary conditions of the 
wavefunction and takes a nonzero value only whenever the given Hermitian operator \thinspace $\mathbf{G}$ \thinspace (entering the theorem) becomes anomalous
due to the non-Hermitian effect. Specifically, by working in position representation, due to symmetry of the integrand, after a 
straightforward integration by parts, the expectation value \thinspace $ \left\langle \bm{\mathcal{O}}_{b} \right\rangle $ \thinspace is always cast  in the form of a boundary integral (assuming real scalar and vector potentials) of a generalized current \thinspace $\mathbf{J}_{G}$ \thinspace flux as
\begin{eqnarray} \label{e7c}  \nonumber 
\left\langle  \bm{\mathcal{O}}_{b} \right\rangle
&=& \frac{i}{\hbar} \left\langle \Psi(t) \right|  \left( {H(\mathbf{r},t)}^+ -H(\mathbf{r},t) \right)\! \mathbf{G} \left|  \Psi(t) \right\rangle 
\\*
&=& \oiint_S \mathbf{J}_{G} \,dS  	
\end{eqnarray}
where the generalized current density \thinspace $\mathbf{J}_{G}$ \thinspace is given by 
\begin{equation} \label{e7d}
\mathbf{J}_{G}=-\frac{1}{2} 
\mathbf{n} \! \cdot \!\!
\left(
( \mathbf{v} \Psi(\mathbf{r},t) )
^{\dagger} 
+ \Psi(\mathbf{r},t)^{\dagger} \mathbf{v}  
\, \right)  
\mathbf{G} \Psi(\mathbf{r},t).
\end{equation}
The wavefunction  $\Psi(\mathbf{r},t)$ entering \thinspace 
Eq.~(\ref{e7d}) \thinspace can be either the electrons' two component spinor wavefunction for spinfull electron (and this will be nontrivially useful in solids with strong spin-orbit interaction) or the scalar wavefunction for spinless electron motion (where the generalized current has the same structure but with the dagger operation $\dagger$ being replaced by the complex conjugation $*$ operation only) and $\mathbf{n}$ is the unit vector locally normal to the surface $S$. For the special case of \thinspace ${ \mathbf{G}=\mathbf{r} }$ \thinspace and either spinlfull or spinless electron motion, by analytically calculating the directional velocity operator \thinspace $ \mathbf{n}\!\cdot\!\mathbf{v}$ \thinspace action on $\mathbf{G} \Psi(\mathbf{r},t)$ \thinspace within \thinspace Eq.~(\ref{e7d}) \thinspace and \thinspace Eq.~(\ref{e7c}), we recover \thinspace Eq.~(\ref{e4}). Alternatively, if we choose  \thinspace $\mathbf{G}$ \thinspace to be the identity operator \thinspace $I$, then \thinspace ${ \mathbf{J}_{G}}$  \thinspace becomes the usual probability current \thinspace ${ \mathbf{J}_{pr}}$ \thinspace and for a closed system \thinspace Eq.~(\ref{e7c}) \thinspace becomes zero, which is consistent with the conservation of total probability (valid for states belonging within the Hilbert space of closed systems).

\subsection{Intrinsic and extrinsic orbital circulation}
In order to define the electrons' intrinsic and extrinsic orbital circulation for an extended and periodic system, we first choose to define a Hermitian circulation operator as 
\begin{equation} \label{e8}
\mathbf{C}=\frac{1}{2}(\mathbf{r} \times \mathbf{v}_{ext} - \mathbf{v}_{ext}\times\mathbf{r})
\end{equation}
namely the electrons' orbital circulation operator that employs the extended velocity operator; it is therefore designed to take into account the inherited anomaly of the position operator when computing circulating currents in periodic systems. The circulation operator always behaves as a Hermitian operator ${\mathbf{C}^+=\mathbf{C}}$  irrespectively of the wavefunctions' boundary conditions as evidenced from Eq.~(\ref{e8}) and Eq.~(\ref{e1}). With the aid of \thinspace Eq.~(\ref{e1}) \textendash \thinspace (\ref{e3}) \thinspace and \thinspace ${\mathbf{r} \times \mathbf{r}=0}$, the circulation operator can be recast in the forms \thinspace ${\displaystyle \mathbf{C}=\frac{i}{2\hbar} \mathbf{r} \times \left( H(\mathbf{r},t)^{+} + H(\mathbf{r},t)\right)  \mathbf{r}}$ \thinspace and \thinspace ${\displaystyle \mathbf{C}=\mathbf{r} \times \mathbf{v} + \frac{1}{2} \mathbf{r} \times \mathbf{v}_b}$. It is interesting to note that in the latter form of \thinspace $\mathbf{C}$ \thinspace the \thinspace $\displaystyle \frac{1}{2} \mathbf{r} \times \ \mathbf{v}_b$ \thinspace term is an anti-Hermitian operator that has imaginary expectation value which exactly cancels any remnant imaginary part of the \thinspace $\mathbf{r} \times \mathbf{v}$ term expectation value. 
Direct calculation gives the orbital circulation operator \thinspace $\mathbf{C}$ \thinspace expectation value form, which is found to be
\begin{eqnarray} \nonumber \label{e9}
\left\langle \Psi(t)\vert \, \mathbf{C} \, \vert\Psi(t)\right\rangle 
&&= \text{Im} [\,i \left\langle \Psi(t)\vert \, \mathbf{r} \times \mathbf{v} \, \vert\Psi(t)\right\rangle\,] \\* 
&&=\iiint_V \mathbf{r} \times \mathbf{J}_{pr}(\mathbf{r},t)dV   
\end{eqnarray}
where the quantum state under consideration $\left|\Psi(t) \right\rangle$ is normalized within the volume $V$ of the system. In spite of the cautious definition of the circulation operator in order to take into account the possible anomaly of the position operator for periodic systems, it is shown in Appendix \ref{a1} that its expectation value  $\left\langle \mathbf{C}\right\rangle$ with respect to a Bloch eigenstate does not quite lead to any theoretical progress as it becomes undefined for an extended system in the thermodynamic limit.

Motivated, however, by classical mechanics,   
either by rigid body dynamics or by continuous medium (hydrodynamical) theories, we find out that the expectation value of the circulation operator $\left\langle \mathbf{C}\right\rangle$ can always be decomposed into two distinct parts. Namely, an intrinsic circulation part $\left\langle \mathbf{C}_{intr}\right\rangle$ that always has an intensive and bulk behavior (with well defined value within Bloch representation in the thermodynamic limit) and an extrinsic circulation part $\left\langle \mathbf{C}_{extr}\right\rangle$ that has an extensive and position origin-dependent behavior (with undefined value within Bloch representation in the thermodynamic limit). The definitions of the intrinsic and extrinsic circulations are given by
\begin{eqnarray} \nonumber \label{e10}
\left\langle \Psi(t)\vert \, \mathbf{C}_{intr} \, \vert\Psi(t)\right\rangle 
&& = \text{Im} [\,i \left\langle \Psi(t)\vert \,
( \mathbf{r} - \left\langle \mathbf{r}\right\rangle ) 
\times \mathbf{v} \, \vert\Psi(t)\right\rangle\,] \\*[2pt]  
&& =\iiint_V (\mathbf{r}-{\left\langle \mathbf{r}\right\rangle}) \times \mathbf{J}_{pr}(\mathbf{r},t)dV   
\end{eqnarray}
and
\begin{eqnarray} \nonumber \label{e11}
\left\langle \Psi(t)\vert \, \mathbf{C}_{extr} \, \vert\Psi(t)\right\rangle 
&&= \text{Im} [\, i \left\langle \Psi(t)\vert \, \left\langle \mathbf{r} \right\rangle 
\times \mathbf{v} \, \vert\Psi(t)\right\rangle \,] 
\\*[2pt] 
&&=\iiint_V {\left\langle \mathbf{r}\right\rangle}\times \mathbf{J}_{pr}(\mathbf{r},t)dV   
\end{eqnarray}
respectively, where $V$ is the volume of the system and \thinspace ${\displaystyle \left\langle \mathbf{r}\right\rangle=\iiint_V \!\mathbf{r}\left|\Psi(\mathbf{r},t)\right|^2\!dV}$ \thinspace is the position operator expectation value that takes an undefined value within Bloch representation in the thermodynamic limit (as shown in Appendix \ref{a1}). 

The intrinsic circulation \thinspace
$\left\langle \mathbf{C}_{intr}\right\rangle$ \thinspace
has no ambiguity and is a position origin-independent quantity whenever the shift of the position origin causes a U(1) transformation for the scalar wavefunction (assuming a spinless electron). The origin-independence is a consequence of the combined transformation (under a shift of the position origin) of the operator \thinspace
${ ((\mathbf{r}-{\left\langle \mathbf{r}\right\rangle}) \times \mathbf{v}) }$ \thinspace and the U(1) transformation of the wavefunction that compensate each other.
For spinfull electrons the velocity operator acquires spin-dependence and, as long as the shift of the position origin can be described by an SU(2) transformation of the spinor 
wavefunction, the intrinsic circulation remains a position origin-independent quantity without any ambiguity.    

The electrons' intrinsic orbital circulation as given by \thinspace Eq.~(\ref{e10}) \thinspace has an inherited boundary contribution which is revealed when taking into account Eq.~(\ref{e1}) and \thinspace
Eq.~(\ref{e4}) \textendash \thinspace (\ref{e5}). In the special case of a   
stationary state \thinspace $\left| \Psi_{n}(t)\right\rangle $ \thinspace the electrons' intrinsic orbital circulation has the explicit boundary dependence given in
\begin{eqnarray} \nonumber \label{e12}
&&
\left\langle \Psi_{n} (t) \vert \, \mathbf{C}_{intr} \, \vert \Psi_{n} (t) \right\rangle = 
\\*[2pt] 
&& \;\;\;\;\;\;\; \iiint_V \mathbf{r} \times \mathbf{J}_{pr(n)} (\mathbf{r})dV 
- 
\left\langle \mathbf{r}\right\rangle_{n} \times 
\oiint_S \mathbf{r} \, (\mathbf{J}_{pr(n)}(\mathbf{r}).d\mathbf{S}).
\nonumber \\*[-6pt] 
\end{eqnarray}

Assuming an extended Bloch eigenstate $\Psi_{n}(\mathbf{r},t,\mathbf{k})$ that obeys PBCs over the boundaries of the system 
(and is normalized within its volume $V$), and in spite of the position operator (undefined) expectation value $\left\langle \mathbf{r}\right\rangle$ that explicitly enters Eq.~(\ref{e10}), we find after a straightforward calculation shown in  Appendix \ref{a1} \thinspace  that the electrons' intrinsic orbital circulation takes a well-defined value at the infinite volume limit \thinspace $V\rightarrow\infty$, \thinspace given by

\begin{eqnarray} \nonumber \label{e13}
&& \left\langle \Psi_{n}(t,\mathbf{k})\vert \, \mathbf{C}_{intr} \, \vert \Psi_{n}(t,\mathbf{k})\right\rangle = \\*[4pt]
&& \;\;\;\;\; \iiint \limits_{V_{cell}} \!\! 
\left( 
\mathbf{r}- 
\left\langle u_{n}(\mathbf{k}) \vert \, \mathbf{r} \, \vert u_{n}(\mathbf{k})\right\rangle_{cell}
\right) 
\times \,
\mathbf{J}_{ pr (n)}(\mathbf{r},\mathbf{k})
dV \nonumber \\*[-12pt]
\end{eqnarray}
with \thinspace ${ u_{n}(\mathbf{r},\mathbf{k}) }$ \thinspace the cell periodic eigenstates, where all space integrals have been truncated (due to symmetry of the integrands) and evaluated within a unit cell of volume $V_{cell}$, the local probability current density being determined with respect to a Bloch eigenstate. It is evident from \thinspace Eq.~(\ref{e13}) \thinspace that the intrinsic circulation is a bulk and intensive quantity of a periodic and extended system. On the contrary, the extrinsic circulation as given from \thinspace Eq.~(\ref{e11}), takes an undefined value for a periodic and extended system (owing to the position operator expectation value); it is therefore not a proper quantity to model any bulk or boundary property of such a periodic and extended system.   
We note that, in deriving \thinspace Eq.~(\ref{e13}) \thinspace we have assumed the normalization convention
${\displaystyle\left\langle \Psi_{n}(\mathbf{k})\vert \Psi_{n}(\mathbf{k})\right\rangle= \left\langle u_{n}(\mathbf{k})\vert u_{n}(\mathbf{k})\right\rangle_{cell}=1}$, \thinspace that is we have assumed a Bloch state in the form ${ \displaystyle \left| \Psi_{n}(\mathbf{k}) \right\rangle =
\frac{1}{\sqrt{N}}
e^{\displaystyle  i \mathbf{k.r} } 
\left| u_{n}(\mathbf{k}) \right\rangle 
}$ where $N$ is the total number of the unit cells enclosed within the volume $V$ of the system.

Summarizing, and with \thinspace Eq.~(\ref{e10}) \thinspace as well as \thinspace Eq.~(\ref{e13}) \thinspace in mind, we can conclude that the quantity  \thinspace
${ (\mathbf{r}-{\left\langle \mathbf{r}\right\rangle}) \times \mathbf{J}_{pr}(\mathbf{r},t) }$ 
\thinspace is a well defined local intrinsic circulation density, even if it is computed with respect to an extended Bloch state in the thermodynamic limit where the electrons' position expectation value acquires an undefined value.

\subsubsection{Physical meaning of the intrinsic orbital circulation}

A physically and intuitively important feature of the intrinsic orbital circulation is that it is a quantity that properly counts the circulating probability micro-currents embodied in the wavefunction's structure. In order to clarify this feature in a simple manner let as  consider two spinless and free electron motions in 3D space: one electron with well defined linear momentum vector \thinspace $\hbar\mathbf{k}$ \thinspace and another one with partially well defined linear momentum vector, e.g. only its \thinspace $z$ \thinspace component $\hbar k_z \mathbf{e}_z$ (with $k_x$ and $k_{y} $ being undetermined). We assume that each electron is in an extended state motion that is normalized within a volume $V$. The free electron motion with well defined linear momentum vector \thinspace $\hbar\mathbf{k}$, hence with a plane wave form for the wavefunction, has a local probability current density that is a homogeneous vectorial quantity proportional to $\hbar \mathbf{k}/m$. On the contrary, the free electron motion with partially well defined linear momentum \thinspace $\hbar k_z \mathbf{e}_z$ \thinspace has a local probability current density that is an inhomogeneous vectorial quantity with a constant \thinspace $z$ \thinspace component proportional to $\hbar k_z/m$. Using \thinspace Eq.~(\ref{e10}), \thinspace we can easily find that the intrinsic orbital circulation of the free electron motion with well defined linear momentum $\hbar \mathbf{k}$ is zero (due to the homogeneous local probability current density), while the intrinsic orbital circulation of the free electron with partially well defined linear momentum \thinspace $\hbar k_z \mathbf{e}_z$ \thinspace is non-vanishing (due to the inhomogeneous local probability current density) and takes contributions only from the \thinspace $x$ \thinspace and \thinspace $y$ \thinspace non-constant components of the local probability current density that may constitute a vortex circulating probability micro-current field on the planes normal to \thinspace $\mathbf{e}_z$ (with free electron vortex state being an example, see below).

Considering such a structured wavefunction, its phase is indeterminate on the dislocation lines (in 3D space) where the modulus of the wavefunction takes a zero value. The intrinsic orbital circulation of the electron as given by \thinspace Eq.~(\ref{e10}) \thinspace becomes zero, namely,  
${\displaystyle \iiint_V \!\mathbf{r} \times \mathbf{J}_{pr}(\mathbf{r},t)dV - \left\langle \mathbf{r}\right\rangle \!\times \!\iiint_V \! \mathbf{J}_{pr}(\mathbf{r},t)dV=0}$ \thinspace whenever, in the simplest scenario, the local probability current density is zero (the gradient of the wavefunction's phase is zero) or whenever the local probability current density is a homogeneous quantity (the gradient of the wavefunction's phase has a constant and well-defined value), therefore the wavefunction is structureless. On the contrary, in structured wavefunctions the electrons' intrinsic orbital circulation is generally not zero and has two competing contributions as given in \thinspace Eq.~\ref{e10}, which are explicitly dependent on the local probability current density field. The bigger the difference of these two competing contributions the bigger the electrons' intrinsic orbital circulation, which occurs for example whenever the internal structure of the wavefunction has such a symmetry that makes some of the components of \thinspace $ \displaystyle \iiint_V \! \mathbf{J}_{pr}(\mathbf{r},t)dV$ \thinspace become zero. The latter symmetry feature is found in the free electron motion that are described by vortex states \cite{{b32},{b33}} where the electron has well defined linear momentum \thinspace $\hbar k_z \mathbf{e}_z$
\thinspace only in the $z$ direction and at the same time has a well defined canonical orbital angular momentum along the same direction (characterized by the azimuthal index $l$). Due to the rotational (azimuthal) symmetry of the wavefunction, the azimuthal component of \thinspace $ \displaystyle \iiint_V \! \mathbf{J}_{pr}(\mathbf{r},t)dV$ \thinspace becomes zero.
\indent
Structured wavefunctions appear naturally in motions under external potentials, e.g. in atomic orbitals with nonzero mechanical angular momentum or in Landau states in a magnetic field. In this respect, we generally expect that the ionic environment will in principle produce structured and extended cell periodic electronic wavefunctions \thinspace $u_{n}(\mathbf{r},\mathbf{k})$, with the dislocation lines being periodically ordered in the bulk owing to the periodicity of \thinspace $u_{n}(\mathbf{r},\mathbf{k})$, while spiraling probability micro-currents around those lines can be taken into account by \thinspace Eq.~(\ref{e10}) \thinspace and \thinspace  Eq.~(\ref{e13}). Intrinsic orbital circulation is the starting point quantity for the microscopic understanding of the orbital magnetization origin and as such will be used in the following to model the orbital magnetization in band theory without the need of any Wannier-localization approximation.

\subsubsection{Physical meaning of the extrinsic orbital circulation}

The extrinsic orbital circulation \thinspace
${ \left\langle \, \mathbf{C}_{extr} \, \right\rangle 
= \text{Im} [\, i  \left\langle \, \mathbf{r} \, \right\rangle 
\times 
\left\langle \, \mathbf{v} \, \right\rangle  \,]  }$ \thinspace 
is an extensive quantity that counts the circulation of the global probability current 
${ \left\langle \, \mathbf{v} \, \right\rangle  }$  with respect to a specific position origin. It does not carry any tractable information about the structure of the wavefunction or the circulating probability micro-currents (due to being a position origin dependent quantity), and has an undefined value within Bloch representation in the thermodynamic limit (owing to the position operator expectation value being undefined).

\subsection{Subtle behaviors and relevant consequences within Bloch representation} \label{s2c}

\begin{table*}
\caption{\label{T1}{ Matrix elements evaluated with respect to Bloch eigenstates in the thermodynamic limit.}}
\begin{ruledtabular}
\begin{tabular}{c   c   c   c    c }
Operator & Matrix element & Value & Origin & Boundary velocity is \\[6pt]
\hline 	
$ {\mathbf{r}} $ &  $ \left\langle \psi_n | \, \mathbf{r} \, | \psi_n  \right\rangle $ 
& undefined & dependent & well defined \\[6pt]
			
$ {\mathbf{r}} $ & $ \left\langle \psi_n | \, \mathbf{r}  \, | \psi_m  \right\rangle \ \ n \neq m $   & well defined & independent & well defined \\[6pt]

${ \mathbf{r} \times \mathbf{v} }$ & 
$ \left\langle \psi_n | \, \mathbf{r} \times \mathbf{v} \, | \psi_n  \right\rangle $
& undefined & dependent & well defined \\[2pt]
			
${ \displaystyle \frac{i}{\hbar} \mathbf{r}\times H_k \sum_m^{HS} \left|  u_m \right\rangle \left\langle u_m \right| \mathbf{r} }$ &
$ \displaystyle \frac{i}{\hbar} \sum_m^{HS} \left\langle u_n | \, 
\mathbf{r} \left|  u_m \right\rangle 
\times E_m
\left\langle u_m \right| \mathbf{r}
\, | u_n  \right\rangle $
& well defined & independent & zero \\[6pt] 
			
${ \mathbf{r} - \left\langle \psi_n \vert \, \mathbf{r} \, \vert \psi_n \right\rangle }$ &
$ \left\langle \psi_m | \, 
( \, \mathbf{r} - \left\langle \psi_n \vert \, \mathbf{r} \, \vert \psi_n \right\rangle )  \, | \psi_m  \right\rangle $
& well defined & independent & well defined \\[6pt] 
			
${ (\mathbf{r} - \left\langle \psi_n \vert \, \mathbf{r} \, \vert \psi_n \right\rangle
) \times \mathbf{v} }$ & 
$ \left\langle \psi_n | \,
(\mathbf{r} - \left\langle \psi_n \vert \, \mathbf{r} \, \vert \psi_n \right\rangle
) \times \mathbf{v}  \, | \psi_n  \right\rangle $
& well defined & independent & well defined \\[6pt] 
			
${ \left\langle \psi_n \vert 
\, \mathbf{r} \, \vert \psi_n \right\rangle
\times \mathbf{v} }$ &
$ \left\langle \psi_n | 
\left\langle \psi_n \vert \, \mathbf{r} \, \vert \psi_n \right\rangle
\times \mathbf{v}
\, | \psi_n  \right\rangle $
& undefined & dependent & well defined \\[4pt] 
\end{tabular}	
\end{ruledtabular} 
\end{table*}

The scope of this subsection is ultimately to facilitate a comparison of our results (derived in later sections) with the literature, and more specifically  (i) to point out the behavior of operators (with respect to their expectation values and position origin dependence) that are commonly used in the modern theory of orbital magnetization, and (ii) to show some subtle consequences that emerge due to implicit Hermiticity assumptions that were silently made during calculations in recent theoretical works \cite{{b2},{b3},{b30}}. 

The expectation value of the position operator with respect to a Bloch eigenstate \thinspace  
${ \left\langle \psi_n (\mathbf{k})
\vert \, \mathbf{r} \, \vert
\psi_n (\mathbf{k})  \right\rangle  }$ \thinspace turns out to be an undefined value in the thermodynamic limit, 
as shown by \thinspace Eq.~(\ref{apa3}) \thinspace 
derived in Appendix \ref{a1}. On the other hand, the corresponding off-diagonal matrix elements  \thinspace  
${ \left\langle \psi_n (\mathbf{k})
\vert \, \mathbf{r} \, \vert
\psi_m (\mathbf{k})  \right\rangle  }$  given by \thinspace Eq.~(\ref{apa2c}) \thinspace  (also derived in Appendix \ref{a1}) remain well defined quantities.  
In this respect we also note that the matrix elements of the operator \thinspace
${ \left( \,
\mathbf{r} - 
\left\langle 
\psi_n(\mathbf{k})
\vert \, \mathbf{r} \, \vert 
\psi_n(\mathbf{k})
\right\rangle 
\right)
}$, \thinspace evaluated with respect to any Bloch state, have a well defined value in the thermodynamic limit. This can be shown by taking the expectation value with respect to \thinspace ${ \left| \psi_m(\mathbf{k'}) \right\rangle }$, that is 
\thinspace
${ \left( \,
\left\langle 
\psi_m(\mathbf{k'})
\vert \, \mathbf{r} \, \vert 
\psi_m(\mathbf{k'})
\right\rangle 
- 
\left\langle 
\psi_n(\mathbf{k})
\vert \, \mathbf{r} \, \vert 
\psi_n(\mathbf{k})
\right\rangle 
\right)
}$,  \thinspace
which by using the 3D analogue of \thinspace
Eq.~(\ref{apa12}), \thinspace shows that the undefined terms cancel each other and, as a result, the expectation  has a well defined value. The same pattern, that is, the undefined terms canceling each other, is what makes the intrinsic circulation \thinspace
${ \text{Im} [ i \left\langle \psi_n (\mathbf{k}) | \,
(\mathbf{r} - \left\langle \psi_n (\mathbf{k}) \vert \, \mathbf{r} \, \vert \psi_n (\mathbf{k}) \right\rangle
) \times \mathbf{v}  \, | \psi_n (\mathbf{k}) \right\rangle ] }$
\thinspace
have a well defined value in the thermodynamic limit.
On the other hand, the real part of the circulation operator expectation value \thinspace
${ \text{Im} [\,i \left\langle \psi_n(\mathbf{k}) \vert \, \mathbf{r} \times \mathbf{v} \, \vert  
\psi_n(\mathbf{k})
\right\rangle\,] }$, is an undefined quantity as shown by
\thinspace Eq.~(\ref{apa10}). 
For what follows it is worth pointing out that, in all the above mentioned calculations, no constraint is assumed with respect to the boundary velocity expectation value.

By then using the Bloch form of the considered state as well as \thinspace 
${ \displaystyle \mathbf{v}=\frac{i}{\hbar}\left[ H(\mathbf{r}),\mathbf{r} \right] }$ \thinspace and \thinspace ${ \mathbf{r} \times \mathbf{r}=0 }$, \thinspace the circulation operator takes the form \thinspace
${ \displaystyle -\frac{1}{\hbar}  \text{Im} [\, \left\langle u_n(\mathbf{k}) \vert \, \mathbf{r} \times H_k(\mathbf{r}, \mathbf{k}) \, \mathbf{r} \, \vert  
u_n(\mathbf{k})
\right\rangle\,] }$ \thinspace
where \thinspace ${ H_{k}(\mathbf{r},\mathbf{k})=e^{ \displaystyle -i \mathbf{k.r}}H(\mathbf{r})e^{ \displaystyle i \mathbf{k.r}}  }$.
Assuming then that the state \thinspace ${ \mathbf{r} \left| 
u_n(\mathbf{k}) \right\rangle 
}$ \thinspace can be expanded in the complete orthonormal basis of the cell periodic eigenstates \thinspace ${ \left| 
u_m(\mathbf{k}) \right\rangle 
}$, \thinspace that is, using the closure relation  
\thinspace 
${ \displaystyle I=\sum_{m}^{\text{HS}} \left| u_{m}(\mathbf{k}) \right\rangle \left\langle u_{m}(\mathbf{k}) \right|  }$ \thinspace 
and acting from the left 
on the above state
\thinspace ${ \mathbf{r} \left| 
u_n(\mathbf{k}) \right\rangle 
}$, \thinspace the circulation operator becomes
\thinspace
${ \displaystyle -\frac{1}{\hbar} 
\sum_m^{\text{HS}} \text{Im} [\, \left\langle u_n(\mathbf{k}) \vert \, \mathbf{r}H_k(\mathbf{r}, \mathbf{k})
 \vert
u_m(\mathbf{k}) \right\rangle 
\times  
\left\langle u_m(\mathbf{k}) \vert
\, \mathbf{r} \, \vert  
u_n(\mathbf{k})
\right\rangle\,] }$ \thinspace which has far reaching consequences. Firstly, the latter operator is now transformed into a well defined quantity due to taking the form 
\thinspace
${ \displaystyle -\frac{1}{\hbar} 
\sum_{ m \neq n}^{\text{HS}} \text{Im} [\, \left\langle u_n(\mathbf{k}) \vert \, \mathbf{r} \, \vert
u_m(\mathbf{k}) \right\rangle \!
E_m(\mathbf{k})
\times  
\left\langle u_m(\mathbf{k}) \vert
\, \mathbf{r} \, \vert  
u_n(\mathbf{k})
\right\rangle\,] }$, \thinspace where we have used
\thinspace
$
\left\langle u_n(\mathbf{k}) \vert
\, \mathbf{r} \, \vert  
u_n(\mathbf{k})
\right\rangle		
\times  \, 
\left\langle u_n(\mathbf{k}) \vert
\, \mathbf{r} \, \vert  
u_n(\mathbf{k})
\right\rangle 
=0
$. 
This is the basic idea behind the theoretical work made in Refs \onlinecite{{b2},{b3}, {b30}} which, however, was performed in a slightly different way. Specifically, these works used a spectral resolution of the Hamiltonian ${ H(\mathbf{r})=IH(\mathbf{r})I }$, where the closure is given by
\thinspace 
${ \displaystyle I=\sum_{m}^{\text{HS}} \left| \phi_{m} \right\rangle \left\langle \phi_{m} \right|  }$ \thinspace
and ${ \left| \phi_m \right\rangle  }$ are the orbitals. As a result of this spectral resolution, the (undefined) diagonal matrix elements of the position operator are excluded from their circulation operator formula.

The subtle consequence of the above two calculations, is that one unintentionally assumes that the state     
\thinspace ${ \mathbf{r} \left| 
\phi_n \right\rangle 
}$ \thinspace belongs within the domain \thinspace of ${ H(\mathbf{r})  }$, that is, certain boundary conditions for \thinspace ${ \phi_n(\mathbf{r} ) }$ \thinspace are assumed which guarantee that the wavefunction \thinspace ${ \mathbf{r} \phi_n(\mathbf{r}) }$ \thinspace also belongs within the domain of definition of the Hamiltonian, and as a result the non-Hermitian boundary velocity \thinspace Eq.~(\ref{e3b}) \thinspace becomes zero. Specifically, the identification
\thinspace ${ 
\displaystyle
\mathbf{r} \left| 
\phi_n \right\rangle 
=
\sum_{m}^{\text{HS}}C_m 
\left| 
\phi_m \right\rangle 
}$ \thinspace is the one that enforces
the state     
\thinspace ${ \mathbf{r} \left| 
\phi_n \right\rangle 
}$ \thinspace to belong within the domain of the Hamiltonian and
the boundary velocity expectation value to become zero.
This is evident when one (i) takes the inner product of \thinspace ${ \mathbf{r} \left| 
\phi_n \right\rangle 
}$ \thinspace
with \thinspace 
${ \displaystyle
\frac{i}{\hbar}\!
\left\langle \phi_n 
\right| \!
\left( {H(\mathbf{r})}^+ - H(\mathbf{r}) \right)}$, \thinspace (ii) uses
Eq.~(\ref{e3}) for the definition of the boundary velocity operator ${ \mathbf{v}_b} $, \thinspace 
and (iii) exploits the fact that the states \thinspace ${ \left| 
\phi_n \right\rangle 
}$ \thinspace
belong within the domain of definition of \thinspace
${ H(\mathbf{r}) }$ \thinspace which finally gives
\begin{eqnarray} \nonumber 
&&\left\langle \phi_n 
\vert \, \mathbf{v}_b \, \vert 
\phi_n \right\rangle 
=
\frac{i}{\hbar}\!
\left\langle \phi_n 
\right| \!
\left( H(\mathbf{r})^+ -H(\mathbf{r}) \right)
\mathbf{r} \left| 
\phi_n \right\rangle 
\nonumber  \\*
&&
\ \ \ \ \ \ \ \ \  \ \ \ \ 
=
\frac{i}{\hbar}\!
\sum_{m}^{\text{HS}}C_m 
\left\langle\phi_n 
\right| \!
\left( H(\mathbf{r})^+ -H(\mathbf{r}) \right) \!
\left| \phi_m \right\rangle=0. 
\nonumber 
\end{eqnarray} 
In this framework, the method of calculation used by \linebreak
Refs \onlinecite{{b2},{b3},{b30}}, enforced on one hand the circulation operator to have a well defined value, but, on the other hand, unwillingly, they induced Hermiticity which sweeps away boundary contributions to the orbital magnetization; in this respect, the conclusion about the irrelevance of the boundary on the orbital magnetization of metals that was made by \thinspace Ref.\onlinecite{b2}, \thinspace although reasonable, is rather unjustified. We also point out that, due to the above mentioned spectral resolution of the Hamiltonian (performed within the circulation operator) the orbitals ${ \left|  \phi_n \right\rangle  }$  that were assumed in Refs \onlinecite{{b2},{b3},{b30}}, \thinspace must have zero standard (bulk) velocity expectation value 
\thinspace  
${ \left\langle \phi_n  \vert \, \mathbf{v} \, \vert \phi_n \right\rangle=0  }$  
\thinspace
owing to the relation \thinspace  
${ \left\langle \phi_n  \vert \, \mathbf{v} \, \vert \phi_n \right\rangle= - \left\langle \phi_n  \vert \, \mathbf{v}_b \, \vert \phi_n \right\rangle  }$  
\thinspace
that holds for any stationary eigenstate of the Hamiltonian.

To demonstrate at a glance the various subtleties hidden in the literature, in Table \ref{T1} we summarize the behavior of certain operators that are related to the modern theory of orbital magnetization: we summarize their values, their position origin dependence, as well as relevant boundary constraints. 
The presented values are results of calculations performed with respect to Bloch eigenstates in the thermodynamic limit.

\subsection{Decomposition of the intrinsic orbital circulation into  local (LC) and \\ itinerant circulation (IC) contributions}

At this point it is useful to make a connection between the one electron's intrinsic circulation as given by \thinspace Eq.~(\ref{e10}) \thinspace and the decomposition of the  
$n \text{th}$ band collective electrons' circulation that was made in a rather ambiguous way, namely, to local circulation (LC) and itinerant circulation (IC) in the seminal work of \thinspace Ref.\onlinecite{b22} \thinspace in order to model the orbital magnetization of normal insulators within a quantum mechanical method.
Therein, they started from the assumption that each electron's eigenstate can be represented by an exponentially localized Wannier function (thus the Bloch states that they used satisfy the periodic gauge \thinspace ${ \displaystyle \left| \Psi_{n}(\mathbf{k+G}) \right\rangle = \displaystyle \left| \Psi_{n}(\mathbf{k}) \right\rangle }$ \thinspace and have zero Chern invariant) and they began their calculation with a collective circulation computed with respect to these Wannier functions, turning at the end of their calculation to the Bloch eigenstates.
In the present work we follow an opposite route, that is we start our calculation from the one electron's intrinsic circulation \thinspace Eq.~(\ref{e10}) \thinspace without any gauge assumptions (restrictions) with respect to the Bloch eigenstates, and using those states as building blocks in the many-body Slater determinant wavefunction we determine analytically the electrons' (ground state) collective orbital magnetization.      
By then taking into account the above mentioned relation between the standard and the boundary velocity for stationary states \thinspace
${ \left\langle\mathbf{v}\right\rangle_{n} 
= - \left\langle\mathbf{v}_{b}\right\rangle_{n} }$,
\thinspace
the electrons' intrinsic circulation with respect to a Bloch eigenstate \thinspace $\Psi_{n}(\mathbf{r},t,\mathbf{k})$ \thinspace is given by
\begin{eqnarray} \nonumber \label{e13b}
\left\langle \Psi_{n} (\mathbf{k}) \vert \, \mathbf{C}_{intr} \, \vert \Psi_{n}(\mathbf{k}) \right\rangle 
&=& 
\text{Im} [\,i \left\langle \Psi_{n}(\mathbf{k}) \vert \, \mathbf{r} \times \mathbf{v} \, \vert \Psi_{n}(\mathbf{k}) \right\rangle\,]
\\*[2pt] 
&+ &
\text{Im} [\,i \left\langle \mathbf{r}\right\rangle_{n} \times
\left\langle \Psi_{n}(\mathbf{k}) \vert \, \mathbf{v}_{b} \, \vert \Psi_{n}(\mathbf{k}) \right\rangle\,].
\nonumber \\* 
\end{eqnarray}
In order to establish the connection with \thinspace Ref.\onlinecite{b22} \thinspace method we assume that the Bloch eigenstates \thinspace $\Psi_{n}(\mathbf{r},t,\mathbf{k})$ \thinspace entering \thinspace Eq.~(\ref{e13b}) \thinspace satisfy the periodic gauge and we expand it as   
\begin{eqnarray} \label{e13c} \nonumber
&& \left| \Psi_{n}(\mathbf{k}) \right\rangle = 
\frac{1}{\sqrt{N}}\sum_{\mathbf{R}} e^{\displaystyle i \mathbf{k.R}}
\left| n, \mathbf{R} \right\rangle
\\*  \nonumber
&& \;\;\;\;\;\;\;\; = 
\frac{1}{\sqrt{N}} \sum_{\mathbf{R}_{I}} e^{\displaystyle i \mathbf{k.R}_{I}}
\left| n, \mathbf{R}_{I} \right\rangle
+
\frac{1}{\sqrt{N}} \sum_{\mathbf{R}_{S}} e^{\displaystyle i \mathbf{k.R}_{S}}
\left| n, \mathbf{R}_{S} \right\rangle
\\*[-8pt]
\end{eqnarray}
where $N$ is the number of primitive cells of the system,  ${ \left| n, \mathbf{R}_{I} \right\rangle }$ is the $n \text{th}$ band Wannier function in the bulk cell $\mathbf{R}$ and ${ \left| n, \mathbf{R}_{S} \right\rangle }$ is the 
$n \text{th}$ surface localized orbital on the surface cell $\mathbf{R}_{S}$.
By then taking into account that the expectation value of the boundary velocity \thinspace ${ \left\langle \Psi_{n}(\mathbf{k}) \vert \, \mathbf{v}_{b} \, \vert \Psi_{n}(\mathbf{k}) \right\rangle }$ \thinspace is determined by a boundary integral, that is, only the boundary localized orbitals 
${ \left| n, \mathbf{R}_{S} \right\rangle }$
enter into the expansion of the expectation value 
\begin{eqnarray} \label{e13d} \nonumber 
&&\left\langle \Psi_{n}(\mathbf{k}) \vert \, \mathbf{v}_{b} \, \vert \Psi_{n}(\mathbf{k}) \right\rangle  
\\* \nonumber
&& \;\;\;\;\;\;\;\;\;\;\;\;\;\;\; = \, \frac{1}{N} 
\sum_{\mathbf{R}_{S'}} \sum_{\mathbf{R}_{S}}
e^{\displaystyle i \mathbf{k.( \mathbf{R}_{S} - \mathbf{R}'_{S} )} }
\left\langle n, \mathbf{R}'_{S} \vert \, \mathbf{v}_{b} \, \vert 
n, \mathbf{R}_{S} \right\rangle,
\\*[-10pt] 
\end{eqnarray}
we calculate the \thinspace $n \text{th}$ \thinspace band collective electrons' intrinsic circulation given by \thinspace 
${ \displaystyle \mathbf{C}_{n(coll)}= 
\frac{V}{(2\pi)^3} \iiint_{BZ}
\left\langle \Psi_{n} (\mathbf{k}) \vert \, \mathbf{C}_{intr} \, \vert \Psi_{n}(\mathbf{k}) \right\rangle d^3k
}$,
\thinspace which takes the form
\begin{eqnarray} \label{e13e} \nonumber
\mathbf{C}_{n(coll)} &=&
 \sum_{\mathbf{R}}
 \left\langle n, \mathbf{R} \vert \, 
 \mathbf{r} \times \mathbf{v} \, \vert 
 n, \mathbf{R} \right\rangle 
\\* \nonumber
&& + 
\sum_{\mathbf{R}'} \sum_{\mathbf{R}}
 \sum_{\mathbf{R}'_{S}} \sum_{\mathbf{R}_{S}}
\, \delta_{\mathbf{R}'+\mathbf{R}'_{S} , \mathbf{R}+\mathbf{R}_{S}}
\left\langle 
n, \mathbf{R}'
\vert
\, \mathbf{r} \,
\vert
n, \mathbf{R}
\right\rangle 
\\* 
&& \;\;\; \times
\left\langle n, \mathbf{R}'_{S}
\vert \, \mathbf{v}_{b} \, \vert 
n, \mathbf{R}_{S} \right\rangle.  
\end{eqnarray}

Assuming then that the crystal has inversion symmetry in the bulk 
${ \left\langle n, -\mathbf{R}_{I} 
\vert \, 
\mathbf{r} 
\, \vert 
n, -\mathbf{R}_{I} \right\rangle
= -
\left\langle n, \mathbf{R}_{I} 
\vert \, 
\mathbf{r} 
\, \vert 
n, \mathbf{R}_{I} \right\rangle
}$,
that is \thinspace
${ \displaystyle \sum_{\mathbf{R}_{I}} 
\left\langle n, \mathbf{R}_{I}  
\vert \, 
\mathbf{r} 
\, \vert 
n, \mathbf{R}_{I} \right\rangle =0 }$,
as well as that for \thinspace ${ \mathbf{R}'_{S} \neq \mathbf{R}_{S} }$ \thinspace the matrix elements
${ \left\langle 
n, \mathbf{R}+\mathbf{R}_{S} - \mathbf{R}'_{S}
\vert
\, \mathbf{r} \,
\vert
n, \mathbf{R}
\right\rangle  
 }$
can be taken as zero, the $n \text{th}$ band electrons' collective intrinsic circulation takes the approximate form
\begin{eqnarray} \label{e13f}
\mathbf{C}_{n(coll)} &=&
\sum_{\mathbf{R}}
\left\langle n, \mathbf{R} \vert \, 
\mathbf{r} \times \mathbf{v} \, \vert 
n, \mathbf{R} \right\rangle 
\\* \nonumber
&& + 
\sum_{\mathbf{R}_{S}}
\left\langle 
n, \mathbf{R}_{S}
\vert
\, \mathbf{r} \,
\vert
n, \mathbf{R}_{S}
\right\rangle 
\times
\left\langle n, \mathbf{R}_{S}
\vert \, \mathbf{v}_{b} \, \vert 
n, \mathbf{R}_{S} \right\rangle 
\end{eqnarray}
where the first term on the right hand side of \thinspace Eq.~(\ref{e13f}) \thinspace gives the electrons' $n \text{th}$ band collective local circulation contribution (LC) and the second term the collective itinerant circulation contribution (IC), as given respectively in Ref.\onlinecite{b22}.

In Ref.\onlinecite{b22} and \onlinecite{b23} they notice that the itinerant circulation (IC) contribution that involves only the surface WFs can always be calculated as a bulk quantity that involves the bulk WFs, and they emphasize that this is quite remarkable and one of their central results. Their finding is explained whenever in the starting \thinspace Eq.~(\ref{e13b}) \thinspace we use the bulk expression \thinspace Eq.~(\ref{e3b}) \thinspace for the boundary velocity expectation value and at the same time replace 
${ 
\left\langle 
H(\mathbf{r}) \Psi_{n}(\mathbf{k}) \vert\, \mathbf{r} \Psi_{n}(\mathbf{k})
\right\rangle 
}$
with its equal 
${ \left\langle \Psi_{n}(\mathbf{k}) \vert \, 
\mathbf{r} H(\mathbf{r})  \Psi_{n}(\mathbf{k}) 
\right\rangle, }$
which is true for all stationary states according to \thinspace Eq.~(\ref{e6}) (and the vanishing of its expectation value). Therefore, with the aid of the extended velocity operator and the intrinsic circulation definitions, we can elucidate and rigorously explain the origin of the heuristic partitioning of the orbital magnetization that was made in Ref.\onlinecite{b22} and \onlinecite{b23}.

\section{Orbital magnetization quantum formulas} \label{s3}

In this section we use the electrons' intrinsic orbital circulation presented in \thinspace Sec.~\ref{s2} \thinspace in order to derive quantum mechanical formulas for the orbital magnetization of non-interacting electrons by accounting for the circulating probability micro-currents. The formulas that we derive are applicable either to conventional or to topological crystalline materials, under periodic or realistic boundary conditions for the electrons' wavefunctions, while any localization assumptions are absent. 

In a system of non-interacting electrons we can define the (single-eigenstate) orbital magnetization \thinspace $\mathbf{M}_n (\mathbf{k})$ \thinspace per electron as
\begin{equation} \label{e14}
\mathbf{M}_n (\mathbf{k})=\frac{\mathbf{m}_{n}(\mathbf{k})}{V}
=\frac{e}{2cV}
\left\langle \Psi_{n}(t,\mathbf{k})\vert \, \mathbf{C}_{intr} \, \vert \Psi_{n}(t,\mathbf{k})\right\rangle
\end{equation}   
where \thinspace $\mathbf{m}_{n}(\mathbf{k})$ \thinspace is the electrons' orbital magnetic moment, $\Psi_{n}(\mathbf{r},t,\mathbf{k})$ \thinspace is a Bloch eigenstate, $V$ is the volume of the system, \thinspace $c$ \thinspace is the speed of light and $e<0$ \thinspace is the electron charge.

\subsection{r-space orbital magnetization quantum formula} 

In the derivation of the  $\mathbf{r}$-space formula we do not take into account the realistic boundary contributions to the orbital magnetization due to the realistic wavefunctions' boundary conditions; rather we provide a formula that has a bulk character. Namely, we calculate the orbital magnetization within PBCs which are imposed on the wavefunction over a \textquotedblleft{terminated}\textquotedblright  \thinspace boundary surface of the (3D) material in the thermodynamic limit.

Using \thinspace Eq.~(\ref{e13}) \thinspace for the electrons' intrinsic orbital circulation and \thinspace Eq.~(\ref{e14}) \thinspace for the orbital magnetization per electron, as well as the collective electrons' ground state magnetization (assumed to be evaluated with respect to a many-body Slater determinant wavefunction) given by \thinspace ${ \displaystyle \mathbf{M}=\frac{1}{(2\pi)^3} \sum_{E_{n} \leq \mu} \iiint_{BZ} \! f_{n}(\mathbf{k},\mu) \,  \mathbf{m}_{n}(\mathbf{k}) \,d^3k }$, \thinspace where $\mu$ is the Fermi energy and $f_{n}(\mathbf{k},\mu)$ is the occupation function, the bulk orbital magnetization of spinless and non-interacting electrons is given by   
\begin{widetext}   
\begin{equation} \label{e15}
\mathbf{M}=\frac{e}{2c(2\pi)^3} 
\sum_{E_{n}\leq \mu} \, \,
\iiint \limits_{BZ} f_{n}(\mathbf{k},\mu) 
\left( \,\,
\iiint \limits_{V_{cell}} \!\! 
\left( 
\, \mathbf{r}-
\left\langle u_{n}(\mathbf{k})\vert \, \mathbf{r} \, \vert u_{n}(\mathbf{k})\right\rangle_{cell}
\right) 
\times 
\mathbf{J}_{ pr (n)}(\mathbf{r},\mathbf{k})
dV
\right)d^3k  
\end{equation}
\end{widetext}
where all expectation value position-integrals are truncated (due to symmetry) and carried out within a primitive cell of volume \thinspace $V_{cell}$ \thinspace as shown in Appendix \ref{a1}. The orbital magnetization \thinspace $\mathbf{r}$-space \thinspace formula is the first major result in this work and the integrand of \thinspace Eq.~(\ref{e15}) \thinspace can be seen as a local orbital magnetization density with respect to real space.    

We re-emphasize that, although the position operator that enters \thinspace Eq.~(\ref{e10}) \thinspace has by itself an undefined expectation value  $ \displaystyle \left\langle \mathbf{r} \right\rangle $ within Bloch representation, its problematic behavior does not show up and it effectively behaves as a well defined operator when it appears within \thinspace Eq.~(\ref{e10}) and subsequently within \thinspace Eq.~(\ref{e15}). Therefore, the position operator does not have to be \textquotedblleft{sandwiched}\textquotedblright \thinspace between the ground-state projector and its complement as done in \thinspace Ref.\onlinecite{b30} \thinspace in order to get a well defined local expression for the electrons' orbital magnetization with respect to periodic and extended states, but this can be realized in a straightforward manner with \thinspace Eq.~(\ref{e15}). 

As evidenced from \thinspace Eq.~(\ref{e15}), the orbital magnetization acquires significant value whenever the difference between the two competing contributions gets as large as possible. We therefore expect that the orbital magnetization will have significant contributions from those states that possess some kind of rotational symmetry within the unit cell that results in ${ \displaystyle \iiint_{V_{cell}}\!\!\mathbf{J}_{ pr (n)}(\mathbf{r},\mathbf{k})dV \rightarrow 0}$. Although we have considered spinless electrons, the orbital magnetization as given by  
\thinspace Eq.~(\ref{e15}) \thinspace is a property that silently carries an explicit spin dependence. In crystals with strong spin-orbit interaction the velocity operator acquires spin dependence \thinspace  ${ \displaystyle \mathbf{v}=\frac{1}{m}\mathbf{\Pi} + \frac{\hbar}{4m^2c^2} \bm{\sigma} \times \nabla{V_{cry}}(\mathbf{r}) }$  \thinspace which is inherited in the local probability current density  \thinspace
${ \displaystyle \mathbf{J}_{pr(n)}(\mathbf{r},\mathbf{k})\!=\! \text{Real}[\Psi_{n}(\mathbf{r},\mathbf{k})^{+} \mathbf{v}\Psi_{n}(\mathbf{r},\mathbf{k})] }$ \thinspace that enters \thinspace Eq.~(\ref{e15}). Therefore, in materials with strong spin-orbit interaction where the spin degree of freedom is essential, the orbital magnetization is directly influenced (apart from the wavefunctions) by the crystal force field that interacts with the electron. 

\subsection{k-space orbital magnetization quantum formula}
In this subsection we derive an orbital magnetization formula that is valid for general boundary conditions for the electrons' wavefunction. Boundary contributions are explicitly taken into account as a consequence of the emergent non-Hermitian effect. 

In order to derive the $\mathbf{k}$-space formula we assume that, each electrons' motion is described from a (generally) extended and stationary form eigenstate (normalized within the volume of the material) in the form \thinspace 
${ 
\left| \Psi_{n}(t,\mathbf{k})\right\rangle = e^{\displaystyle -\frac{1}{\hbar} E_{n}(\mathbf{k})t} e^{\displaystyle i\mathbf{k.r}} \left|u_{n}(\mathbf{k})\right\rangle }$ \thinspace and no  Wannier-localization approximation is involved. The bulk values of each (position representation) wavefunction \thinspace $ u_{n}(\mathbf{r},\mathbf{k})$ \thinspace are assumed to be periodic (with respect to any direct lattice vector translation $\mathbf{R}$), while we relax this property near the boundaries of the material in order to take into account the realistic boundary contributions.
We derive below a \thinspace $\mathbf{k}$-space \thinspace orbital magnetic formula \thinspace $\mathbf{m}_n (\mathbf{k})$ \thinspace for each electron, starting from
\begin{equation} \label{e16}
\mathbf{m}_n (\mathbf{k})=
\frac{e}{2c}
\text{Im} [\, i \left\langle \Psi_{n}(t,\mathbf{k})\vert \, (\mathbf{r}-{\left\langle \mathbf{r}\right\rangle}_{n}) \times \mathbf{v} \, \vert\Psi_{n}(t,\mathbf{k})\right\rangle \, ]
\end{equation}
(\textit{cf}. Eqs (\ref{e10}) and (\ref{e14})), and by straightforward generalization we provide the collective orbital magnetization formula for non-interacting electrons calculated with respect to a many-body Slater determinant wavefunction. The formula we derive will explicitly incorporate \thinspace
$\mathbf{k}$ derivatives (with the thermodynamic limit assumed), thus we are cautious from the very beginning against possible $\mathbf{k}$-dependent ambiguities of our final result. For this reason we consider the dynamical phase as well as an arbitrary $\mathbf{k}$-dependent phase (due to gauge freedom) for the wavefunctions from the very beginning of our derivation. Therefore, the Bloch type quantum eigenstates that we consider (for each electron) have the form \thinspace
${ \left|\Psi_{n}(t,\mathbf{k})\right\rangle = 
e^{ \displaystyle i \mathbf{k.r}} \,
e^{ \displaystyle i \Theta_{n}(t,\mathbf{k}) }
\left| u_{n}(\mathbf{k})\right\rangle }$,
where \thinspace $\Theta_{n}(t,\mathbf{k})$ \thinspace is the dynamical phase  
\thinspace augmented by an additional $\mathbf{k}$-dependent gauge phase. The \thinspace $\Theta_{n}(t,\mathbf{k})$ \thinspace phase has explicit form, for a static $H$, given by \thinspace     
${ \displaystyle \Theta_{n}(t,\mathbf{k})= -\frac{1}{\hbar} E_{n}(\mathbf{k})t + \Lambda(\mathbf{k}) }$.

By taking into account the above Bloch type eigenstate for each electron, the standard velocity operator as given by \thinspace Eq.~(\ref{e2}), and the Schr\"{o}dinger equation that evolves the quantum eigenstate, the action of the standard velocity operator on the Bloch type eigenstate is given, as analytically derived in Appendix \ref{a2}, by   
\begin{widetext}
\begin{equation} \label{e17}
\mathbf{v} \left|\Psi_{n}(t,\mathbf{k})\right\rangle  =
-\frac{1}{\hbar} \, 
e^{ \displaystyle i \mathbf{k.r}} 
e^{ \displaystyle i \Theta_{n}(t,\mathbf{k}) }
\left( 
H_{k}(\mathbf{r},\mathbf{k}) - E_{n}(\mathbf{k})  
\right)  
\left| \partial_{\mathbf{k}}  u_{n}(\mathbf{k})\right\rangle    
+ \frac{1}{\hbar} \partial_{\mathbf{k}}E_{n}(\mathbf{k})  \left|\Psi_{n}(t,\mathbf{k})\right\rangle. 
\end{equation}	
where \thinspace ${ H_{k}(\mathbf{r},\mathbf{k})=e^{ \displaystyle -i \mathbf{k.r}}H(\mathbf{r})e^{ \displaystyle i \mathbf{k.r}}  }$. In view of Eq.~(\ref{e16}), the orbital magnetic moment for each electron is given by
\begin{equation} \label{e18}
\mathbf{m}_n (\mathbf{k}) =
- \frac{e}{2c \hbar}
\text{Im} [\, i 
\left\langle u_{n}(\mathbf{k})\vert \, (\mathbf{r}-{\left\langle \mathbf{r}\right\rangle}_{n}) 
\, \times \,
\left( 
H_{k}(\mathbf{r},\mathbf{k}) - E_{n}(\mathbf{k})  
\right)  
\vert \partial_{\mathbf{k}}  u_{n}(\mathbf{k})\right\rangle \, ]. 
\end{equation}
\end{widetext}

It is now helpful to see the origin of the electron's orbital magnetic moment
\thinspace Eq.~(\ref{e16}) \thinspace and
\thinspace Eq.~(\ref{e18}); a comparison with the semi-classical counterpart of a localized wave packet \cite{b25} \thinspace will be made at the end of this subsection after Eq.~(\ref{e27}). In virtue of \thinspace Eq.~(\ref{e16}) \thinspace and
\thinspace Eq.~(\ref{e13}), \thinspace 
the orbital magnetic moment
is always a well defined quantity even if the wavefunction is an extended one and the volume ${ V }$ of the system infinite. It is a quantity that emerges due to the circulating probability micro-currents embodied in the wavefunction's (bulk as well as boundary) structure. For free electron and plane wavefunction, namely, a wavefunction with a well defined crystal momentum ${ \hbar \mathbf{k} }$, the electron's orbital magnetic moment \thinspace Eq.~(\ref{e16}) \thinspace becomes zero. In virtue now of \thinspace Eq.~(\ref{e18}), \thinspace where \thinspace Eq.~(\ref{e17}) \thinspace has been used, although the orbital magnetic moment holds its above mentioned physical origin, is now also explicitly dependent on the remnant non-Hermitian boundary term \thinspace 
${ 
( 
H_{k}(\mathbf{r},\mathbf{k}) - E_{n}(\mathbf{k})  
) \! 
\left|  
\partial_{\mathbf{k}}  u_{n}(\mathbf{k})
\right\rangle 
}$ \thinspace
of the Hellmann-Feynman theorem, a fact first noticed in Ref. {\onlinecite{b16}}.
Specifically, by taking the inner product of
\thinspace Eq.~(\ref{e17}), \thinspace
with \thinspace 
${ \left\langle \Psi_{n}(t,\mathbf{k}) \right|  }$, \thinspace
the electrons standard velocity expectation value is found to be
\begin{eqnarray} \label{e18b} \nonumber
&&
\left\langle u_{n}(\mathbf{k})
\vert \mathbf{v} \vert u_{n}(\mathbf{k})\right\rangle 
=
\frac{1}{\hbar} \partial_{\mathbf{k}}E_{n}(\mathbf{k})
\\*  \nonumber   
&&
\; \; \; \; \; \; \; \; \; \; \; \; \; \; \; \; \; \; \; \; \; \; \; 
-\frac{1}{\hbar} \, 
\left\langle 
u_{n}(\mathbf{k}) \vert 
\left( 
H_{k}(\mathbf{r},\mathbf{k}) - E_{n}(\mathbf{k})  
\right)  
\vert \partial_{\mathbf{k}}  u_{n}(\mathbf{k})\right\rangle,
\\*  [-2pt]
\end{eqnarray}
where the second term on the right side of \thinspace Eq.~(\ref{e18b}) is precisely the non-Hermitian boundary term of  \thinspace Ref. \onlinecite{b16}, which emerges 
due to the momentum gradient operator \thinspace
${ \partial_{\mathbf{k}} }$ \thinspace that becomes anomalous. In this respect, despite the fact that the electron's orbital magnetic moment \thinspace Eq.~(\ref{e16}) \thinspace is an intensive quantity, when we transform it into a  
$\mathbf{k}$-derivative formula, \thinspace Eq.~(\ref{e18}), \thinspace 
this is dominated by the remnant boundary term
${ 
( 
H_{k}(\mathbf{r},\mathbf{k}) - E_{n}(\mathbf{k})  
) \! 
\left|  
\partial_{\mathbf{k}}  u_{n}(\mathbf{k})
\right\rangle 
}$.

We then express the action of the operator \thinspace ${ \mathbf{r} - \left\langle \mathbf{r} \right\rangle_{n}  }$ \thinspace on 
the eigenstate \thinspace
${ \left| u_{n}(\mathbf{k})\right\rangle }$ \thinspace 
as a ${ \mathbf{k}}$-derivative formula,
and then substitute the result in \thinspace Eq.~(\ref{e18}). This is done by taking into account that the time-independent eigenstate 
${ \left| u_{n}(\mathbf{k})\right\rangle }$ can be recast in the form
${
\left| u_{n}(\mathbf{k})\right\rangle =
e^{ \displaystyle -i \mathbf{k.r}}\, e^{ \displaystyle -i \Lambda(\mathbf{k})} 
\left| \Psi_{n}(\mathbf{k})\right\rangle
}$,
where the time-dependence has been eliminated. In this manner, the action of the position operator on the eigenstate \thinspace
${ \left| u_{n}(\mathbf{k})\right\rangle }$ \thinspace is expressed as a \thinspace $\mathbf{k}$-derivative given by 
\begin{eqnarray} \label{e18c} \nonumber 
\mathbf{r} \left| u_{n}(\mathbf{k}) \right\rangle &=& 
i \left| \partial_{\mathbf{k}} u_{n}(\mathbf{k}) \right\rangle
-\partial_{\mathbf{k}} \Lambda(\mathbf{k}) 
\left| u_{n}(\mathbf{k}) \right\rangle  
\nonumber \\*
&&
-i \, e^{ \displaystyle -i \mathbf{k.r}}\, e^{ \displaystyle -i \Lambda(\mathbf{k})} 
\left| \partial_{\mathbf{k}}  \Psi_{n}(\mathbf{k})\right\rangle.
\end{eqnarray}
Accordingly, the expectation value of the position operator \thinspace $\mathbf{r}$ \thinspace with respect to the eigenstate \thinspace 
${ \left| u_{n}(\mathbf{k})\right\rangle }$ \thinspace takes the form
\begin{equation} \label{e18d} 
\left\langle u_{n}(\mathbf{k}) \vert \, \mathbf{r} \, \vert u_{n}(\mathbf{k}) \right\rangle = 
{\mathbf{A}}_{nn}(\mathbf{k}) 
-\partial_{\mathbf{k}} \Lambda_{n}(\mathbf{k}) 
-i \left\langle \Psi_{n}(\mathbf{k}) \vert \partial_{\mathbf{k}} \Psi_{n}(\mathbf{k})\right\rangle,
\end{equation}
where \thinspace ${ {\mathbf{A}}_{nn}(\mathbf{k})=i\left\langle u_{n}(\mathbf{k}) \vert \partial_{\mathbf{k}} u_{n}(\mathbf{k}) \right\rangle }$ \thinspace is the Abelian Berry connection. By acting with
\thinspace Eq.~(\ref{e18d}) \thinspace 
\thinspace on
${ \left| u_{n}(\mathbf{k}) \right\rangle }$ \thinspace and then subtracting the product from \thinspace Eq.~(\ref{e18c}) \thinspace we obtain the identity
	
\begin{eqnarray} \label{e18f} \nonumber
&&\left( \,
\mathbf{r} - \left\langle \mathbf{r} \right\rangle_{n} 
\right)
\left| u_{n}(\mathbf{k})\right\rangle  
=  
\left( 
\, i \partial_{\mathbf{k}} - {\mathbf{A}}_{nn}(\mathbf{k})
\right) 
\left| u_{n}(\mathbf{k})\right\rangle
\nonumber \\*
&&
\ \ \ \ \ \ \ \  \ \ \ \ \ \ \ \ \ \ \  \ \ \ \ \ \ 
-
\, i \, e^{ \displaystyle 
-i \mathbf{k.r}} \, e^{ \displaystyle -i \Lambda(\mathbf{k})} 
\left| \partial_{\mathbf{k}}  \Psi_{n}(\mathbf{k})\right\rangle
\nonumber \\*
&& 
\ \ \ \ \ \ \ \ \ \ \
+ \, i \, e^{ \displaystyle 
-i \mathbf{k.r}} \, e^{ \displaystyle -i \Lambda(\mathbf{k})} 
\left\langle \Psi_{n}(\mathbf{k}) \vert \partial_{\mathbf{k}} \Psi_{n}(\mathbf{k})\right\rangle
\left| \Psi_{n}(\mathbf{k})\right\rangle.
\nonumber \\*[-2pt] 
\end{eqnarray}

The first two terms on the right hand side of \thinspace Eq.~(\ref{e18f}) \thinspace can be recast in the form, 
\begin{equation} \label{e20}
\left( 
\, i \partial_{\mathbf{k}} - {\mathbf{A}}_{nn}(\mathbf{k})
\right)
\left|  
u_n(\mathbf{k}) \right\rangle 
= i \widetilde{{\partial}_{\mathbf{k}}} 
\left|  u_n(\mathbf{k}) \right\rangle 
\end{equation}
where
\begin{equation} \label{e21}
\widetilde{{\partial}_{\mathbf{k}}}=
\partial_{\mathbf{k}}
\; + \;
i\mathbf{A}_{nn}(\mathbf{k})
\end{equation}
is the one-band covariant derivative  (of the ${ \textit{n}_{\text{th}} }$ band) \cite{b33b}
that will explicitly enter the final many-body orbital 
magnetization formula as an emerging operator, and as such has never shown up in the literature of modern theory of orbital magnetization. 
On the contrary, in the orbital magnetization modern theory, they implement by heuristic argument covariant derivatives \cite{{b31},{b34},{b35}} 
in order to make their final orbital magnetization formulas gauge invariant.

In this fashion, 
\thinspace Eq.~(\ref{e18f}) \thinspace takes the form
\begin{eqnarray} \label{e18g} \nonumber
&&\left( \,
\mathbf{r} - \left\langle \mathbf{r} \right\rangle_{n} 
\right)
\left| u_{n}(\mathbf{k})\right\rangle  
=  
i \left|  \widetilde{{\partial}_{\mathbf{k}}} 
 u_n(\mathbf{k}) \right\rangle 
\nonumber \\*
&&
\ \ \ \ \ \ \ \  \ \ \ \ \ \ \ \ \ \ \  \ \ \ \ \ \ 
-
\, i \, e^{ \displaystyle 
	-i \mathbf{k.r}} \, e^{ \displaystyle -i \Lambda(\mathbf{k})} 
\left| \partial_{\mathbf{k}}  \Psi_{n}(\mathbf{k})\right\rangle
\nonumber \\*
&& 
\ \ \ \ \ \ \ \ \ \ \
+ \, i \, e^{ \displaystyle 
	-i \mathbf{k.r}} \, e^{ \displaystyle -i \Lambda(\mathbf{k})} 
\left\langle \Psi_{n}(\mathbf{k}) \vert \partial_{\mathbf{k}} \Psi_{n}(\mathbf{k})\right\rangle
\left| \Psi_{n}(\mathbf{k})\right\rangle.
\nonumber \\*[-2pt] 
\end{eqnarray}

We then expand the state \thinspace
${ 
\left| \partial_{\mathbf{k}}  \Psi_{n}(\mathbf{k})\right\rangle }$ \thinspace \thinspace of \thinspace Eq.~(\ref{e18g}) \thinspace on the complete basis of the Bloch eigenstates \thinspace 
${ \left| \psi_{m}(\mathbf{k'}) \right\rangle }$ \thinspace
by using the identity operator 
\thinspace 
${ \displaystyle I=
\sum_{m,  \mathbf{k}'}^{HS}
\left| \psi_{m}(\mathbf{k}') \right\rangle 
\!
\left\langle \psi_{m}(\mathbf{k}') \right|  }$,
\thinspace 
that is, we substitute 
\thinspace
$ \displaystyle 
\left| \partial_{\mathbf{k}}  \Psi_{n}(\mathbf{k})\right\rangle 
=
\sum_{m, \mathbf{k}' }
\left\langle \psi_{m}(\mathbf{k'}) 
\vert
\partial_{\mathbf{k}} 
\psi_{n}(\mathbf{k})
\right\rangle 
\left| \psi_{m}(\mathbf{k'}) \right\rangle 
$ . 
By then taking into account that the operator \thinspace   ${ \left(  \mathbf{r}- \left\langle \mathbf{r} \right\rangle_{n}  \right) }$  \thinspace is by definition Hermitian (without the need of any specific boundary conditions to be imposed) owing to the position representation that we are working in \thinspace ($\mathbf{r}^+\!=\mathbf{r}$) \thinspace and to the reality of the position operator expectation value \thinspace
$\left\langle \mathbf{r} \right\rangle_{n}$,
we let it act on the left to the eigenstate $ \left\langle u_{n}(\mathbf{k})\right| $ in
\thinspace Eq.~(\ref{e18}), which is carried out by taking the Hermitian conjugate of  
\thinspace Eq.~(\ref{e18g}) \thinspace
and then plugging it into \thinspace Eq.~(\ref{e18}). In this respect and as
analytically shown in Appendix \ref{apb22},  \thinspace Eq.~(\ref{e18}) \thinspace 
takes the form (\ref{apb10j}), namely
\begin{widetext}
\begin{eqnarray} \label{e18h} \nonumber
\mathbf{m}_n (\mathbf{k}) &=&
- \frac{e}{2c \hbar }
\text{Im} 
\left[ \, 
\left\langle \widetilde{{\partial}_{\mathbf{k}}}u_{n}(\mathbf{k})
\vert 
\, \times \,
\left( 
H_{k}(\mathbf{r},\mathbf{k}) - E_{n}(\mathbf{k})  
\right)  
\vert 	
{\partial}_{\mathbf{k}}
u_{n}(\mathbf{k})\right\rangle 
\, \right]
\nonumber \\*[4pt] 
&& - 
\frac{e}{2c \hbar}
\sum_{m \neq n} 
\text{Im} 
\left[  \,
\left\langle \Psi_{n}(\mathbf{k}) \vert \partial_{\mathbf{k}} \Psi_{m}(\mathbf{k})\right\rangle
\times 
\left\langle u_{m}(\mathbf{k}) 
\vert \,
H_{k}(\mathbf{r},\mathbf{k}) - E_{n}(\mathbf{k})  
\, \vert
{\partial}_{\mathbf{k}}
u_{n}(\mathbf{k})\right\rangle 
\, \right], 
\end{eqnarray}
which, by using the identity \thinspace
${ 
\left( 
H_{k}(\mathbf{r},\mathbf{k}) - E_{n}(\mathbf{k})  
\right) \!  
\left|  
\partial_{\mathbf{k}}
u_{n}(\mathbf{k})\right\rangle 	
=	
\left( 
H_{k}(\mathbf{r},\mathbf{k}) - E_{n}(\mathbf{k})  
\right) \!  
\left| 
\widetilde{{\partial}_{\mathbf{k}}}	u_{n}(\mathbf{k})\right\rangle 
}$, \thinspace  takes the form
\begin{eqnarray} \label{e18i} \nonumber
\mathbf{m}_n (\mathbf{k}) &=&
- \frac{e}{2c \hbar }
\text{Im} 
\left[ \, 
\left\langle \widetilde{{\partial}_{\mathbf{k}}}u_{n}(\mathbf{k})
\vert 
\, \times \,
\left( 
H_{k}(\mathbf{r},\mathbf{k}) - E_{n}(\mathbf{k})  
\right)  
\vert 
\widetilde{{\partial}_{\mathbf{k}}}u_{n}
\right\rangle 
\, \right]
\nonumber \\*[4pt] 
&& - 
\frac{e}{2c \hbar}
\sum_{m \neq n} 
\text{Im} 
\left[  \,
\left\langle \Psi_{n}(\mathbf{k}) \vert \partial_{\mathbf{k}} \Psi_{m}(\mathbf{k})\right\rangle
\times 
\left\langle u_{m}(\mathbf{k}) 
\vert \,
H_{k}(\mathbf{r},\mathbf{k}) - E_{n}(\mathbf{k})  
\, \vert
{\partial}_{\mathbf{k}}u_{n}
\right\rangle 
\, \right]. 
\end{eqnarray}
\end{widetext}

Now, some further analysis is in order concerning the off-diagonal elements in Eq. (38). As rigorously shown in Appendix \ref{a3},
by deriving an off-diagonal Hellmann-Feynman theorem that takes into account non-Hermitian corrections, the off-diagonal matrix elements \thinspace
${ \left\langle \Psi_{n}(\mathbf{k}) \vert \partial_{\mathbf{k}} \Psi_{m}(\mathbf{k})\right\rangle }$  \thinspace are found to be emergent non-Hermitian boundary quantities, that are given by

\begin{equation} \label{e18k} 
\left\langle \psi_n(\mathbf{k}) \vert
\partial_{\mathbf{k}}
\psi_{m}(\mathbf{k}) \right\rangle 
= \frac{ {\mathbf{S}}_{nm}(\mathbf{k}) }
{ ( E_n(\mathbf{k}) - E_m(\mathbf{k}) ) }
\end{equation}
where the matrix elements of the non-Hermitian term \thinspace ${\mathbf{S}}_{nm}(\mathbf{k})$
\thinspace are always transformed after an integration by parts (due to symmetry of the integrands) into a boundary quantity that is given by
\begin{equation} \label{e18l} 
\mathbf{S}_{nm}(\mathbf{k})
=
\frac{i \hbar}{2} \oiint_S 
\mathbf{n}\!\cdot\! 
\left( \,
( \mathbf{v} \, \psi_{n} )^{\displaystyle *} 
+ \psi_{n}^{\displaystyle *} \, \mathbf{v}  
\, \right) \! 
\, \partial_{\mathbf{k}} \psi_{m}
\, dS,	
\end{equation}
where ${{\mathbf{S}}_{nm}(\mathbf{k})  }$ is defined as
\begin{eqnarray} \label{a18l} \nonumber
\mathbf{S}_{nm}(\mathbf{k})
&=&
\left\langle H(\mathbf{r}) \psi_{n}(\mathbf{k}) \vert  
\partial_{\mathbf{k}} \psi_{m}(\mathbf{k}) \right\rangle 
-
\left\langle \psi_{n}(\mathbf{k}) \vert  
H(\mathbf{r}) \partial_{\mathbf{k}} \psi_{m}(\mathbf{k}) \right\rangle
\nonumber  \\*  
&=&
\left\langle 
\Psi_{n}(\mathbf{k}) \vert  
\left( 
H(\mathbf{r})^{+} - H(\mathbf{r})
\, \right)  
\partial_{\mathbf{k}} \Psi_{m}(\mathbf{k}) 
\right\rangle.
\end{eqnarray}
By then using the considered  Bloch eigenstate
${ \left|\Psi_{m}(\mathbf{k})\right\rangle = 
e^{ \displaystyle i \mathbf{k.r}} \,
e^{ \displaystyle i \Lambda(\mathbf{k}) }
\left| u_{m}(\mathbf{k})\right\rangle }$ \thinspace
in the non-Hermitian boundary term expression \thinspace ${\mathbf{S}}_{nm}(\mathbf{k})$, we transform the non-Hermitian boundary term and express it in a form that is evaluated only by the use of the cell periodic eigenstates. Straightforward calculation shows that
\begin{equation} \label{apc11}
{\mathbf{S}}_{nm}(\mathbf{k})=
\hbar \left\langle u_{n}(\mathbf{k}) \vert \, {\mathbf{v}}_{b} \, \vert u_{m}(\mathbf{k}) \right\rangle
+
\left\langle u_{n}(\mathbf{k}) \vert \, \mathbf{k}_{b} \, \vert u_{m}(\mathbf{k}) \right\rangle,
\end{equation}
where, 
\begin{eqnarray} \label{epc12} \nonumber 
\left\langle u_{n}(\mathbf{k}) \vert \, {\mathbf{v}}_{b} \, \vert u_{m}(\mathbf{k}) \right\rangle 
&=&
\frac{i}{\hbar} 
( 
\left\langle 
H_k(\mathbf{r}, \mathbf{k})u_{n}(\mathbf{k})  \vert\, \mathbf{r} u_{m}(\mathbf{k})
\right\rangle 
\\* \nonumber
&&
-
\left\langle
u_{n}(\mathbf{k}) \vert \, 
H_k(\mathbf{r}, \mathbf{k}) \mathbf{r} 
u_{m}(\mathbf{k})
\right\rangle 
) 
\\*
\end{eqnarray} 
are the off-diagonal matrix elements of the boundary velocity operator \thinspace ${ \mathbf{v}_b }$ \thinspace defined as 
\begin{equation} \label{apc14}
\mathbf{v}_{b} 
=
\frac{i}{\hbar} 
\left( \, 
H_{k}(\mathbf{r},\mathbf{k})^{+} - H_{k}(\mathbf{r},\mathbf{k})
\, \right)  
\mathbf{r},
\end{equation} 
while
\begin{eqnarray} \label{apc15} \nonumber 
\left\langle u_{n}(\mathbf{k}) \vert \, \mathbf{k}_{b} \, \vert u_{m}(\mathbf{k}) \right\rangle
&=&
\left\langle 
H_k(\mathbf{r}, \mathbf{k})u_{n}(\mathbf{k})  \vert\, \partial_{\mathbf{k}} u_{m}(\mathbf{k})
\right\rangle 
\\* \nonumber
&&
-
\left\langle
u_{n}(\mathbf{k}) \vert \, 
H_k(\mathbf{r}, \mathbf{k}) 
\partial_{\mathbf{k}}
u_{m}(\mathbf{k})
\right\rangle
\\*[-4pt] 
\end{eqnarray} 
are the off-diagonal matrix elements of the boundary momentum gradient \thinspace \textquotedblleft{operator}\textquotedblright  \thinspace defined as
\begin{equation} \label{apc13}
\mathbf{k}_{b} 
=
\left( \, 
H_{k}(\mathbf{r},\mathbf{k})^{+} - H_{k}(\mathbf{r},\mathbf{k})
\, \right)  
\partial_{\mathbf{k}}\,,
\end{equation} 
where the Hamiltonian $H_{k}(\mathbf{r},\mathbf{k})$ is the standard \thinspace ${ H_{k}(\mathbf{r},\mathbf{k})=e^{ \displaystyle -i \mathbf{k.r}}H(\mathbf{r})e^{ \displaystyle i \mathbf{k.r}}  }$
and \thinspace ${u_m = u_m(\mathbf{r},\mathbf{k})}$
\thinspace are the cell-periodic eigenfunctions.

In position representation and after an integration by parts, both the above off-diagonal matrix elements are always transformed (due to symmetry of the integrands) to boundary quantities given by 
\begin{eqnarray} \label{e25} \nonumber
\left\langle u_{n}(\mathbf{k}) \vert \, {\mathbf{v}}_{b} \, \vert u_{m}(\mathbf{k}) \right\rangle 
&=& 
-\frac{1}{2} \oiint_S 
\mathbf{n}\!\cdot\! 
\left( \,
( \mathbf{v} \, u_{n} )^{\displaystyle *} 
+ u_{n}^{\displaystyle *} \, \mathbf{v}  
\, \right) \! 
\, \mathbf{r} \, u_{m}
\, dS  
\nonumber \\*[4pt] 
&=&  
-\frac{1}{2}\oiint_S 
\mathbf{r}
\left( 
(\mathbf{v} \, u_{n})^{\displaystyle *} u_{m}
+ u_{n}^{\displaystyle *}  \mathbf{v} \, u_{m} 
\, \right) \! \cdot \! d\mathbf{S} 
\nonumber  \\*
&& + \frac{i\hbar}{2m} \oiint_S 
u_{n}^{\displaystyle *} \, u_{m} \, d\mathbf{S}  
\end{eqnarray}
(which comes out from Eq. (\ref{e7b})),
and  
\begin{eqnarray} \label{e26} \nonumber
\left\langle u_{n}(\mathbf{k}) \vert \, \mathbf{k}_{b} \, \vert u_{m}(\mathbf{k}) \right\rangle &=&
\frac{i \hbar}{2} \oiint_S 
\mathbf{n}\!\cdot\! 
\left( \,
( \mathbf{v} \, u_{n} )^{\displaystyle *} 
+ u_{n}^{\displaystyle *} \, \mathbf{v}  
\, \right) \! 
\, \partial_{\mathbf{k}}u_{m}
\, dS
\\*[-6pt] 	
\end{eqnarray}
respectively, where \thinspace ${u_m = u_m(\mathbf{r},\mathbf{k})}$
\thinspace are cell-periodic in the bulk. 
We note that \thinspace Eq.~(\ref{e25}) \thinspace and \thinspace Eq.~(\ref{e26}) \thinspace are not zero {\it only whenever} \thinspace the position operator \thinspace $\mathbf{r}$ \thinspace and the momentum gradient operator \
$\partial_{\mathbf{k}}$ \thinspace become anomalous respectively. For example, for bulk localized states (defined as the ones that the wavefunction \thinspace ${ u_m(\mathbf{r},\mathbf{k})}$
\thinspace and all of its derivatives are zero over the materials boundaries) both operators behave as normal operators and have zero matrix elements,  \thinspace 
${ \left\langle u_{n}(\mathbf{k}) \vert \, {\mathbf{v}}_{b} \, \vert u_{m}(\mathbf{k}) \right\rangle=0 }$
\thinspace and 
\thinspace
${ \left\langle u_{n}(\mathbf{k}) \vert \, \mathbf{k}_{b} \, \vert u_{m}(\mathbf{k}) \right\rangle=0  }$
\thinspace respectively. On the other hand, for extended states that satisfy PBSs over the material boundaries   
\thinspace 
${ \left\langle u_{n}(\mathbf{k}) \vert \, {\mathbf{v}}_{b} \, \vert u_{m}(\mathbf{k}) \right\rangle \neq 0 }$  \thinspace while 
\thinspace
${ \left\langle u_{n}(\mathbf{k}) \vert \, \mathbf{k}_{b} \, \vert u_{m}(\mathbf{k}) \right\rangle=0  }$.
We also point out that the matrix elements \thinspace
${\left\langle u_{n}(\mathbf{k}) \vert \, {\mathbf{v}}_{b} \, \vert u_{m}(\mathbf{k}) \right\rangle  }$
\thinspace and \thinspace
\thinspace
${ \left\langle u_{n}(\mathbf{k}) \vert \, \mathbf{k}_{b} \, \vert u_{m}(\mathbf{k}) \right\rangle  }$, thus also their sum \thinspace
${ \mathbf{S}_{nm}(\mathbf{k}) }$, \thinspace can equally be computed as bulk quantities whenever the integrations by parts are not performed.

Using \thinspace Eq.~(\ref{e18i}) \thinspace and \thinspace Eq.~(\ref{e18k}) \thinspace we find the second major result in this work, namely the $\mathbf{k}$-space orbital magnetic moment of each electron \thinspace $\mathbf{m}_n (\mathbf{k})$ \thinspace that is given by
\begin{widetext}
\begin{eqnarray} \label{e27} \nonumber
\mathbf{m}_n (\mathbf{k}) &=&
- \frac{e}{2c \hbar }
\text{Im} 
\left[ \, 
\left\langle \widetilde{{\partial}_{\mathbf{k}}}u_{n}(\mathbf{k})
\vert 
\, \times \,
\left( 
H_{k}(\mathbf{r},\mathbf{k}) - E_{n}(\mathbf{k})  
\right)  
\vert 
\widetilde{{\partial}_{\mathbf{k}}}
u_{n}(\mathbf{k})\right\rangle 
\, \right]
\nonumber \\*[4pt] 
&& - 
\frac{e}{2c \hbar}
\sum_{m \neq n} 
\text{Im} 
\left[  \,
\frac{1}{\left( E_n(\mathbf{k})-E_m(\mathbf{k})\right)}
\, \, {\mathbf{S}}_{nm}(\mathbf{k})
\times 
\left\langle u_{m}(\mathbf{k}) 
\vert \,
H_{k}(\mathbf{r},\mathbf{k}) - E_{n}(\mathbf{k})  
\, \vert
{\partial}_{\mathbf{k}}
u_{n}(\mathbf{k})\right\rangle 
\, \right]. 
\end{eqnarray}
\end{widetext}

It is worth noticing that, 
due to the off-diagonal Hellmann-Feynman theorem
\thinspace Eq.~(\ref{apc6d}), \thinspace
the combination of the off-diagonal matrix elements
${ ( {\mathbf{A}}_{nm}(\mathbf{k}) 
- \left\langle u_{n}(\mathbf{k}) \vert \, \mathbf{r} \, \vert u_{m}(\mathbf{k}) \right\rangle ) }$,
where 
\thinspace ${ {\mathbf{A}}_{mn}(\mathbf{k})=i\left\langle u_{m}(\mathbf{k}) \vert \partial_{\mathbf{k}} u_{n}(\mathbf{k}) \right\rangle }$ \thinspace
is the non-Abelian Berry connection,  turns out to be a  boundary-dependent quantity that emerges due to the non-Hermitian effect and is given by
\begin{eqnarray}  \label{e22b} \nonumber
i \left\langle \psi_n(\mathbf{k}) \vert
\partial_{\mathbf{k}}
\psi_{m}(\mathbf{k}) \right\rangle 
&=&
( {\mathbf{A}}_{nm}(\mathbf{k}) 
- \left\langle u_{n}(\mathbf{k}) \vert \, \mathbf{r} \, \vert u_{m}(\mathbf{k}) \right\rangle )
\nonumber \\*
& = & i \frac{ {\mathbf{S}}_{nm}(\mathbf{k})} 
{\left( E_n(\mathbf{k})-E_m(\mathbf{k})\right)},
\end{eqnarray}
which can be viewed as a new result in the Berry phase literature showing the role of the non-Hermitian effect on Berry curvatures.

By using \thinspace Eq.~(\ref{e27}) \thinspace we can now make a comparison 
with the semi-classical electron's orbital magnetic moment given by Ref. \onlinecite{b25}. \thinspace In that framework, the electron's state \thinspace ${ \left| W_{o} \right\rangle  }$, \thinspace is a localized wave packet state, composed of one band Bloch states. The wave packet is sharply centered in ${ \mathbf{k} }$-space around the wave vector \thinspace ${ \mathbf{k}_{c} }$ \thinspace and its center of mass is well defined and given by ${ \left\langle  W_o \vert \, \mathbf{r} \, \vert  W_o \right\rangle = \mathbf{r}_{c} }$. Due to the self rotation of the wave packet around its center of mass, they found that the electron acquires an intrinsic orbital magnetic moment given by
\thinspace 
$ \displaystyle
\mathbf{m}_n (\mathbf{k}_{c}) =
\frac{ie}{2c \hbar} \!
\left\langle {\partial}_{\mathbf{k}_c} u_{n}(\mathbf{k}_c)
\vert 
\times 
\left( 
H_{k_{c}}(\mathbf{k}_c) - E_{n}(\mathbf{k}_c)  
\right)  
\vert 
{\partial}_{\mathbf{k}_c}
u_{n}(\mathbf{k}_c)\right\rangle 
$   where they have use the convention ${ e>0 }$. If we assume that the electron's state is a bulk state, as well as that the electron completely avoids the boundaries of the system where it is enclosed, hence the wavefunction ${ u_n(\mathbf{r},\mathbf{k}) }$ and all of its derivatives are zero over the boundaries, then, all non-Hermitian boundary terms become zero. In this respect, letting \thinspace ${ \mathbf{S}_{nm}(\mathbf{k})=0 }$ \thinspace 
in \thinspace Eq.~(\ref{e27}) \thinspace
we find
\thinspace
$ \displaystyle
\mathbf{m}_n (\mathbf{k})=
- \frac{e}{2c \hbar }
\text{Im} 
\left[ 
\left\langle \widetilde{{\partial}_{\mathbf{k}}}u_{n}(\mathbf{k})
\vert 
\times
\left( 
H_{k}(\mathbf{r},\mathbf{k}) - E_{n}(\mathbf{k})  
\right)  
\vert 
{\partial}_{\mathbf{k}}
u_{n}(\mathbf{k})\right\rangle  
\right]
$
\thinspace
where we have used
\thinspace
$
\left( 
H_{k}(\mathbf{r},\mathbf{k}) - E_{n}(\mathbf{k})  
\right) \!  
\left| 
\widetilde{{\partial}_{\mathbf{k}}}
u_{n}(\mathbf{k})\right\rangle
=
\left( 
H_{k}(\mathbf{r},\mathbf{k}) - E_{n}(\mathbf{k})  
\right) \!  
\left|  
\partial_{\mathbf{k}}
u_{n}(\mathbf{k})\right\rangle 	 
$. \thinspace 
Furthermore, by using the one-band covariant derivative \thinspace Eq.~(\ref{e21}), \thinspace we make the replacement
${  
\left\langle 
\widetilde{{\partial}_{\mathbf{k}}}
u_{n}(\mathbf{k})
\right| 
=
\left\langle 
\partial_{\mathbf{k}}u_{n}(\mathbf{k})
\right| 
+
\left\langle 
u_{n}(\mathbf{k})
\vert
\partial_{\mathbf{k}}u_{n}(\mathbf{k})
\right\rangle \!
\left\langle 
u_{n}(\mathbf{k})
\right|  
}$  \thinspace 
in the above approximated \thinspace ${ \mathbf{m}_n (\mathbf{k}) }$ \thinspace which results into

\begin{widetext}
\begin{eqnarray}  \nonumber
\mathbf{m}_n (\mathbf{k})
&=&
- \frac{e}{2c \hbar }
\text{Im} 
\left[ 
\left\langle \partial_{\mathbf{k}} u_{n}(\mathbf{k})
\vert 
\times
\left( 
H_{k}(\mathbf{r},\mathbf{k}) - E_{n}(\mathbf{k})  
\right)  
\vert 
{\partial}_{\mathbf{k}}
u_{n}(\mathbf{k})\right\rangle  
\right]
\\* \nonumber
&&- 
\frac{e}{2c \hbar}
\text{Im} 
\left[
\left\langle 
u_{n}(\mathbf{k})
\vert
\partial_{\mathbf{k}}u_{n}(\mathbf{k})
\right\rangle
\times
\left\langle u_{n}(\mathbf{k})
\vert 
\left( 
H_{k}(\mathbf{r},\mathbf{k}) - E_{n}(\mathbf{k})  
\right)  
\vert 
{\partial}_{\mathbf{k}}
u_{n}(\mathbf{k})\right\rangle  
\right].
\end{eqnarray}
The term \thinspace 
${ \left\langle 
u_{n}(\mathbf{k}) \vert 
\left( 
H_{k}(\mathbf{r},\mathbf{k}) - E_{n}(\mathbf{k})  
\right)  
\vert \partial_{\mathbf{k}}  u_{n}(\mathbf{k})\right\rangle }$ in the above equation, is also a non-Hermitian boundary term, which by assumption is also zero (due to the wavefunction ${ u_n(\mathbf{r},\mathbf{k}) }$ and of its derivatives being zero over the boundaries). In this respect, for the assumed bulk states, defined as the ones that the electron completely avoids the boundaries, the electron's magnetic moment is given by

\begin{equation} \label{e27b} 
\mathbf{m}_n (\mathbf{k}) \vert_{Localized \; State}^{Bulk}
=
- \frac{e}{2c \hbar }
\text{Im} 
\left[ 
\left\langle \partial_{\mathbf{k}} u_{n}(\mathbf{k})
\vert 
\times
\left( 
H_{k}(\mathbf{r},\mathbf{k}) - E_{n}(\mathbf{k})  
\right)  
\vert 
{\partial}_{\mathbf{k}}
u_{n}(\mathbf{k})\right\rangle  
\right]
\end{equation}
\\
which is a form that  has the same structure as the real part of semi-classical electron's orbital magnetic moment \thinspace 
${ \text{Re} [ \mathbf{m}_n (\mathbf{k}_{c})]
=  \text{Im} [ i \, \mathbf{m}_n (\mathbf{k}_{c})] }$ \thinspace
of \thinspace Ref. \onlinecite{b25}.

\end{widetext}
It is evident that, the semi-classical electron's orbital magnetic moment \thinspace ${ \mathbf{m}_n (\mathbf{k}_{c}) }$ \thinspace does not explicitly take into account contributions from the realistic boundaries of a material, and as a consequence, the electrons' magnetization formulas that are derived by semi-classical means do not take into account such boundary contributions. We expect that our more general result \thinspace Eq.~(\ref{e27}) \thinspace will be able to provide such contributions.

Moreover, our own result for the electron's orbital magnetic moment formula \thinspace Eq.~(\ref{e27}) \thinspace satisfies the two basic invariant properties, namely it is invariant with respect to gauge transformations of the form \thinspace ${ u_n(\mathbf{k})\rightarrow    
e^{ \displaystyle i f_n(\mathbf{k})}  	u_n(\mathbf{k}) }$ \thinspace and with respect to a shift of the zero of the Hamiltonian \thinspace ${ H_{k}(\mathbf{r},\mathbf{k})\rightarrow  H_{k}(\mathbf{r},\mathbf{k}) + \epsilon }$; \linebreak 
we expect therefore that the many body orbital magnetization formula that we derive further below will share the same invariant properties.
Also, as will be explicitly shown in the next subsection, a boundary contribution that is encoded by the one-band covariant derivative is hidden within the first term of the right hand side of our \thinspace Eq.~(\ref{e27}) \thinspace and it is attributed to the emerging momentum gradient \ operator \ $\partial_{\mathbf{k}}$ \ anomaly.

Finally, let us in what follows use our  \thinspace Eq.~(\ref{e27}) (or Eq.~(\ref{e18i}))  to provide a general result for the total orbital magnetization and apply it to particular cases, by always keeping an eye on corresponding results in the literature. (Our total final result is \thinspace Eq.~(\ref{e31}) \thinspace  below.) Let us, however, first start with the simplest one-band case.

\subsubsection{One-band formula}
In the one-band formula we assume that each combination of the 
off-diagonal matrix elements \thinspace
${
( {\mathbf{A}}_{nm}(\mathbf{k}) 
- \left\langle u_{n}(\mathbf{k}) \vert \, \mathbf{r} \, \vert u_{m}(\mathbf{k}) \right\rangle )
}$ \thinspace 
can be neglected,  due to
\[
\displaystyle  \frac{ {\mathbf{S}}_{nm}(\mathbf{k})} 
{\left( E_n(\mathbf{k})-E_m(\mathbf{k})\right)} \rightarrow 0 \]
which is a good approximation for conventional insulators with large band gap and negligible boundary contributions as evidenced from \thinspace Eq.~(\ref{e18l})
\thinspace and
\thinspace Eq.~(\ref{apc11}).
If we assume that in Eq. (49) each of the non-Hermitian effect terms \thinspace  
${ {\mathbf{S}}_{nm}(\mathbf{k}) }$ \thinspace is zero, then \thinspace
${
\hbar \left\langle u_{n}(\mathbf{k}) \vert \, {\mathbf{v}}_{b} \, \vert u_{m}(\mathbf{k}) \right\rangle
=
-
\left\langle u_{n}(\mathbf{k}) \vert \, \mathbf{k}_{b} \, \vert u_{m}(\mathbf{k}) \right\rangle
 }$
must be satisfied, which in the simplest scenario is fulfilled whenever the electrons completely avoid the boundaries of the material and at the same time no band-crossings exist in the Brillouin zone. 

In a different point of view, the assumption of zero value for the off-diagonal matrix elements, namely, \thinspace
${ \left( 
\mathbf{A}_{nm}(\mathbf{k}) 
- \left\langle u_{n}(\mathbf{k}) \vert \, \mathbf{r} \, \vert u_{m}(\mathbf{k}) \right\rangle 
 \right) 
= 0 }$  ({\it cf}. \thinspace Eq.~(\ref{e22b}))
\thinspace
can be attributed to adiabatically deformed Bloch eigenstates \thinspace  $ \left|  \Psi_{m}(\mathbf{k}) \right\rangle  $ \thinspace with respect to crystal momentum differentiation, that  is, \thinspace
${ \displaystyle \left| 
\partial_{\mathbf{k}} \Psi_{m}(\mathbf{k}) 
\right\rangle 
=
\left\langle 
\Psi_{m}(\mathbf{k}) 
\vert 
\partial_{\mathbf{k}} \Psi_{m}(\mathbf{k}) 
\right\rangle 
\left| \Psi_{m}(\mathbf{k}) \right\rangle
}$.  \thinspace
The latter equality is satisfied whenever each one of the off-diagonal amplitudes 
${\left\langle 
\Psi_{n}(\mathbf{k}) 
\vert 
\partial_{\mathbf{k}} \Psi_{m}(\mathbf{k})
\right\rangle  
}$
is zero which defines the restriction \thinspace
${\left\langle 
\Psi_{n}(\mathbf{k}) 
\vert 
\partial_{\mathbf{k}} \Psi_{m}(\mathbf{k}) 
\right\rangle = 0  }$ 
\thinspace for ${ n \neq m }$. 
Substituting  \thinspace
${ \Psi_{m}(\mathbf{r},\mathbf{k})
=e^{\displaystyle i \mathbf{k.r}}
u_{m}(\mathbf{r},\mathbf{k})
}$   \thinspace  
in the latter adiabatically deformed restriction we find the former assumption of zero value for each one of the off-diagonal matrix elements \thinspace
${ \left( 
\mathbf{A}_{nm}(\mathbf{k}) 
- \left\langle u_{n}(\mathbf{k}) \vert \, \mathbf{r} \, \vert u_{m}(\mathbf{k}) \right\rangle 
\right) 
 }$.

In this respect and within the one-band (adiabatically deformed) approximation, the many-body electron orbital magnetization
(Eq.~(\ref{e27})) is given by
\begin{widetext}
\begin{equation} \label{e28}
\mathbf{M}=- \frac{e}{2c \hbar (2\pi)^3 }
\sum_{E_{n} \leq \mu} \iiint_{BZ} \! f_{n}(\mathbf{k},\mu) \, 
\text{Im} 
\left[ \, 
\left\langle \widetilde{{\partial}_{\mathbf{k}}}u_{n}(\mathbf{k})
\vert 
\, \times \,
\left( 
H_{k}(\mathbf{r},\mathbf{k}) - E_{n}(\mathbf{k})  
\right)  
\vert 
\widetilde{{\partial}_{\mathbf{k}}}
u_{n}(\mathbf{k})\right\rangle 
\, \right] 
\,d^3k 
\end{equation}
\end{widetext} 
which satisfies the two basic invariant properties, namely, it is gauge invariant and invariant with respect to a shift of the zero of the Hamiltonian. 

Although we have apparently dropped out any boundary contributions of the orbital magnetization by approximating the off-diagonal matrix elements  
\thinspace $\mathbf{S}_{nm}(\mathbf{k}) $ \thinspace
values as zero, there still exists an explicit boundary contribution within \thinspace Eq.~(\ref{e28}) \thinspace which is attributed to the one-band covariant derivative. Specifically, if we use the definition of the one-band covariant derivative as given by \thinspace Eq.~(\ref{e21}) \thinspace we can recast the integrand of \thinspace Eq.~(\ref{e28}) \thinspace in the following form  
\begin{eqnarray} \label{e29} \nonumber
&& \left\langle  \widetilde{{\partial}_{\mathbf{k}}}u_{n}(\mathbf{k})
\vert 
\, \times   \,
\left( 
H_{k}(\mathbf{r},\mathbf{k}) - E_{n}(\mathbf{k})  
\right)  
\vert 
\widetilde{{\partial}_{\mathbf{k}}}
u_{n}(\mathbf{k})\right\rangle 
\nonumber \\*[4pt] 
 && \;\;\;\;\;\;\;\;\;\; =  
\left\langle \partial_{\mathbf{k}} u_{n}(\mathbf{k})
\vert 
\, \times \,
\left( 
H_{k}(\mathbf{r},\mathbf{k}) - E_{n}(\mathbf{k})  
\right)  
\vert 
{\partial}_{\mathbf{k}}
u_{n}(\mathbf{k})\right\rangle
\nonumber \\*[4pt]
&& \;\;\;\;\;\;\;\;\;\;\;\;\; 
- i \, \mathbf{A}_{nn}(\mathbf{k}) 
\times 
\left\langle u_{n}(\mathbf{k})
\vert \, 
H_{k}(\mathbf{r},\mathbf{k}) - E_{n}(\mathbf{k})   
\, \vert 
{\partial}_{\mathbf{k}}
u_{n}(\mathbf{k})\right\rangle.
\nonumber \\*[-1pt]
\end{eqnarray}
The second term on the right hand side of \thinspace Eq.~(\ref{e29}) \thinspace gives a non-zero boundary contribution to the orbital magnetization only whenever the non-Hermitian effect with respect to the momentum gradient operator emerges, that is
\begin{eqnarray} \label{e30} \nonumber
&& \left\langle u_{n}(\mathbf{k})
\vert \, 
H_{k}(\mathbf{r},\mathbf{k}) - E_{n}(\mathbf{k})   
\, \vert 
{\partial}_{\mathbf{k}}
u_{n}(\mathbf{k})\right\rangle=  
\\*[4pt]  \nonumber
 && \;\;\;\;\;\;\;\;\;\;\;\;\;\;\;\;\;\;\;\;\;\;\;  
 -
\left\langle u_{n}(\mathbf{k})
\vert \, 
H_{k}(\mathbf{r},\mathbf{k})^{+} - H_{k}(\mathbf{r},\mathbf{k})   
\, \vert 
{\partial}_{\mathbf{k}}
u_{n}(\mathbf{k})\right\rangle
\\*[4pt]  \nonumber
&& \;\;\;\;\;\;\;\;\;\;\;\;\;\;\;\;\;
=
-
\frac{i \hbar}{2} \oiint_S 
\mathbf{n}\!\cdot\! 
\left( \,
( \mathbf{v} \, u_{n} )^{\displaystyle *} 
+ u_{n}^{\displaystyle *} \,\mathbf{v}  
\, \right) \! 
\, \partial_{\mathbf{k}}u_{n}
\, dS
\, \neq \, 0
\\*[-4pt]
\end{eqnarray}
where \thinspace ${u_n = u_n(\mathbf{r},\mathbf{k})}$
\thinspace are the cell-periodic eigenfunctions. If we further assume within a stricter approximation that the position operator \thinspace $\mathbf{r}$ \thinspace and the momentum gradient operator \ $\partial_{\mathbf{k}}$ \thinspace are separately normal operators, that is the expectation value of the boundary momentum gradient operator \thinspace   
${ \left\langle u_{n}(\mathbf{k}) \vert \, \mathbf{k}_{b} \, \vert u_{n}(\mathbf{k}) \right\rangle }$ 
\thinspace
is zero, \thinspace 
${ 
\left\langle u_{n}(\mathbf{k})
\vert \, 
H_{k}(\mathbf{r},\mathbf{k}) - E_{n}(\mathbf{k})   
\, \vert 
{\partial}_{\mathbf{k}}
u_{n}(\mathbf{k})\right\rangle
=0
}$, 
we can replace the covariant derivative entering \thinspace Eq.~(\ref{e28}) \thinspace with the normal derivative that yields the orbital magnetization formula that was derived in \thinspace Ref.\onlinecite{b22} \thinspace but with the correct opposite sign between the Hamiltonian operator and the energy.
Alternatively, if one assumes from the beginning a solid with one band denoted by $n$ then, the sum in the second term on the right side of
\thinspace Eq.~(\ref{e18i}), \thinspace will not be present due to the one band closure relation 
\thinspace 
${ \displaystyle 
I=
\iiint_{BZ} \!\! d^3k'	
\left| \psi_{n}(\mathbf{k'}) \right\rangle \left\langle \psi_{n}(\mathbf{k'}) \right|  }$, \thinspace
that must be used in  \thinspace Eq.~(\ref{apb10f}) \thinspace and subsequently in  \thinspace Eq.~(\ref{apb10h}) \thinspace leading to \thinspace Eq.~(\ref{e18h}).

\subsubsection{Many-band formula}

In the many-band formula we don't a priori make any assumption with respect to the behavior of 
the position operator \thinspace $\mathbf{r}$ \thinspace and the momentum gradient \thinspace operator \thinspace $\partial_{\mathbf{k}}$, \thinspace thus no restrictions for the wavefunctions' boundary conditions are made; the goal is to derive a general formula applicable to non-interacting electrons within topological materials, insulators or semimetals. In this respect, and because of Eq.~(\ref{e27}), the many-band orbital magnetization formula of non-interacting electrons within a periodic solid is given by
\begin{widetext}	
\begin{eqnarray} \label{e31}  \nonumber
\mathbf{M} & = & - \frac{e}{2c \hbar (2\pi)^3 }
\sum_{E_{n} \leq \mu} \iiint_{BZ} \! f_{n}(\mathbf{k},\mu) \, 
\text{Im} 
\left[ \, 
\left\langle \widetilde{{\partial}_{\mathbf{k}}}u_{n}(\mathbf{k})
\vert 
\, \times \,
\left( 
H_{k}(\mathbf{r},\mathbf{k}) - E_{n}(\mathbf{k})  
\right)  
\vert 
\widetilde{{\partial}_{\mathbf{k}}}
u_{n}(\mathbf{k})\right\rangle 
\, \right] 
d^3k
\nonumber  \\*[4pt]
&& - 
\frac{e}{2c \hbar (2\pi)^3 }
\sum_{E_{n} \leq \mu} \, \, \sum_{m \neq n} \iiint_{BZ} \! f_{n}(\mathbf{k},\mu) \, 
\text{Im} 
\left[ \,\frac{1}{\left( E_n(\mathbf{k})-E_m(\mathbf{k})\right)}
\, \,
{\mathbf{S}}_{nm}(\mathbf{k})
\times 
\left\langle u_{m}(\mathbf{k}) 
\vert \,
H_{k}(\mathbf{r},\mathbf{k}) - E_{n}(\mathbf{k})  
\, \vert
{\partial}_{\mathbf{k}}
u_{n}(\mathbf{k})\right\rangle 
\, \right] d^3k
\nonumber \\*[-4pt] 
\end{eqnarray}
\end{widetext}
which is valid for arbitrary boundary conditions on the wavefunctions \thinspace $u_m(\mathbf{r},\mathbf{k})$.
Orbital magnetization many-band formula \thinspace Eq.~(\ref{e31}) \thinspace is the major result of this work; it rigorously provides within a quantum mechanical theoretical framework, and
without any Wannier-localization approximation or heuristic extension\cite{b23}, the manner in which one could generally model the orbital magnetization of periodic topological solids.

The energy differences in the denominator of the second term on the right hand side of \thinspace Eq.~(\ref{e31}) \thinspace captures the possible local (in momentum space) 
gigantic orbital magnetization contributions in the vicinity of band crossings. These gigantic orbital magnetization contributions are predicted to occur only whenever band crossings exist along with an imbalance of electron accumulation at the opposite boundary surfaces of the material that creates a Hall voltage. 

In order to verify the need of the presence of a Hall voltage, we will show that within PBSs for the wavefunctions (thus with no electron accumulation occurring) no gigantic local contribution of the orbital magnetization is possible even if the material is topological. Within PBCs the momentum gradient operator \thinspace $\partial_\mathbf{k}$ \thinspace does not break the domain of definition of the Hamiltonian, that is the wavefunctions \thinspace
${ u_n(\mathbf{r},\mathbf{k}) }$ \thinspace and \thinspace 
${ \partial_\mathbf{k} u_n(\mathbf{r},\mathbf{k}) }$ \thinspace both satisfy periodic boundary conditions in $\mathbf{r}$-space. 
Indeed, the latter periodicity of \thinspace 
${ \partial_\mathbf{k} u_n(\mathbf{r},\mathbf{k}) }$ \thinspace
can be deduced from the periodicity \thinspace 
${ u_n(\mathbf{r+L},\mathbf{k}) = u_n(\mathbf{r},\mathbf{k}) }$, where ${\mathbf{L}}$ is the length of the material, by differentiating both sides with respect to the momentum ${\mathbf{k}}$ (which is treated as an independent parameter in the assumed thermodynamic limit) that gives \thinspace   
${ \partial_\mathbf{k} u_n(\mathbf{r+L},\mathbf{k}) = \partial_\mathbf{k} u_n(\mathbf{r},\mathbf{k}) }$. 
In this fashion, each one of the matrix elements \thinspace  
${ \left\langle u_{n}(\mathbf{k}) \vert \, \mathbf{k}_{b} \, \vert u_{m}(\mathbf{k}) \right\rangle }$ \thinspace
is zero due to symmetry, and any emergence of the non-Hermitian effect
owing to the momentum gradient
operator \thinspace $\partial_{\mathbf{k}}$ anomaly is prohibited. 
We point out that the absence of this non-Hermitian effect is invariant with respect to twisted
boundary conditions of the form \thinspace
${ u_n(\mathbf{r+L},\mathbf{k}) = e^{\displaystyle i f(\mathbf{L},\mathbf{k})} u_n(\mathbf{r},\mathbf{k}) }$
\thinspace
as long as the system is closed.

By recasting the 
${ 
\left\langle u_{m}(\mathbf{k}) 
\vert \,
H_{k}(\mathbf{r},\mathbf{k}) - E_{n}(\mathbf{k})  
\, \vert
{\partial}_{\mathbf{k}}
u_{n}(\mathbf{k})\right\rangle
}$  
term entering the right hand side of  \thinspace Eq.~(\ref{e31}) \thinspace
in the form
\begin{eqnarray} \label{e32} \nonumber
&& \left\langle u_{m}(\mathbf{k}) 
\vert \,
H_{k}(\mathbf{r},\mathbf{k}) - E_{n}(\mathbf{k})  
\, \vert
{\partial}_{\mathbf{k}}
u_{n}(\mathbf{k})\right\rangle = 
\nonumber \\*[4pt]
&& 
\left( 
E_m(\mathbf{k})-E_n(\mathbf{k})
\right) \!
\left\langle u_{m}(\mathbf{k}) \vert \partial_{\mathbf{k}} u_{n}(\mathbf{k}) \right\rangle
-
\left\langle u_{m}(\mathbf{k}) \vert \, \mathbf{k}_{b} \, \vert u_{n}(\mathbf{k}) \right\rangle, 
\nonumber \\*[-2pt]
\end{eqnarray}
as well as by taking into account the non-Hermitian boundary term as given by \thinspace Eq.~(\ref{apc11})
\[
{\mathbf{S}}_{nm}(\mathbf{k})=
\hbar \left\langle u_{n}(\mathbf{k}) \vert \, {\mathbf{v}}_{b} \, \vert u_{m}(\mathbf{k}) \right\rangle
+
\left\langle u_{n}(\mathbf{k}) \vert \, \mathbf{k}_{b} \, \vert u_{m}(\mathbf{k}) \right\rangle,  
\]
then, under periodic boundary conditions the boundary momentum gradient 
\thinspace \textquotedblleft{operator}\textquotedblright  \ 
\thinspace $\mathbf{k}_{b}$ \thinspace matrix elements given by 
\thinspace  Eq.~(\ref{e26})  \thinspace are zero
\thinspace ${ \left\langle u_{m}(\mathbf{k}) \vert \, \mathbf{k}_{b} \, \vert u_{n}(\mathbf{k}) \right\rangle=0 }$, \thinspace and the multi-band and unrestricted orbital magnetization formula \thinspace Eq.~(\ref{e31}) \thinspace takes the form

\begin{widetext}
\begin{eqnarray} \label{e33}  \nonumber
\mathbf{M} & = & - \frac{e}{2c \hbar (2\pi)^3 }
\sum_{E_{n} \leq \mu} \iiint_{BZ} \! f_{n}(\mathbf{k},\mu) \, 
\text{Im} 
\left[ \, 
\left\langle {\partial}_{\mathbf{k}}u_{n}(\mathbf{k})
\vert 
\, \times \,
\left( 
H_{k}(\mathbf{r},\mathbf{k}) - E_{n}(\mathbf{k})  
\right)  
\vert 
{\partial}_{\mathbf{k}}
u_{n}(\mathbf{k})\right\rangle 
\, \right] 
d^3k
\nonumber  \\*[4pt]
&& + 
\frac{e}{2c (2\pi)^3 }
\sum_{E_{n} \leq \mu} \, \, \sum_{m \neq n} \iiint_{BZ} \! f_{n}(\mathbf{k},\mu) \, 
\text{Im} 
\left[ \,
\left\langle u_{n}(\mathbf{k}) \vert \, {\mathbf{v}}_{b} \, \vert u_{m}(\mathbf{k}) \right\rangle
\times 
\left\langle u_{m}(\mathbf{k})  
\vert
{\partial}_{\mathbf{k}}
u_{n}(\mathbf{k})\right\rangle 
\, \right] d^3k
\end{eqnarray}
\end{widetext}
where we have also replaced the one-band covariant derivative with the normal one 
\begin{eqnarray} \label{e29a} \nonumber
&& \left\langle  \widetilde{{\partial}_{\mathbf{k}}}u_{n}(\mathbf{k})
\vert 
\, \times   \,
\left( 
H_{k}(\mathbf{r},\mathbf{k}) - E_{n}(\mathbf{k})  
\right)  
\vert 
\widetilde{{\partial}_{\mathbf{k}}}
u_{n}(\mathbf{k})\right\rangle 
\nonumber \\*[4pt]  \nonumber 
&& \;\;\;\;\;\;\;\;\;\; =  
\left\langle \partial_{\mathbf{k}} u_{n}(\mathbf{k})
\vert 
\, \times \,
\left( 
H_{k}(\mathbf{r},\mathbf{k}) - E_{n}(\mathbf{k})  
\right)  
\vert 
{\partial}_{\mathbf{k}}
u_{n}(\mathbf{k})\right\rangle
\end{eqnarray}
due to \thinspace ${ \left\langle u_{n}(\mathbf{k}) \vert \, \mathbf{k}_{b} \, \vert u_{n}(\mathbf{k}) \right\rangle=0 }$ \thinspace
in accordance with \thinspace Eq.~(\ref{e29}) \thinspace and \thinspace Eq.~(\ref{e30}).

It is now clear from \thinspace Eq.~(\ref{e33}) \thinspace that, whenever a Hall voltage is zero owing to periodic boundary conditions, the orbital magnetization cannot acquire local gigantic values, even if the material is topological with non-trivial band structure crossings, while whenever imbalance of electron charge is formed, local gigantic orbital magnetization contributions near the band crossings are generically expected to occur.

It is also interesting to point out that, whenever the material's realistic boundary conditions are periodic, 
by expanding the cell periodic functions in a Fourier series over all reciprocal lattice vectors ${ \mathbf{G} }$, namely, ${ u_n(\mathbf{r}, \mathbf{k})
= \sum_{\mathbf{G}} C_{n} (\mathbf{k},\mathbf{G}) e^{\displaystyle -i\mathbf{G}\!\cdot\! \mathbf{r}} 
}$, it is evident  that  ${\partial}_{\mathbf{k}}
u_{n}(\mathbf{r}, \mathbf{k})$ is periodic in space (as well as ${ u_n(\mathbf{r}, \mathbf{k}) }$ and the Hamiltonian  ${ H_{k}(\mathbf{r},\mathbf{k}) }$). By then using the normalization convention  \thinspace
${\displaystyle\left\langle \Psi_{n}(t, \mathbf{k})\vert \Psi_{n}(t, \mathbf{k})\right\rangle= \left\langle u_{n}(\mathbf{k})\vert u_{n}(\mathbf{k})\right\rangle_{cell}=1}$, \thinspace that is, assume a Bloch state in the form ${ \displaystyle \left| \Psi_{n}(t, \mathbf{k}) \right\rangle =
\frac{1}{\sqrt{N}} \,
e^{\displaystyle -\frac{1}{\hbar} E_{n}(\mathbf{k})t} e^{\displaystyle i\mathbf{k.r}} \left|u_{n}(\mathbf{k})\right\rangle
}$, we replace  \thinspace 
${ \displaystyle 
\left|u_{n}(\mathbf{k})\right\rangle 
\rightarrow 
\frac{1}{\sqrt{N}} \left|u_{n}(\mathbf{k})\right\rangle
}$ \thinspace
in all terms in 
\thinspace Eq.~(\ref{e33}) \thinspace
(the initially assumed eigenstate was normalized over the volume ${ V }$ without taking into account the cell normalization convention, therefore it defers by a factor ${\displaystyle 
\frac{1}{\sqrt{N}} }$)
and exploiting the symmetry of the integrands, the orbital magnetization formula it truncates into the form

\begin{widetext}
\begin{eqnarray} \label{e34}  \nonumber
\mathbf{M} & = & - \frac{e}{2c \hbar (2\pi)^3 }
\sum_{E_{n} \leq \mu} \iiint_{BZ} \! f_{n}(\mathbf{k},\mu) \, 
\text{Im} 
\left[ \, 
\left\langle {\partial}_{\mathbf{k}}u_{n}(\mathbf{k})
\vert 
\, \times \,
\left( 
H_{k}(\mathbf{r},\mathbf{k}) - E_{n}(\mathbf{k})  
\right)  
\vert 
{\partial}_{\mathbf{k}}
u_{n}(\mathbf{k})\right\rangle_{cell} 
\, \right] 
d^3k
\nonumber  \\*[4pt]
&& + 
\frac{e}{2c (2\pi)^3 }
\sum_{E_{n} \leq \mu} \, \, \sum_{m \neq n} \iiint_{BZ} \! f_{n}(\mathbf{k},\mu) \, 
\text{Im} 
\left[ \,
\left\langle u_{n}(\mathbf{k}) \vert \, {\mathbf{v}}_{b} \, \vert u_{m}(\mathbf{k}) \right\rangle_{cell}
\times 
\left\langle u_{m}(\mathbf{k})  
\vert
{\partial}_{\mathbf{k}}
u_{n}(\mathbf{k})\right\rangle_{cell} 
\, \right] d^3k
\end{eqnarray}
\end{widetext}
where all space integrals are performed within one primitive cell, and the off-diagonal matrix elements of the boundary velocity are given by

\begin{eqnarray} \label{e35} \nonumber
\left\langle u_{n}(\mathbf{k}) \vert \, {\mathbf{v}}_{b} \, \vert u_{m}(\mathbf{k}) \right\rangle 
= 
-\frac{1}{2}\oiint_{cell}
\! \! \! \! \! 
\mathbf{r}
\left( 
(\mathbf{v} \, u_{n})^{\displaystyle *} u_{m}
+ u_{n}^{\displaystyle *}  \mathbf{v} \, u_{m} 
 \right) \! \cdot \! d\mathbf{S} 
\nonumber  \\*[-4pt]
\end{eqnarray}
in accordance with \thinspace Eq.~(\ref{e25}), \thinspace where we have taken into account that \thinspace
${ u_n(\mathbf{r}, \mathbf{k}) }$  \thinspace are periodic over the unit cell boundaries.
\thinspace Eq.~(\ref{e34}) \thinspace
can be thought as the ${ \mathbf{k} }$-space analog of 
\thinspace Eq.~(\ref{e15}).

It is worth comparing (i) the orbital magnetization formula of periodic solids that is given by \thinspace Eq.~(\ref{e33}) \thinspace with the one that was proposed in \thinspace Ref.\onlinecite{b23} \thinspace by a heuristic argument, as well as, (ii) compare
the general orbital magnetization formula 
\thinspace Eq.~(\ref{e31}) \thinspace with the one derived in \thinspace Ref. \onlinecite{b2}, \thinspace where they propose a theoretical approach to discriminate the separate contributions of the total magnetization, that is, the one contribution coming from the bulk states and the other coming from the surface states. 

(i) Orbital magnetization formula 
\thinspace Eq.~(\ref{e33}), \thinspace is relaxed from any Wannier localization approximation as well as from the periodic gauge approximation, and it is therefore valid for topological materials as long as the electrons' wavefunctions satisfy periodic boundary conditions (zero Hall voltage) over the materials boundaries. The heuristic extension of the orbital magnetization formula\cite{b22} by an additional term, assumed to be proportional to the chemical potential, that was made in  
\thinspace Ref.\onlinecite{b23} \thinspace in order to model the orbital magnetization of Chern insulators and metals,
is rigorously given by the second term of the right hand side of \thinspace Eq.~(\ref{e33}). This term has explicit boundary contributions due to the off-diagonal matrix elements of the boundary velocity operator \thinspace $ \mathbf{v}_{b} $ \thinspace which are not zero due to the emerging non-Hermitian effect of the position operator \thinspace $\mathbf{r}$ \thinspace that becomes anomalous within periodic boundary conditions, as should be expected.

(ii) In \thinspace Ref. \onlinecite{b2} \thinspace they use the standard circulation operator together with the spectral resolution of the Hamiltonian, and as a result of this spectral resolution, the (undefined) diagonal matrix elements of the position operator are excluded from the circulation operator formula; one can therefore evaluate the standard circulation operator expectation value (even in the thermodynamic limit within PBCs), hence one can calculate the orbital magnetization. Due to the spectral resolution within the circulation operator, the assumed orbitals must satisfy \thinspace  
${ \left\langle \phi_n  \vert \, \mathbf{v} \, \vert \phi_n \right\rangle= - \left\langle \phi_n  \vert \, \mathbf{v}_b \, \vert \phi_n \right\rangle =0  }$  
\thinspace
owing to \thinspace ${ \left\langle \phi_n  \vert \, \mathbf{v}_b \, \vert \phi_n \right\rangle=0  }$, \thinspace therefore,  the orbitals that are taken into account indeed describe bound bulk states. Then,   \thinspace Ref. \onlinecite{b2} \thinspace  
extracted their result from the semi-classical orbital magnetization formula given in
\thinspace Ref. \onlinecite{b24}, \thinspace and they stated that the remaining part gives the boundary contribution of the orbital magnetization. The theoretical method that they use rests on the argument that the semi-classical orbital magnetization formula given in 
\thinspace Ref. \onlinecite{b24} \thinspace
correctly gives the total (bulk and boundary) orbital magnetization of non interacting electrons. We argue that this may not be entirely true due to the  approximations that are made during the derivation of the \thinspace Ref. \onlinecite{b24} \thinspace orbital magnetization formula. First, as we have shown in the derivation of \thinspace Eq.~(\ref{e27b}), \thinspace the structure of the semi-classical electron's orbital magnetic moment can be attained by the unrestricted quantum formula \thinspace Eq.~(\ref{e27}) \thinspace
whenever the electron's state is a localized bulk state, that is, when the electron completely avoids the boundaries of the system where it is enclosed. Therefore, the orbital magnetization that is evaluated only by taking into account the electron's semi-classical orbital magnetic moment, does not account for magnetization contributions coming from all possible states, i.e. does not take into account topologically non trivial extended states. On the other hand, the semi-classical orbital magnetization formula given in \thinspace Ref. \onlinecite{b24}, \thinspace namely as the derivative of the electrons' total energy with respect to the magnetic field (at zero magnetic field), besides the contribution coming from the electrons' intrinsic orbital moment, it also acquires two extra terms that come up due to the modified density of states. One is attributed to the explicit magnetic field dependence of the density of states and the other is due to the resulting change in the Fermi volume. 
The two extra terms cannot carry any topologically non trivial information, on one hand due to the localized wavepacket employed, and on the other hand due to the explicit assumption \thinspace  ${ \bm{\nabla}_{\! k} \! \cdot \! {\mathbf\Omega}_n({\mathbf{k}})=0 }$, where \thinspace  ${ {\mathbf\Omega}_n({\mathbf{k}}) }$ is the Berry curvature, that was made in \thinspace Ref. \onlinecite{b24} \thinspace for deriving the modified density of states (hence Berry type of monopoles, crucial for the non trivial topology, were ignored). In this framework, these two extra terms most probably represent corrections to the semi-classical orbital magnetization formula and do not carry any topologically non trivial information. We argue therefore that, although the method followed in \thinspace Ref. \onlinecite{b2} \thinspace is reasonable, the findings do not represent the orbital magnetization of topologically non trivial surface states, but they instead provide trivial corrections to the orbital magnetization for non localized states. This discussion here is given so that our results \thinspace Eq.~(\ref{e31})
\thinspace or \thinspace Eq.~(\ref{e33})
\thinspace can be directly compared with the state of the art results.

\section{Conclusions} \label{s4}
We have reconsidered the modern theory of orbital magnetization through careful definition of additional quantities that rigorously and analytically take into account the boundary contributions to the observable. 
These contributions are shown to originate from non-Hermitian effects that emerge whenever the
position operator $\mathbf{r}$ and the momentum gradient operator \thinspace $\partial_{\mathbf{k}}$ \thinspace (that enter the Ehrenfest and the Hellmann-Feynman theorems respectively) become anomalous, in the sense that they break the domain of definition of the Hamiltonian operator. 
In this theoretical framework, we have first extended the standard velocity operator definition in order to incorporate the anomaly of the position operator that is inherited in band theory, which results to an explicit boundary velocity contribution. 

Using the extended velocity, we have defined the electrons' intrinsic orbital circulation within Bloch representation which we have shown that is an intensive and well defined quantity of periodic solids that properly counts the circulating micro-currents embodied in the wavefunctions' bulk and boundary structure. Using the defined electrons' intrinsic circulation, we have made a rigorous connection between the $n \text{th}$ band electrons' collective intrinsic circulation and the local (LC) and itinerant circulation (IC) contributions, that are used within Wannier-localization and periodic gauge approximation in the modern theory of orbital magnetization \cite{{b18},{b22},{b23}}.

With these concepts in hand, we have been able to rigorously reconsider the modern theory of orbital magnetization and derive quantum mechanical expressions for the orbital magnetization of non-interacting electrons that move within extended and topological solids (insulators or semimetals),
without any Wannier-localization approximation \cite{{b18},{b22},{b34}} or heuristic extension \cite{b23} been made. 

We have rigorously shown that, in the one-band approximation $\mathbf{k}$-space formula,
a one-band covariant derivative enters the magnetization formula as an emerging operator 
due to the non-Hermitian effect that is attributed to the anomaly of the momentum gradient operator 
\thinspace $\partial_{\mathbf{k}}$; the one-band covariant derivative can be replaced by the normal derivative only whenever PBCs are satisfied. 

In the many-band and unrestricted \thinspace $\mathbf{k}$-space formula of the orbital magnetization, the non-Hermitian effect has been shown to contribute an additional boundary term that originates from the anomalies of the position operator $\mathbf{r}$ and the momentum gradient operator 
\thinspace $\partial_{\mathbf{k}}$. This additional boundary term, is expected to give local gigantic orbital magnetization contributions in the vicinity of band crossings in topological materials (insulators or semimetals) whenever band crossings exist along with Hall voltage due to imbalance of electron accumulation at the opposite boundaries of the materials.  These local gigantic 
orbital magnetization contributions are encoded by the emerging non-Hermitian effect of the momentum  
gradient operator
\thinspace $\partial_{\mathbf{k}}$ \thinspace that becomes anomalous whenever PBCs for the electrons' wavefunctions are broken. 
On the contrary, whenever Hall voltage is zero and the electrons' 
wavefunctions satisfy PBCs, the momentum  
gradient operator 
\thinspace $\partial_{\mathbf{k}}$ \thinspace has a well defined behavior and, as a consequence, gigantic boundary contributions are not possible. 
By making a comparison between our derived formula and the one that had heuristically been given (in order to model the orbital magnetization of Chern insulators and metals) in \thinspace Ref.\onlinecite{b23}, \thinspace 
we have shown the (previously unnoticed) property that, within periodic boundary conditions, the orbital magnetization has explicit boundary contributions encoded by the off-diagonal matrix elements of the boundary velocity operator (which are not zero due to the emerging non-Hermitian effect of the position operator that becomes anomalous within PBCs).
Finally, we point out that, all boundary contributions that emerge due to non-Hermitian effects can equally be calculated as bulk properties whenever the integrations by parts are not performed.   
\\
\\   




\appendix 

\section{Position expectation value 
$\left\langle \mathbf{r} \right\rangle_{n} $, displacement $\Delta\left\langle \mathbf{r} \right\rangle$, orbital circulation $\left\langle \mathbf{C}\right\rangle_{n} $, and intrinsic orbital circulation $\left\langle \mathbf{C}_{intr}\right\rangle_{n} $} \label{a1} 

In the following all calculations are performed for one electron states within Bloch representation.

\subsection{Explicit calculation of $\left\langle \mathbf{r} \right\rangle_{n} $}
For simplicity and without loss of generality we assume 1D configuration (while the generalization to 3D is straightforward). We assume a closed system ${ \left\langle \Psi (t)\vert \Psi(t)\right\rangle=1}$ of length $L_{x}$ with PBSs for the wavefunction over the edges. We calculate the electrons position expectation value ${ \left\langle \Psi (t)\vert \, x \, \vert \Psi(t)\right\rangle}$ with respect to a Bloch eigenstate ${ \displaystyle \left|\Psi_{n}(t,k)\right\rangle= \frac{1}{\sqrt{N_x}} e^{\displaystyle -\frac{i}{\hbar} E_{n}(k)t} e^{\displaystyle ikx} \left| u_{n}(k)\right\rangle }$. The length $L_{x}$ of the system is equal to ${L_x = N_x \alpha_x}$, where $\alpha_x$ is the primitive cell length and $N_x$ the number of the primitive cells of the system. The Bloch state  ${ \left|\Psi_{n}(t,k)\right\rangle }$ is normalized within  the length $L_x$ of the system, thus ${ \left\langle \Psi_{n}(k) \vert \Psi_{n}(k) \right\rangle = \left\langle u_{n}(k) \vert u_{n}(k)\right\rangle_{cell} =1 }$, where $\left\langle u_{n}(k) \vert u_{n}(k)\right\rangle_{cell}$ is calculated within one primitive cell and a normalization constant is assumed to be absorbed in the cell periodic state $\left| u_n(k)\right\rangle$. The electrons' position expectation value is a quantity that is position origin-dependent and is given by 
\begin{equation} \label{apa1}
\left\langle \psi_{n}(k) \vert \, x \, \vert \psi_{n}(k) \right\rangle = \frac{1}{N_x} \int_{0}^{L_x} \!\!\! x \left| u_{n}(x,k)\right|^2 dx
\end{equation}
where the lower limit of the space integration is the starting point of the 1D system that coincides with the position-origin. 

Using the periodicity of the cell periodic states \thinspace $u_{n}(x,k)$ \thinspace we \textquotedblleft{transfer}\textquotedblright \thinspace all $(N_{x}-1)$ primitives cells in the position of the $1^{
\text{st}}$ primitive cell (adjacent to the position origin), which gives
\begin{widetext}	
\begin{eqnarray} \label{apa2}
\left\langle \psi_{n}(k) \vert \, x \, \vert \psi_{n}(k) \right\rangle &=& \int_{0}^{\alpha_x} \left| u_{n}(x,k)\right|^2 x \, dx 
\,
+ 
\,
\frac{1}{N_x} 
\, 
\alpha_{x} 
\left( 
1+2+3+...+(N_{x}-1) 
\right)  
\int_{0}^{\alpha_x} \left| u_{n}(x,k)\right|^2 dx  
\nonumber \\*
&=& \left\langle u_{n}(k) \vert \, x \, \vert u_{n}(k) \right\rangle_{cell} 
\,
+ 
\,
\,
\frac{1}{N_x}
\alpha_{x} N_{x} \frac{(N_x -1)}{2} \left\langle u_{n}(k) \vert u_{n}(k) \right\rangle_{cell}
\nonumber \\*
\end{eqnarray}
\end{widetext}
where the \textit{cell} subscript denotes that the space integrals are evaluated within the primitive cell located at the systems edge. Using the normalization condition \thinspace Eq.~(\ref{apa2}) \thinspace takes the form
\begin{equation} \label{apa3}
\left\langle \psi_{n}(k) \vert \, x \, \vert \psi_{n}(k) \right\rangle = \frac{\alpha_x N_x}{2} \, + \, \left( 
\left\langle u_{n}(k) \vert \, x \, \vert u_{n}(k) \right\rangle_{cell}
- \frac{\alpha_x}{2} \right).
\end{equation}

Performing analogous calculation as that in \thinspace Eq.~(\ref{apa2}), \thinspace we evaluate the off-diagonal matrix elements of the position operator for ${ n \neq m }$ which gives
\begin{eqnarray} \label{apa2b} \nonumber
\left\langle \psi_{n}(k) \vert \, x \,  \vert \psi_{m}(k) \right\rangle &=& 
\left\langle u_{n}(k) \vert \, x  \, \vert u_{m}(k) \right\rangle_{cell} 
\,
\\*  \nonumber
&&+ 
\,
\alpha_{x} \frac{(N_x -1)}{2} \left\langle u_{n}(k) \vert u_{m}(k) \right\rangle_{cell},
\\* 
\end{eqnarray}
where using \thinspace  
${ \displaystyle \left\langle \psi_{n}(k) \vert \psi_{m}(k) \right\rangle =
\left\langle u_{n}(k) \vert u_{m}(k) \right\rangle_{cell} = \delta_{nm} }$, we finally find that the off-diagonal matrix elements are given by
\begin{equation}  \label{apa2c}
\left\langle \psi_{n}(k) \vert \, x \, \vert \psi_{m}(k) \right\rangle =
\left\langle u_{n}(k) \vert \, x \, \vert u_{m}(k) \right\rangle_{cell}.
\end{equation}
Therefore, in the system's infinite length limit $L_{x}\rightarrow \infty$, \thinspace the number of primitive cells enclosed within the system also becomes infinite $N_{x}\rightarrow \infty$, and as a result, the electrons' position expectation value \thinspace Eq.~(\ref{apa3}) \thinspace takes an undefined value (due to the first term of the right hand side), in contrast to the off-diagonal position matrix elements \thinspace 
Eq.~(\ref{apa2c}) \thinspace which they return a well defined value. 

\subsection{Explicit calculation of \thinspace $\Delta\!\left\langle \mathbf{r} \right\rangle $}

We assume a position periodic and closed system of length $L_x$. We will calculate the electrons' displacement after a finite time interval $T$ in the limit of infinite length ${L_{x}\rightarrow \infty}$ and show that is a well-defined quantity. We assume that the electron is in an extended and time-dependent Bloch type state at every instant, that is,  
${ \displaystyle
\left|\Psi(t,k(t))\right\rangle=
\frac{1}{\sqrt{N_x}}
e^{\displaystyle ik(t)x} \left| u(t,k(t))\right\rangle }$, where the state $ \left| u(t,k(t))\right\rangle $ has arbitrary time-dependence and is cell-periodic at every instant, as well as ${ \left|\Psi(t,k(t))\right\rangle }$ is normalized to unity at every instant  ${ 
\left\langle \Psi(t,k(t)) \vert \Psi(t,k(t))\right\rangle =
\left\langle u(t,k(t)) \vert u(t,k(t))\right\rangle_{cell}=1
 }$.

Using similar reasoning as in \thinspace Eq.~(\ref{apa3}) \thinspace we find that the electrons' displacement is given by 
\begin{widetext}
\begin{eqnarray} \label{apa4} \nonumber
\Delta \left\langle \, x \,\right\rangle 
=
\Delta \left\langle \Psi(t,k(t)) \vert
\, x \,
\vert \Psi(t,k(t))\right\rangle 
& = & 
\frac{1}{N_x} 
\left\langle u(t+T,k(t+T)) \vert \, x \, \vert u(t+T,k(t+T)) \right\rangle 
- 
\frac{1}{N_x} 
\left\langle u(t,k) \vert \, x \, \vert u(t,k) \right\rangle \\*[6pt]
& = & 
\left\langle u(t+T,k(t+T)) \vert \, x \, \vert u(t+T,k(t+T)) \right\rangle_{cell}
-\left\langle u(t,k(t)) \vert \, x \, \vert u(t,k(t)) \right\rangle_{cell}
\nonumber \\*
\end{eqnarray}
\end{widetext}
where the undefined terms \thinspace ${ \displaystyle \frac{\alpha_x N_x}{2}  }$ \thinspace canceled each one another. In this fashion, Eq.(\ref{apa4}) \thinspace takes the form
\begin{equation} \label{apa5}
\Delta \left\langle x \right\rangle = \int_{t}^{t+T} \displaystyle \frac{d}{dt'}\left\langle u(t',k(t')) \vert \, x \,  \vert u(t',k(t')) \right\rangle_{cell}dt',
\end{equation}
where, by using the extended velocity operator 
$\text{\large{v}}_{ext}$ defined in \thinspace Eq.~(\ref{e6}) \thinspace in the main text, it turns out that the electron displacement in a position periodic system has to be evaluated as  
\begin{equation} \label{apa6}
\Delta \left\langle x \right\rangle =\int_{t}^{t+T} \displaystyle \left\langle u(t',k(t')) \vert \, \text{\large{v}}_{ext} \,  \vert u(t',k(t')) \right\rangle_{cell}dt'.
\end{equation}

\subsection{Explicit calculation of $\left\langle \mathbf{C}\right\rangle_{n} $}

We calculate the electrons' circulation operator expectation value $\mathbf{\left\langle C \right\rangle }$ given by \thinspace Eq.~(\ref{e9}) \thinspace of the main text with respect to a Bloch eigenstate \thinspace ${ \displaystyle 
\left|\Psi_{n}(t,\mathbf{k})\right\rangle=\frac{1}{\sqrt{N}} \, e^{\displaystyle -\frac{i}{\hbar} E_{n}(\mathbf{k})t} e^{\displaystyle i \mathbf{k.r}} \left| u_{n}(\mathbf{k})\right\rangle }$ that satisfies PBSs over the edges. For simplicity we assume a 2D system while the generalization to 3D is straightforward. The system has length  ${L_x = N_x \alpha_x}$ in the $x$ direction and ${L_{y} = N_{y} \alpha_{y}}$  in the normal $y$ direction, where  ${ N=N_x N_y }$ is the total number of primitive cells within the system and $\alpha_x \alpha_y $  is the area of the primitive cell. The  Bloch eigenstate is normalized within the area $ L_x L_y $, therefore a normalization constant is assumed to be absorbed within the cell periodic states, 
${ \left\langle \Psi_{n}(\mathbf{k}) \vert \Psi_{n}(\mathbf{k}) \right\rangle = \left\langle u_{n}(\mathbf k) \vert u_{n}(\mathbf k)\right\rangle_{cell} =1 }$. The electrons' circulation is given by
\begin{eqnarray} \label{apa7} \nonumber
&&\left\langle 
\Psi_{n}(\mathbf{k})
\vert 
\, \mathbf{C} \, 
\vert 
\Psi_{n}(\mathbf{k})
\right\rangle =
\nonumber \\*
&& 
\ \ \ \ \ \ \ \ \ \ \ \  \ \ 
\frac{1}{N_x N_y}
\int_{0}^{N_{x} \alpha_{x}} \!\!\! \int_{0}^{N_{y} \alpha_{y}} \!\!\! \mathbf{r} \times \mathbf{J}_{pr(n)}(x,y, \mathbf{k}) dx dy
\nonumber \\*[-4pt]
\end{eqnarray}
where the local probability current density, is evaluated with respect to the cell periodic eigenstate ${  u_{n}(\mathbf{r}, \mathbf{k})  }$ and is a cell-periodic quantity. We first carry out the integral \thinspace ${ \displaystyle 
\frac{1}{N_y}
\int_{0}^{N_{y} \alpha_{y}} \!\! \mathbf{r} \times \mathbf{J}_{pr(n)}(x, y, \mathbf{k})dy }$, 
where we exploit the periodicity of the local probability current density and  \textquotedblleft{transfer}\textquotedblright \thinspace $(N_{y}-1)$ primitives cells along the $y$ direction on the ${ y=0 }$ line which gives

\begin{widetext}
\begin{eqnarray} \label{apa8} \nonumber
\frac{1}{N_y}
\int_{0}^{N_{y} \alpha_{y}} \!\! \mathbf{r} \times \mathbf{J}_{pr(n)}(x, y, \mathbf{k})dy 
&=& 
\int_{0}^{ \alpha_{y}} \!\! \mathbf{r} \times \mathbf{J}_{pr(n)}(x, y, \mathbf{k})dy 
\,
+ 
\,
\frac{1}{N_y}
\left(  
1+2+...+(N_{y} -1) 
\right)  
\bm{\alpha}_{y} \times 
\!\! \int_{0}^{ \alpha_{y}} \!\! \mathbf{J}_{pr(n)}(x, y, \mathbf{k})dy	 
\nonumber \\*
&=& 
\int_{0}^{ \alpha_{y}} \!\! \mathbf{r} \times \mathbf{J}_{pr(n)}(x, y, \mathbf{k})dy
\, 
+
\frac{1}{N_y}
\,
N_{y} \frac{(N_{y}-1)}{2} \bm{\alpha}_{y} \times \! \int_{0}^{ \alpha_{y}} \!\!  
\mathbf{J}_{pr(n)}(x, y, \mathbf{k})dy.
\end{eqnarray} 
Exploiting the periodicity of the local probability current, we perform analogous calculation for the integral along the $x$ direction which gives  
\begin{eqnarray} \label{apa9} \nonumber
\frac{1}{N_x N_y}
\int_{0}^{N_{x} \alpha_{x}} \!\!\!\!\!dx  \left(  \int_{0}^{N_{\psi} \alpha_{\psi}} \!\!\!\!\! \mathbf{r} \times \mathbf{J}_{pr(n)}(x,\psi, \mathbf{k})dy \right)&=& \int_{0}^{\alpha_{x}} \!\!\! \int_{0}^{ \alpha_{y}} \!\! \mathbf{r} \times \mathbf{J}_{pr(n)}(x, y, \mathbf{k}) \, dx dy  
\\*
& & + \, \left( \frac{(N_{x}-1)}{2} \bm{\alpha}_{x} + \frac{(N_{y}-1)}{2} \bm{\alpha}_{y} \right) \times  
\int_{0}^{\alpha_{x}} \!\!\! \int_{0}^{ \alpha_{y}} \!\! \mathbf{J}_{pr(n)}(x, y, \mathbf{k}) \, dx dy.  
\nonumber \\*[-6pt]
\end{eqnarray}
Eq.~(\ref{apa7}) \thinspace with the aid of Eq.~(\ref{apa9}) \thinspace finally takes the form
\begin{eqnarray} \label{apa10} \nonumber
\left\langle 
\Psi_{n}(\mathbf{k})
\vert 
\, \mathbf{C} \, 
\vert 
\Psi_{n}(\mathbf{k})
\right\rangle = 
\int_{0}^{\alpha_{x}} \!\!\! \int_{0}^{ \alpha_{y}} \!\! \mathbf{r} \times   
\mathbf{J}_{pr(n)}(\mathbf{r}, \mathbf{k})
dx dy 
\, + \, 
\left( \frac{(N_{x}-1)}{2} \bm{\alpha}_{x} + \frac{(N_{y}-1)}{2} \bm{\alpha}_{y} \right) 
\times  
\int_{0}^{\alpha_{x}} \!\!\! \int_{0}^{ \alpha_{y}} \!\! \mathbf{J}_{pr(n)}(\mathbf{r}, \mathbf{k})
dx dy
\nonumber \\*[-4pt]
\end{eqnarray}
\end{widetext}
where all space integrals are taken within one primitive cell adjacent to a system's edge and located at the position origin. The first term on the right hand side of \thinspace Eq.~(\ref{apa10}) \thinspace is always a well-defined quantity even in the thermodynamic limit. Therefore, in the thermodynamic limit the electrons' circulation becomes infinite due to the second term of the right hand side of \thinspace Eq.~(\ref{apa10}.)

\subsection{Explicit calculation of $\left\langle \mathbf{C}_{intr}\right\rangle_{n} $}
We calculate the electrons' intrinsic circulation ${\left\langle \mathbf{C}_{intr} \right\rangle }$ given by \thinspace Eq.~(\ref{e10}) \thinspace of the main text with respect to a Bloch eigenstate \thinspace ${ \displaystyle \left|\Psi_{n}(t,\mathbf{k})\right\rangle=\frac{1}{\sqrt{N}}
e^{\displaystyle -\frac{i}{\hbar} E_{n}(\mathbf{k})t} e^{\displaystyle i \mathbf{k.r}} \left| u_{n}(\mathbf{k})\right\rangle }$  \thinspace in a 2D system identical to the one of the previous subsection. Therefore, we have to calculate

\begin{widetext}
\begin{equation} \label{apa11} 
\left\langle 
\Psi_{n}(\mathbf{k})
\vert 
\, \mathbf{C}_{intr} \, 
\vert 
\Psi_{n}(\mathbf{k})
\right\rangle_{n}
= 
\frac{1}{N_x N_y}
\int_{0}^{N_{x} \alpha_{x}} \!\!\! \int_{0}^{N_{y} \alpha_{y}} \!\!\! \mathbf{r} 
\times 
\mathbf{J}_{pr(n)}(x, y, \mathbf{k})
dx dy  
\,
- 
\, 
\left\langle \mathbf{r} \right\rangle_{n} \times \!
\frac{1}{N_x N_y}
\int_{0}^{N_{x} \alpha_{x}} \!\!\! \int_{0}^{N_{y} \alpha_{y}}  \!\! \mathbf{J}_{pr(n)}(x, y, \mathbf{k}) dx dy. 
\end{equation}
The first term of the right hand side of \thinspace Eq.~(\ref{apa11}) \thinspace is given by \thinspace Eq.~(\ref{apa10}). The electrons' position expectation value \thinspace $ \left\langle \mathbf{r} \right\rangle_{n}$ \thinspace is given by the 2D generalization of \thinspace Eq.~(\ref{apa2}), \thinspace that is,
\begin{equation} \label{apa12} 
\left\langle \mathbf{r} \right\rangle_{n} = 
\left( 
\frac{(N_{x}-1)}{2} \bm{\alpha}_{x} + \frac{(N_{y}-1)}{2} \bm{\alpha}_{y} 
\right) 
+ \, \,
\left\langle u_{n}(\mathbf{k}) \vert \, \mathbf{r} \, \vert u_{n}(\mathbf{k}) \right\rangle_{cell}
\end{equation}
and the space integral of the local probability current density is easily truncated within one primitive cell adjacent to a system's edge (located at the position origin) due to the cell periodicity of the probability current, thus giving
\begin{equation} \label{apa13} 
\frac{1}{N_x N_y}
\int_{0}^{N_{x} \alpha_{x}} \!\!\! \int_{0}^{N_{y} \alpha_{y}}  \!  \mathbf{J}_{pr(n)}(x, y, \mathbf{k}) \, dx dy  
= \int_{0}^{\alpha_{x}} \!\!\! \int_{0}^{ \alpha_{y}}  \! \mathbf{J}_{pr(n)}(x, y, \mathbf{k})
\, dx dy \nonumber  
\end{equation}
Substituting \thinspace Eq.~(\ref{apa10}) \thinspace and \thinspace Eq.~(\ref{apa12}) \textendash \thinspace (\ref{apa13}) \thinspace into \thinspace Eq.~(\ref{apa11}) \thinspace we finally obtain
\begin{equation} \label{apa14}
\left\langle 
\Psi_{n}(\mathbf{k})
\vert 
\, \mathbf{C}_{intr} \, 
\vert 
\Psi_{n}(\mathbf{k})
\right\rangle
=
\int_{0}^{\alpha_{x}} \!\!\! \int_{0}^{ \alpha_{y}} \!\! 
\left( \, \mathbf{r}-
\left\langle u_{n}(\mathbf{k})\vert \, \mathbf{r} \, \vert u_{n}(\mathbf{k})\right\rangle_{cell}
\right) 
\times 
\mathbf{J}_{pr(n)}(x, y, \mathbf{k})
dx dy
\end{equation} 
\\
where the two terms \thinspace
${ \displaystyle 
\pm
\left( \frac{(N_{x}-1)}{2} \bm{\alpha}_{x} + \frac{(N_{y}-1)}{2} \bm{\alpha}_{y} \right) 
\times  
\int_{0}^{\alpha_{x}} \!\!\! \int_{0}^{ \alpha_{y}} \!\! \mathbf{J}_{pr(n)}(\mathbf{r}, \mathbf{k})
dx dy }$,
each one undefined in the thermodynamic limit,
have canceled each other. 
\end{widetext}

\section{Action of the velocity operator $\mathbf{v}$ on a Bloch eigenstate $\left|{\Psi_n (t, \mathbf{k})}\right\rangle$ and of the operator \thinspace $\left( \mathbf{r} - \left\langle \mathbf{r} \right\rangle_{n} \right)$ on a cell periodic state $\left| u_{n}(\mathbf{k})\right\rangle $} \label{a2}

\subsection{Action of \thinspace $\mathbf{v}$ \thinspace on a Bloch eigenstate $\left| \Psi_n (t,\mathbf{k}) \right\rangle$}
 
At first we derive a general $\mathbf{k}$-derivative formula that gives the action of the standard velocity operator \thinspace Eq.~(\ref{e2}) \thinspace on a Bloch type state of the form  
${ \left|\Psi(t,\mathbf{k})\right\rangle=e^{\displaystyle i \mathbf{k.r}} \left| \Phi(t,\mathbf{k}) \right\rangle }$, where $\mathbf{k}$ is a static wave vector (assumed to take continuous values).  This is accomplished by taking into account the specific Bloch type form of the state 
${ \left|\Psi(t,\mathbf{k})\right\rangle }$ as well as the time evolution of the state by \thinspace ${ \displaystyle i \hbar \frac{d}{dt} \left|\Psi(t,\mathbf{k})\right\rangle =
H(\mathbf{r}) \left|\Psi(t,\mathbf{k})\right\rangle }$, \thinspace that is governed by a static Hamiltonian $H(\mathbf{r})$.

Under these conditions, the action of the position operator on the state \thinspace ${ \left|\Psi(t,\mathbf{k})\right\rangle }$ \thinspace can be expressed as
\begin{equation} \label{apb1}
\mathbf{r} \left|  \Psi(t,\mathbf{k})\right\rangle =
-i \left| \partial_{\mathbf{k}}\Psi(t,\mathbf{k})\right\rangle 
+ 
i e^{ \displaystyle i \mathbf{k.r}} \left|  \partial_{\mathbf{k}} \Phi(t,\mathbf{k})\right\rangle.
\end{equation}
Acting on both sides of \thinspace Eq.~(\ref{apb1}) \thinspace with the Hamiltonian $H(\mathbf{r})$  of the system, and taking into account that the Hamiltonian does not depended on the wavevector, that is \thinspace ${ [H(\mathbf{r}),\partial_{\mathbf{k}}\,]=0 }$, we find
\begin{eqnarray} \label{apb2}  \nonumber
H(\mathbf{r}) \mathbf{r} \left|  \Psi(t,\mathbf{k})\right\rangle &=& 
-i \partial_{\mathbf{k}} \left( H(\mathbf{r}) \left| \Psi(t,\mathbf{k})\right\rangle\right)  \\* 
&& +
i e^{ \displaystyle i \mathbf{k.r}} H_{k}(\mathbf{r},\mathbf{k}) \left|  \partial_{\mathbf{k}} \Phi(t,\mathbf{k})\right\rangle 
\end{eqnarray}
where the Hamiltonian $H_{k}(\mathbf{r},\mathbf{k})$ is defined by \thinspace ${ H_{k}(\mathbf{r},\mathbf{k})=e^{ \displaystyle -i \mathbf{k.r}}H(\mathbf{r})e^{ \displaystyle i \mathbf{k.r}}  }$. The term \thinspace ${ \partial_{\mathbf{k}} \left( H(\mathbf{r}) \left| \Psi(t,\mathbf{k})\right\rangle\right)  }$ \thinspace of the right hand side of \thinspace Eq.~(\ref{apb2}) \thinspace can be recast in the form
\begin{eqnarray} \label{apb3} \nonumber
\partial_{\mathbf{k}} \left( H(\mathbf{r}) \left| \Psi(t,\mathbf{k})\right\rangle \right) & = &  
i \, \mathbf{r} \, H(\mathbf{r}) \left|\Psi(t,\mathbf{k})\right\rangle  \\*
& & + \,
e^{ \displaystyle i \mathbf{k.r}} \, i \hbar \displaystyle  \frac{d}{dt} \left| \partial_{\mathbf{k}}  \Phi(t,\mathbf{k})\right\rangle
\end{eqnarray}
where we have used
\[ H(\mathbf{r}) \left| \Psi(t,\mathbf{k})\right\rangle =
i \hbar \frac{d}{dt} \left|\Psi(t,\mathbf{k})\right\rangle =
e^{ \displaystyle i \mathbf{k.r}} \,
i \hbar \frac{d}{dt} \left|\Phi(t,\mathbf{k})\right\rangle, \] 
as well as
\begin{eqnarray*}
 \partial_{\mathbf{k}} \left( e^{ \displaystyle i \mathbf{k.r}} \,
i \hbar \frac{d}{dt} \left|\Phi(t,\mathbf{k})\right\rangle
\right) & = & -\hbar \, e^{ \displaystyle i \mathbf{k.r}} \mathbf{r} \frac{d}{dt} \left|\Phi(t,\mathbf{k})\right\rangle  \\*
&& + \, e^{ \displaystyle i \mathbf{k.r}} \, i \hbar \frac{d}{dt} \left| \partial_{\mathbf{k}}  \Phi(t,\mathbf{k})\right\rangle   \\*
\end{eqnarray*}
and
\begin{eqnarray*}
- \hbar e^{ \displaystyle i \mathbf{k.r}} \mathbf{r} \frac{d}{dt} \left|\Phi(t,\mathbf{k})\right\rangle  & = &
 \mathbf{r} \, i^2 \hbar \frac{d}{dt} \left(  e^{ \displaystyle i \mathbf{k.r}} \left|\Phi(t,\mathbf{k})\right\rangle \right)   \\ 
 & = & i \, \mathbf{r} \, H(\mathbf{r})  \left|\Psi(t,\mathbf{k})\right\rangle.
\end{eqnarray*}
Substituting \thinspace Eq.~(\ref{apb3}) \thinspace into \thinspace Eq.~(\ref{apb2}) \thinspace we find that the action of commutator ${ \left[ H(\mathbf{r}),\mathbf{r} \right] }$ on the Bloch type state 
$ \left|\Psi(t,\mathbf{k})\right\rangle $  is given by 
\begin{equation} \label{apb4}
\left[ H(\mathbf{r}),\mathbf{r} \right] \left|\Psi(t,\mathbf{k})\right\rangle =
i \, e^{ \displaystyle i \mathbf{k.r}} 
\left( 
H_{k}(\mathbf{r},\mathbf{k}) - i \hbar \displaystyle  \frac{d}{dt}
\right) 
 \left| \partial_{\mathbf{k}}  \Phi(t,\mathbf{k})\right\rangle.
\end{equation}

The action of the commutator on a stationary Bloch type state of the form \thinspace
${ \left|\Psi_{n}(t,\mathbf{k})\right\rangle = 
e^{ \displaystyle i \mathbf{k.r}} \,
e^{ \displaystyle i \Theta_{n}(t,\mathbf{k}) }
 \left| u_{n}(\mathbf{k})\right\rangle }$ 
where $\Theta_{n}(t,\mathbf{k})$ is the dynamical phase  
\thinspace with an additional $\mathbf{k}$-dependent gauge phase, that is,     
${ \displaystyle \Theta_{n}(t,\mathbf{k})= -\frac{1}{\hbar} E_{n}(\mathbf{k})t + \Lambda_{n}(\mathbf{k}) }$ 
\thinspace can be calculated by replacing \thinspace  
${ \left| \Phi(t,\mathbf{k})\right\rangle = 
e^{ \displaystyle i \Theta_{n}(t,\mathbf{k}) 
} \left| u_{n}(\mathbf{k})\right\rangle }$
within \thinspace Eq.~(\ref{apb4}). This gives
\begin{widetext}
\begin{equation} \label{apb5} 
\left[ H(\mathbf{r}),\mathbf{r} \right] \left|\Psi_{n}(t,\mathbf{k})\right\rangle  =
i \, e^{ \displaystyle i \mathbf{k.r}} e^{ \displaystyle i \Theta_{n}(t,\mathbf{k}) }
\left( 
H_{k}(\mathbf{r},\mathbf{k}) + \hbar \displaystyle  \frac{d}{dt} \Theta_{n}(t,\mathbf{k})   
\right)  
\left| \partial_{\mathbf{k}}  u_{n}(\mathbf{k})\right\rangle    
+ 
i \hbar \left(  \partial_{\mathbf{k}}\frac{d}{dt} \Theta_{n}(t,\mathbf{k}) \right)  \left|\Psi_{n}(t,\mathbf{k})\right\rangle 
\end{equation}
where we have used that
${ \displaystyle \frac{d}{dt} \left| u_{n}(\mathbf{k})\right\rangle = 0 }$,
as well as \thinspace
${ \left( 
H_{k}(\mathbf{r},\mathbf{k}) + \hbar \displaystyle  \frac{d}{dt} \Theta_{n}(t,\mathbf{k})
\right) 
\left|  u_{n}(\mathbf{k})\right\rangle = 0 }$.  
From Eq.~(\ref{apb5}) we can deduce that the action of the standard velocity operator $\mathbf{v}$ on a stationary Bloch type state is given from
\begin{equation} \label{apb6}
\mathbf{v} \left|\Psi_{n}(t,\mathbf{k})\right\rangle  =
-\frac{1}{\hbar} \, e^{ \displaystyle i \mathbf{k.r}} e^{ \displaystyle i \Theta_{n}(t,\mathbf{k}) }
\left( 
H_{k}(\mathbf{r},\mathbf{k}) - E_{n}(\mathbf{k})  
\right)  
\left| \partial_{\mathbf{k}}  u_{n}(\mathbf{k})\right\rangle    
+ \frac{1}{\hbar} \partial_{\mathbf{k}}E_{n}(\mathbf{k})  \left|\Psi_{n}(t,\mathbf{k})\right\rangle.
\end{equation}

\subsection{Action of   \thinspace   $\left( \mathbf{r}- \left\langle \mathbf{r} \right\rangle \right) $  \thinspace on a cell-periodic  eigenstate  $\left| u_n (\mathbf{k})\right\rangle$}    \label{apb22}
We assume a Bloch type eigenstate in the form \thinspace
${ \left|\Psi_{n}(t,\mathbf{k})\right\rangle = 
	e^{ \displaystyle i \mathbf{k.r}} \,
	e^{ \displaystyle i \Theta_{n}(t,\mathbf{k}) }
	\left| u_{n}(\mathbf{k})\right\rangle }$ 
where $\Theta_{n}(t,\mathbf{k})$ is the dynamical phase  
\thinspace with an additional $\mathbf{k}$-dependent gauge phase, that is,     
${ \displaystyle \Theta_{n}(t,\mathbf{k})= -\frac{1}{\hbar} E_{n}(\mathbf{k})t + \Lambda(\mathbf{k}) }$. The time-independent eigenstate 
${ \left| u_{n}(\mathbf{k})\right\rangle }$ can be recast in the form
\begin{equation} \label{apb7}
\left| u_{n}(\mathbf{k})\right\rangle =
e^{ \displaystyle -i \mathbf{k.r}}\, e^{ \displaystyle -i \Lambda(\mathbf{k})} 
\left| \Psi_{n}(\mathbf{k})\right\rangle
\end{equation}
where the time-dependence has been eliminated as expected.

In the position representation and by using \thinspace Eq.~(\ref{apb7}), the action of the position operator on the eigenstate \thinspace
${ \left| u_{n}(\mathbf{k})\right\rangle }$ \thinspace can be transformed to a \thinspace $\mathbf{k}$-derivative identity given by 
\begin{equation} \label{apb8} 
\mathbf{r} \left| u_{n}(\mathbf{k}) \right\rangle = 
i \left| \partial_{\mathbf{k}} u_{n}(\mathbf{k}) \right\rangle
-\partial_{\mathbf{k}} \Lambda(\mathbf{k}) 
\left| u_{n}(\mathbf{k}) \right\rangle  -i \, e^{ \displaystyle -i \mathbf{k.r}}\, e^{ \displaystyle -i \Lambda(\mathbf{k})} 
\left| \partial_{\mathbf{k}}  \Psi_{n}(\mathbf{k})\right\rangle.
\end{equation}

Accordingly, the expectation value of the position operator \thinspace $\mathbf{r}$ \thinspace with respect to the eigenstate \thinspace 
${ \left| u_{n}(\mathbf{k})\right\rangle }$ \thinspace takes with the aid of \thinspace Eq.~(\ref{apb8}) \thinspace the form
\begin{equation} \label{apb10} 
\left\langle u_{n}(\mathbf{k}) \vert \, \mathbf{r} \, \vert u_{n}(\mathbf{k}) \right\rangle = 
{\mathbf{A}}_{nn}(\mathbf{k}) 
-\partial_{\mathbf{k}} \Lambda_{n}(\mathbf{k}) 
-i \left\langle \Psi_{n}(\mathbf{k}) \vert \partial_{\mathbf{k}} \Psi_{n}(\mathbf{k})\right\rangle,
\end{equation}
where \thinspace ${ {\mathbf{A}}_{nn}(\mathbf{k})=i\left\langle u_{n}(\mathbf{k}) \vert \partial_{\mathbf{k}} u_{n}(\mathbf{k}) \right\rangle }$ \thinspace is the Abelian Berry connection. By acting with
\thinspace Eq.~(\ref{apb10}) \thinspace 
\thinspace on
${ \left| u_{n}(\mathbf{k}) \right\rangle }$ \thinspace and then subtracting the product from \thinspace Eq.~(\ref{apb8}) \thinspace we find the identity
\begin{equation} \label{apb10b}
\left( \,
\mathbf{r}- \left\langle \mathbf{r} \right\rangle_{n} 
\right)
\left| u_{n}(\mathbf{k})\right\rangle  
=  
\left( 
\, i \partial_{\mathbf{k}} - {\mathbf{A}}_{nn}(\mathbf{k})
\right) 
\left| u_{n}(\mathbf{k})\right\rangle
- 
\,i \, e^{ \displaystyle -i \mathbf{k.r}}\, e^{ \displaystyle -i \Lambda(\mathbf{k})} 
\left( \
\left| \partial_{\mathbf{k}}  \Psi_{n}(\mathbf{k})\right\rangle
-
\left\langle \Psi_{n}(\mathbf{k}) \vert \partial_{\mathbf{k}} \Psi_{n}(\mathbf{k})\right\rangle
\left| \Psi_{n}(\mathbf{k})\right\rangle
\, \right). 
\end{equation}
By then using the one-band covariant derivative definition, namely, \  
${
 i \widetilde{{\partial}_{\mathbf{k}}}
\left|  
u_n(\mathbf{k}) \right\rangle 
=	
\left( 
\, i \partial_{\mathbf{k}} - {\mathbf{A}}_{nn}(\mathbf{k})
\right) \!
\left|  
u_n(\mathbf{k}) \right\rangle 
}$,
where \thinspace ${\widetilde{{\partial}_{\mathbf{k}}} }$ \thinspace is given by \  
${
\widetilde{{\partial}_{\mathbf{k}}}=
\partial_{\mathbf{k}}
\; + \;
i\mathbf{A}_{nn}(\mathbf{k})
}$, \thinspace
\thinspace Eq.~(\ref{apb10b}) \thinspace
takes the form
\begin{equation} \label{apb10d}
\left( \,
\mathbf{r}- \left\langle \mathbf{r} \right\rangle_{n} 
\right)
\left| u_{n}(\mathbf{k})\right\rangle  
=i  
\left| \widetilde{{\partial}_{\mathbf{k}}} u_{n}(\mathbf{k})\right\rangle
- 
\,i \, e^{ \displaystyle -i \mathbf{k.r}}\, e^{ \displaystyle -i \Lambda(\mathbf{k})} 
\left| \partial_{\mathbf{k}}  \Psi_{n}(\mathbf{k})\right\rangle
+
i \left\langle \Psi_{n}(\mathbf{k}) \vert \partial_{\mathbf{k}} \Psi_{n}(\mathbf{k})\right\rangle
\left| u_{n}(\mathbf{k})\right\rangle. 
\end{equation}

We then expand the state \thinspace
${ 
\left| \partial_{\mathbf{k}}  \Psi_{n}(\mathbf{k})\right\rangle }$ \thinspace on the complete basis of the Bloch eigenstates \thinspace 
${ \left| \psi_{m}(\mathbf{k'}) \right\rangle }$ \thinspace
by using the closure relation 
\thinspace 
${ \displaystyle I=
\sum_{m, \mathbf{k}' }^{HS}
\left| \psi_{m}(\mathbf{k}') \right\rangle 
\!
\left\langle \psi_{m}(\mathbf{k}') \right|  }$,
\thinspace 
that is, we substitute
\thinspace
${ \displaystyle 
\left| \partial_{\mathbf{k}}  \Psi_{n}(\mathbf{k})\right\rangle 
=
\sum_{m, \mathbf{k}'}^{HS}
\left\langle \psi_{m}(\mathbf{k'}) 
\vert
\partial_{\mathbf{k}} 
\psi_{n}(\mathbf{k})
\right\rangle 
\left| \psi_{m}(\mathbf{k'}) \right\rangle 
}$
\thinspace 
which gives 
\begin{eqnarray} \label{apb10f} \nonumber
\left( \,
\mathbf{r}- \left\langle \mathbf{r} \right\rangle_{n} 
\right)
\left| u_{n}(\mathbf{k})\right\rangle  
&=&i  
\left| \widetilde{{\partial}_{\mathbf{k}}} u_{n}
(\mathbf{k})\right\rangle
- 
i
\sum_{m, \mathbf{k}' }
\left\langle \psi_{m}(\mathbf{k'}) 
\vert
\partial_{\mathbf{k}} 
\psi_{n}(\mathbf{k})
\right\rangle  
\, 
e^{ \displaystyle i (\Lambda(\mathbf{k'})-\Lambda(\mathbf{k}))} 
e^{ \displaystyle i \mathbf{(k'-k).r}} \, 
\left| u_{m}(\mathbf{k'}) \right\rangle
\nonumber \\*   [4pt]
&& +
i \left\langle \Psi_{n}(\mathbf{k}) \vert \partial_{\mathbf{k}} \Psi_{n}(\mathbf{k})\right\rangle
\left| u_{n}(\mathbf{k})\right\rangle,
\end{eqnarray}
and then use the Hermitian conjugate of 
\thinspace Eq.~(\ref{apb10f}) \thinspace
to evaluate the orbital magnetic moment of the electron, namely
\begin{equation} \label{apb10g}
\mathbf{m}_n (\mathbf{k}) =
- \frac{e}{2c \hbar}
\text{Im} [\, i 
\left\langle u_{n}(\mathbf{k})\vert \, (\mathbf{r}-{\left\langle \mathbf{r}\right\rangle}_{n}) 
\, \times \,
\left( 
H_{k}(\mathbf{r},\mathbf{k}) - E_{n}(\mathbf{k})  
\right)  
\vert \partial_{\mathbf{k}}  u_{n}(\mathbf{k})\right\rangle \, ]. 
\end{equation}
This way,
\thinspace Eq.~(\ref{apb10g}) \thinspace takes the form
\begin{eqnarray} \label{apb10h} \nonumber
\mathbf{m}_n (\mathbf{k}) 
&=&
- \frac{e}{2c \hbar }
\text{Im} 
\left[ \, 
\left\langle \widetilde{{\partial}_{\mathbf{k}}}u_{n}(\mathbf{k})
\vert 
\, \times \,
\left( 
H_{k}(\mathbf{r},\mathbf{k}) - E_{n}(\mathbf{k})  
\right)  
\vert 	
{\partial}_{\mathbf{k}}
u_{n}(\mathbf{k})\right\rangle 
\, \right]
\nonumber \\*[6pt] 
&& 
+ 
\frac{e}{2c \hbar}
\text{Im}
\left[   \,
\sum_{m, \mathbf{k}' }^{HS}
\left\langle \Psi_{m}(\mathbf{k'}) \vert \partial_{\mathbf{k}} \Psi_{n}(\mathbf{k})\right\rangle^{\displaystyle *}
\times \
e^{ \displaystyle i (\Lambda(\mathbf{k})-\Lambda(\mathbf{k'}))} 
\left\langle u_{m}(\mathbf{k}') 
\vert \,
e^{ \displaystyle i \mathbf{(k-k').r}} 
\, 
( 
H_{k}(\mathbf{r},\mathbf{k}) - E_{n}(\mathbf{k})  
) 
\, \vert
{\partial}_{\mathbf{k}}
u_{n}(\mathbf{k})\right\rangle 
\,  \right] 
\nonumber \\*  [4pt]
&&  
-
\frac{e}{2c \hbar}
\text{Im} 
\left[  \,
\left\langle \Psi_{n}(\mathbf{k}) \vert \partial_{\mathbf{k}} \Psi_{n}(\mathbf{k})\right\rangle^{\displaystyle *}
\times 
\left\langle u_{n}(\mathbf{k}) 
\vert \,
H_{k}(\mathbf{r},\mathbf{k}) - E_{n}(\mathbf{k})  
\, \vert
{\partial}_{\mathbf{k}}
u_{n}(\mathbf{k})\right\rangle 
\, \right]. 
\end{eqnarray}
By then assuming that the states, \thinspace ${\partial}_{\mathbf{k}}
u_{n}(\mathbf{r}, \mathbf{k})$  and 
 ${ u_m(\mathbf{r}, \mathbf{k}') }$, as well as the Hamiltonian  ${ H_{k}(\mathbf{r},\mathbf{k}) }$,  are invariant  with respect to real-space translations by $ \mathbf{R} $   (with  $ \mathbf{R} $ being  the real-space lattice vectors),  the quantity 
${
 \left\langle u_{m}(\mathbf{k}') 
 \vert \,
 e^{ \displaystyle i \mathbf{(k-k').r}} 
 \,
 ( 
 H_{k}(\mathbf{r},\mathbf{k}) - E_{n}(\mathbf{k})  
 ) 
 \, \vert
 {\partial}_{\mathbf{k}}
 u_{n}(\mathbf{k})\right\rangle  
}$
truncates into a unit cell expression given from
\begin{eqnarray}       \nonumber   \label{ap24}
&&
\!  \!  \!  \!  \!  \!  \!  \!  \!  \!  \!  \!  \!  \!  \!  \!  \!  \!  \!  \!  \!  \!  \!  \!  \!  \!  \!  \!  \!  \!  \!  \!  \!  \!  \!   
 \left\langle u_{m}(\mathbf{k}') 
\vert \,
e^{ \displaystyle i \mathbf{(k-k').r}} 
\,
( 
H_{k}(\mathbf{r},\mathbf{k}) - E_{n}(\mathbf{k})  
) 
\, \vert
{\partial}_{\mathbf{k}}
u_{n}(\mathbf{k})\right\rangle
=
\\*   \nonumber
&& 
\, \, \, \, \, \, \, \,  \, \, \, \, \, \, \, \,  \, \, \, \, \, \, \, \,  \, \, \, \, \, \, \, \,  \, \, 
=
\sum_{\mathbf{R}}
e^{ \displaystyle i \mathbf{(k-k').R}}
\,  
\left\langle u_{m}(\mathbf{k}') 
\vert \,
e^{ \displaystyle i \mathbf{(k-k').r}} 
\,
( 
H_{k}(\mathbf{r},\mathbf{k}) - E_{n}(\mathbf{k})  
) 
\, \vert
{\partial}_{\mathbf{k}}
u_{n}(\mathbf{k})\right\rangle_{cell}
\\*   
&& 
\, \, \, \, \, \, \, \,  \, \, \, \, \, \, \, \,  \, \, \, \, \, \, \, \,  \, \, \, \, \, \, \, \,  \, \, 
=
N
 \delta_{\mathbf{k}', \mathbf{k}}
\,  
\left\langle u_{m}(\mathbf{k}') 
\vert 
\,
e^{ \displaystyle i \mathbf{(k-k').r}} 
\,
( 
H_{k}(\mathbf{r},\mathbf{k}) - E_{n}(\mathbf{k})  
) 
\, \vert
{\partial}_{\mathbf{k}}
u_{n}(\mathbf{k})\right\rangle_{cell},
\end{eqnarray}
where $N$  is the total number of unit cells enclosed by the volume $ V $ of the system,   while \thinspace  $\mathbf{k}' $ \thinspace and  \thinspace $\mathbf{k} $  \thinspace are assumed to lie in the first Brillouin zone.
By replacing \thinspace   Eq.~(\ref{ap24}) \thinspace 
into
\thinspace   Eq.~(\ref{apb10h}) \thinspace 
we find
\begin{eqnarray} \label{apb10i} \nonumber
\mathbf{m}_n (\mathbf{k})& =&
- \frac{e}{2c \hbar }
\text{Im} 
\left[  \, 
\left\langle \widetilde{{\partial}_{\mathbf{k}}}u_{n}(\mathbf{k})
\vert 
\, \times \,
\left( 
H_{k}(\mathbf{r},\mathbf{k}) - E_{n}(\mathbf{k})  
\right)  
\vert 	
{\partial}_{\mathbf{k}}
u_{n}(\mathbf{k})\right\rangle 
\, \right] 
\nonumber \\*[6pt]   \nonumber
&&
\! \! \! \!  \! \! \! \!  \! \! \! \!  \! \! \! \!  \! \! \! \!  \! \! \! \!  
+ 
\frac{e}{2c \hbar}
\text{Im}
\left[   \,
\sum_{m, \mathbf{k}'}^{HS}
\left\langle 
\partial_{\mathbf{k}} \Psi_{n}(\mathbf{k})
\vert
\Psi_{m}(\mathbf{k}')
\right\rangle
\times  \
e^{ \displaystyle i (\Lambda(\mathbf{k})-\Lambda(\mathbf{k'}))} 
\,
N \ \delta_{\mathbf{k}', \mathbf{k}}
\,  
\left\langle u_{m}(\mathbf{k}') 
\vert 
\,
e^{ \displaystyle i \mathbf{(k-k').r}} 
\,
( \, 
H_{k}(\mathbf{r},\mathbf{k}) - E_{n}(\mathbf{k}) )  
\, \vert
{\partial}_{\mathbf{k}}
u_{n}(\mathbf{k})\right\rangle_{cell} 
\, \right] 
\nonumber \\*  [4pt]
&&  
-
\frac{e}{2c \hbar}
\text{Im} 
\left[  \,
\left\langle
\partial_{\mathbf{k}} \Psi_{n}(\mathbf{k}) 
\vert 
\Psi_{n}(\mathbf{k})
\right\rangle
\times 
\left\langle u_{n}(\mathbf{k}) 
\vert \,
H_{k}(\mathbf{r},\mathbf{k}) - E_{n}(\mathbf{k})  
\, \vert
{\partial}_{\mathbf{k}}
u_{n}(\mathbf{k})\right\rangle 
\frac{}{} \right],
\end{eqnarray}
\\
which finally gives
\begin{eqnarray} \label{apb10j} \nonumber
\mathbf{m}_n (\mathbf{k}) &=&
- \frac{e}{2c \hbar }
\text{Im} 
\left[ \, 
\left\langle \widetilde{{\partial}_{\mathbf{k}}}u_{n}(\mathbf{k})
\vert 
\, \times \,
\left( 
H_{k}(\mathbf{r},\mathbf{k}) - E_{n}(\mathbf{k})  
\right)  
\vert 	
{\partial}_{\mathbf{k}}
u_{n}(\mathbf{k})\right\rangle 
\, \right]
\nonumber \\*[4pt] 
&& -
\frac{e}{2c \hbar}
\text{Im}
\left[  \,
\sum_{m \neq n }^{\text{HS}} 
\left\langle 
\Psi_{n}(\mathbf{k})
\vert
\partial_{\mathbf{k}}
\Psi_{m}(\mathbf{k})
\right\rangle
\times
\left\langle u_{m}(\mathbf{k}) 
\vert \,
( \, 
H_{k}(\mathbf{r},\mathbf{k}) - E_{n}(\mathbf{k})  
\, \vert
{\partial}_{\mathbf{k}}
u_{n}(\mathbf{k})\right\rangle 
\, \right],
\nonumber \\*  
\end{eqnarray}
where we have used \thinspace
${ \left\langle 
\partial_{\mathbf{k}} \Psi_{n}(\mathbf{k})
\vert \Psi_{m}(\mathbf{k})
\right\rangle
= -
\left\langle 
\Psi_{n}(\mathbf{k})
\vert 
\partial_{\mathbf{k}} 
\Psi_{m}(\mathbf{k})
\right\rangle
}$
\thinspace  that is valid due to ${ m \neq n}$,  \vspace{2pt}  as well as,
${
N 
\left\langle u_{m}(\mathbf{k}) 
\vert \,
( \, 
H_{k}(\mathbf{r},\mathbf{k}) - E_{n}(\mathbf{k})  
\, \vert
{\partial}_{\mathbf{k}}
u_{n}(\mathbf{k})\right\rangle_{cell}
=
\left\langle u_{m}(\mathbf{k}) 
\vert \,
( \, 
H_{k}(\mathbf{r},\mathbf{k}) - E_{n}(\mathbf{k})  
\, \vert
{\partial}_{\mathbf{k}}
u_{n}(\mathbf{k})\right\rangle
}$.
Eq.~(\ref{apb10j}) \thinspace is \thinspace Eq.~(\ref{e18h}) \thinspace of   \vspace{2pt}  the main text.
As a final step, we find an expression for \thinspace
${ \left\langle \psi_{n}(\mathbf{k}) 
\vert \partial_{\mathbf{k}} 
\psi_{m}(\mathbf{k})
\right\rangle  }$, \thinspace
provided that $ n \neq m $, and then replace it in the sum of \thinspace  Eq.~(\ref{apb10j}). This is accomplished by the off-diagonal Hellmann-Feynman theorem that we derive in Appendix \ref{a3}.
\end{widetext}

\section{ Derivation of the off-diagonal 
Hellmann-Feynman theorem and the matrix elements 
${ \left\langle \psi_{n}(\mathbf{k}) 
\vert \partial_{\mathbf{k}} 
\psi_{m}(\mathbf{k})
\right\rangle  }$
} \label{a3}

We develop an off-diagonal Hellmann-Feynman theorem by starting from the eigenvalue equation
\begin{equation} \label{apc1}
\, ( H(\mathbf{r}) - E_m(\mathbf{k}) ) \,
\left| 
\psi_{m}(\mathbf{k}) \right\rangle 
= 0.
\end{equation}
where ${ H(\mathbf{r}) }$ is the initial system's Hamiltonian. Specifically, for the purpose of calculations of this work, we use the initial Hamiltonian of the system which does not depend 
on the wavevector ${ \mathbf{k} }$, \thinspace that is, ${ \partial_{\mathbf{k}}H(\mathbf{r})=0 }$. 
The result that we derive, is easily extended to include a Hamiltonian that has explicit parameter dependence by simply adding to it the term that has the derivative of the Hamiltonian with respect to the parameter.  

By assuming that the crystal momentum takes continuous values, 
we act with the momentum gradient operator \thinspace $\partial_{\mathbf{k}}$ \thinspace on  Eq.~(\ref{apc1}) obtaining
\begin{equation} \label{apc2}
- \partial_{\mathbf{k}}
E_m(\mathbf{k}) \left|  
\psi_{m}(\mathbf{k}) \right\rangle  
\, + \,
( H(\mathbf{r}) - E_m(\mathbf{k}) )  \left|  
\partial_{\mathbf{k}}
\psi_{m}(\mathbf{k}) \right\rangle 
= 0.
\end{equation}
and then take the inner product of \thinspace Eq.~(\ref{apc2}) \thinspace with \thinspace ${ \left\langle \psi_n(\mathbf{k}) \right|  }$  \thinspace which gives 
\begin{equation} \label{apc3}
- \partial_{\mathbf{k}}
E_m(\mathbf{k}) \delta_{nm}  
\, + \,
\left\langle \psi_n(\mathbf{k}) \vert
\, 
( H(\mathbf{r}) - E_m(\mathbf{k}) )  \vert 
\partial_{\mathbf{k}}
\psi_{m}(\mathbf{k}) \right\rangle 
= 0.
\end{equation}

We now take into account a possible anomaly of the momentum gradient operator due to the non-Hermitian effect, that emerges whenever the gradient operator \thinspace $\partial_{\mathbf{k}}$ \thinspace breaks the domain of definition \thinspace ${ \mathit{D}_H }$ \thinspace of the Hamiltonian \thinspace $H(\mathbf{r})$. 
In this framework, the wavefunctions \thinspace $ \psi_{m}(\mathbf{r},\mathbf{k}) $ \thinspace and 
\thinspace  
$ \partial_{\mathbf{k}} \psi_{m}(\mathbf{r},\mathbf{k}) $
\thinspace 
fulfill different boundary conditions over the edges of the system, and as a result they don't belong within the same domain of definition, that is, 
 \thinspace 
${ \psi_{m}(\mathbf{r},\mathbf{k}) \in \mathit{D}_H }$ \thinspace while
\thinspace ${ \partial_{\mathbf{k}} \psi_{m}(\mathbf{r},\mathbf{k}) \notin \mathit{D}_H }$. Therefore, whenever the non-Hermitian effect emerges, the term \thinspace 
${ \left\langle \psi_{n}(\mathbf{k}) \vert 
\left( 
H(\mathbf{r}) - E_m(\mathbf{k})
\right) 
\vert \partial_{\mathbf{k}} \psi_{m}(\mathbf{k}) \right\rangle
}$ \thinspace entering \thinspace Eq.~(\ref{apc3}) \thinspace is not zero as a result of the following non-trivial inequality 
\begin{eqnarray*}  
\left\langle H_{k}(\mathbf{r},\mathbf{k}) \psi_{n}(\mathbf{k}) \vert  
\partial_{\mathbf{k}} \psi_{m}(\mathbf{k}) \right\rangle 
& = &
\left\langle \psi_{n}(\mathbf{k}) \vert  
H_{k}(\mathbf{r},\mathbf{k})^{+} \partial_{\mathbf{k}} \psi_{m}(\mathbf{k}) \right\rangle
\\*
& \neq & 
\left\langle \psi_{n}(\mathbf{k}) \vert  
H_{k}(\mathbf{r},\mathbf{k}) \, \partial_{\mathbf{k}} \psi_{m}(\mathbf{k}) \right\rangle.
\end{eqnarray*}
We treat this non-Hermitian effect by expressing the term
\thinspace  
$ \left\langle \psi_{n}(\mathbf{k}) \vert  
H(\mathbf{r}) \, \partial_{\mathbf{k}} \psi_{m}(\mathbf{k}) \right\rangle $   
\thinspace as
\begin{eqnarray} \label{apc4} \nonumber 
\left\langle \psi_{n}(\mathbf{k}) \vert  
H(\mathbf{r}) \, \partial_{\mathbf{k}} \psi_{m}(\mathbf{k}) \right\rangle 
& =  &
\left\langle H(\mathbf{r}) \psi_{n}(\mathbf{k}) \vert  
\partial_{\mathbf{k}} \psi_{m}(\mathbf{k}) \right\rangle 
\nonumber  \\*
&& - \,
{\mathbf{S}}_{nm}(\mathbf{k})
\end{eqnarray}
where the \thinspace ${\mathbf{S}}_{nm}(\mathbf{k})$
\thinspace term represents the non-Hermitian effect 
and is a boundary quantity. Its explicit boundary integral form is given below. In this respect, by taking into account \thinspace Eq.~(\ref{apc4}),
\thinspace Eq.~(\ref{apc3}) \thinspace takes the form
\begin{eqnarray} \label{apc6} \nonumber
\partial_{\mathbf{k}} 
E_m(\mathbf{k}) \, \delta_{nm}  
& = & 
( E_n(\mathbf{k}) - E_m(\mathbf{k}) ) 
\left\langle \psi_n(\mathbf{k}) \vert
\partial_{\mathbf{k}}
\psi_{m}(\mathbf{k}) \right\rangle 
\nonumber  \\*
&& - \,
{\mathbf{S}}_{nm}(\mathbf{k}),
\end{eqnarray}
which for \thinspace ${ n \neq m}$ \thinspace gives \thinspace
${ \left\langle \psi_{n}(\mathbf{k}) 
\vert \partial_{\mathbf{k}} 
\psi_{m}(\mathbf{k})
\right\rangle  }$ \thinspace as a function of \thinspace ${\mathbf{S}}_{nm}(\mathbf{k})$, \thinspace given by
\begin{equation} \label{apc6d} 
\left\langle \psi_n(\mathbf{k}) \vert
\partial_{\mathbf{k}}
\psi_{m}(\mathbf{k}) \right\rangle 
= \frac{ {\mathbf{S}}_{nm}(\mathbf{k}) }
{ ( E_n(\mathbf{k}) - E_m(\mathbf{k}) ) }.
\end{equation}

We now give the explicit integral form of  \thinspace ${\mathbf{S}}_{nm}(\mathbf{k})$. Specifically, (i) by using 
\thinspace Eq.~(\ref{apc4}) \thinspace
as the definition of the \thinspace ${\mathbf{S}}_{nm}(\mathbf{k})$,
\thinspace (ii) by working in the position representation, and (iii) after an integration by parts (assuming a 3D system), the matrix elements of the non-Hermitian term \thinspace ${\mathbf{S}}_{nm}(\mathbf{k})$
\thinspace are always transformed, due to symmetry of the integrands, into a boundary quantity that is given by 
\begin{equation} \label{apc6c} 
\mathbf{S}_{nm}(\mathbf{k})
=
\frac{i \hbar}{2} \oiint_S 
\mathbf{n}\!\cdot\! 
\left( \,
( \mathbf{v} \, \psi_{n} )^{\displaystyle *} 
+ \psi_{n}^{\displaystyle *} \, \mathbf{v}  
\, \right) \! 
\, \partial_{\mathbf{k}} \psi_{m}
\, dS,	
\end{equation}
\linebreak
where \thinspace ${\psi_m = \psi_m(\mathbf{r},\mathbf{k})}$
\thinspace are the Bloch eigenfunctions, $\mathbf{v}$ is the standard velocity operator and $\mathbf{n}$ is the unit vector that is locally normal to the surface $S$. The corresponding abstract form of   
\thinspace $ {\mathbf{S}}_{nm}(\mathbf{k}) $
\thinspace is given by
\begin{equation} \label{apc5} \nonumber
{\mathbf{S}}_{nm}(\mathbf{k})
= 
\left\langle 
\Psi_{n}(\mathbf{k}) \vert  
\left( 
H(\mathbf{r})^{+} - H(\mathbf{r})
\, \right)  
\partial_{\mathbf{k}} \Psi_{m}(\mathbf{k}) 
\right\rangle.
\end{equation}

It is now intuitively useful to give the extension of  \thinspace Eq.~(\ref{apc6}) \thinspace  to the one that includes the explicit  dependence of the Hamiltonian on a static parameter, in order to show the necessity of a non-Hermitian boundary term that solves a \textquotedblleft paradox\textquotedblright  \thinspace concerning the band theory.
First we present the \textquotedblleft paradox\textquotedblright \thinspace and then we show how this is resolved by taking into account the non-Hermitian term \thinspace ${\mathbf{S}}_{nm}(\mathbf{k})$.
When one uses the cell periodic eigenstates and applies the Hellmann-Feynman theorem into the equation \thinspace 
${ \left\langle u_n(\mathbf{k}) \vert
\, 
H_k(\mathbf{r}, \mathbf{k}) \vert 
u_{n}(\mathbf{k}) \right\rangle
=
E_n(\mathbf{k}) }$, \thinspace one finds the standard velocity expectation value with respect to the dispersion relation derivative, that is \thinspace 
${ \left\langle u_n(\mathbf{k}) \vert
\, 
\partial_{\mathbf{k}} H_k(\mathbf{r}, \mathbf{k}) \vert 
u_{n}(\mathbf{k}) \right\rangle
=
\partial_{\mathbf{k}} E_n(\mathbf{k}) \neq 0 }$. On the other hand, if one uses the Bloch eigenstates, that is applies the Hellmann-Feynman theorem into the equation \thinspace 
${ \left\langle \psi_n(\mathbf{k}) \vert
\, 	
H(\mathbf{r}) \vert 
\psi_{n}(\mathbf{k}) \right\rangle	
=
E_n(\mathbf{k}) }$, \thinspace
one deduces that 
${ \partial_{\mathbf{k}} E_n(\mathbf{k}) = 0  }$. These sorts of subtleties are attributed to non-Hermitian boundary terms that are not properly taken into account. Specifically, by assuming a Hamiltonian \thinspace ${ H(\mathbf{r}, \mathbf{R}) }$, \thinspace where ${ \mathbf{R} }$ is a general parameter, then    
\thinspace Eq.~(\ref{apc6}) \thinspace takes the form
\begin{eqnarray} \label{apc6a} \nonumber
&& \partial_{\mathbf{R}} 
E_m(\mathbf{R}) \, \delta_{nm}  
=
( E_n(\mathbf{R}) - E_m(\mathbf{R}) ) 
\left\langle \psi_n(\mathbf{R}) \vert
\partial_{\mathbf{R}}
\psi_{m}(\mathbf{R}) \right\rangle 
\nonumber  \\*[3pt] 
&& 
\ \ \ \ \ \ \ \ \ \ \ \ \ \ \ \ \
+
\left\langle \psi_n(\mathbf{R}) \vert
\, \partial_{\mathbf{R}}H(\mathbf{r}, \mathbf{R}) \,
\vert
\psi_{m}(\mathbf{R}) \right\rangle 
- \,
{\mathbf{S}}_{nm}(\mathbf{R})
\nonumber \\*
\end{eqnarray}
where \thinspace 
${ \left| \psi_n(\mathbf{R}) \right\rangle }$ \thinspace are the eigenstates of the Hamiltonian. 
By way of an example, using
the diagonal form of \thinspace Eq.~(\ref{apc6}) \thinspace
and assuming \thinspace ${ \mathbf{R}\equiv \mathbf{k}}$ \thinspace as well as an initial Hamiltonian  ${ H(\mathbf{r}) }$ we find \thinspace 
${ \partial_{\mathbf{k}} 
E_n(\mathbf{k}) 
=- \,
{\mathbf{S}}_{nn}(\mathbf{k}) }$. 
\thinspace In this manner, one 
will deduce that the bands are always flat (or equivalently that the group velocity is always zero) if the non-Hermitian boundary contribution is not taken into account, which will lead to an apparent 
\textquotedblleft paradox\textquotedblright.  \thinspace
Using now the Bloch form eigenstate 
${ \left|\Psi_{n}(\mathbf{k})\right\rangle = 
e^{ \displaystyle i \mathbf{k.r}} \,
e^{ \displaystyle i \Lambda(\mathbf{k}) }
\left| u_{n}(\mathbf{k})\right\rangle }$
into the boundary term \thinspace 
${ {\mathbf{S}}_{nn}(\mathbf{k})
=
\left\langle H(\mathbf{r}) \psi_{n}(\mathbf{k}) \vert  
\partial_{\mathbf{k}} \psi_{n}(\mathbf{k}) \right\rangle 
-
\left\langle \psi_{n}(\mathbf{k}) \vert  
H(\mathbf{r}) \partial_{\mathbf{k}} \psi_{n}(\mathbf{k}) \right\rangle 
}$, as well as by taking into account 
\thinspace 
Eq.~(\ref{e18b})
\thinspace 
and the explicit form of the boundary velocity definition \thinspace Eq.~(\ref{e3}) \thinspace 
of the main text,
the relation between the 
boundary velocity and the standard (group) velocity for stationary states \thinspace 
${ \left\langle\mathbf{v}_{b}\right\rangle_n
=-	
\left\langle\mathbf{v}\right\rangle_n}$
\thinspace
is restored and 
the \textquotedblleft paradox\textquotedblright  \thinspace is resolved.


\begin{thebibliography}{36}
\bibitem{b33b} The one-band covariant derivative \thinspace ${ \widetilde{{\partial}_{\mathbf{k}}} }$ \thinspace can alternatively  be defined as  \thinspace  ${ \widetilde{{\partial}_{\mathbf{k}}}=
(\, 1- \left| u_n(\mathbf{k})\right\rangle \! \left\langle u_n(\mathbf{k})\right| \, ) \, {{\partial}_{\mathbf{k}}}  }$ 
\thinspace without changing any of the final results.
\end{thebibliography}
\end{document}